\documentclass[11pt,preprint]{aastex}

\newcommand\en{\end{equation}}

\shorttitle{Silicate Evolution}
\shortauthors{Sicilia-Aguilar et al.}

\begin{document}

\title{Silicate Dust in Evolved Protoplanetary Disks: Growth, Sedimentation, and Accretion }

\author{Aurora Sicilia-Aguilar\altaffilmark{1}, Lee W. Hartmann\altaffilmark{2}, Dan Watson\altaffilmark{3}, }
\author{Chris Bohac\altaffilmark{3}, Thomas Henning\altaffilmark{1},Jeroen Bouwman\altaffilmark{1}}

\altaffiltext{1}{Max-Planck-Institut f\"{u}r Astronomie, K\"{o}nigstuhl 17, 69117 Heidelberg, Germany}
\altaffiltext{2}{University of Michigan, 830 Dennison 500 Church St., Ann Arbor, MI 48109}
\altaffiltext{3}{Department of Physics and Astronomy, University of Rochester, Rochester, NY 14627-0171}

\email{sicilia@mpia.de}

\begin{abstract}

We present the Spitzer IRS spectra for 33 young stars in Tr 37 and NGC 7160. The sample 
includes the high- and intermediate-mass stars with MIPS 24$\mu$m excess, the only
known active accretor in the 12 Myr-old cluster NGC 7160, and 19 low-mass stars with disks 
in the 4 Myr-old cluster Tr 37. We examine the 10$\mu$m silicate feature, present 
in the whole sample of low-mass star and in 3 of the high- and intermediate-mass targets,
and we find that PAH emission is detectable only in the Herbig Be star. We analyze the
composition and size of the warm photospheric silicate grains by fitting the
10 $\mu$m silicate feature, and study the possible correlations between the silicate
characteristics and the stellar and disk properties (age, SED slope, accretion rate, 
spectral type). We find  indications of dust settling with age and of the effect of turbulent 
enrichment of the disk atmosphere with large grains. Crystalline grains are only small 
contributors to the total silicate mass in all disks, and do not seem to correlate with any 
other property, except maybe binarity. We also observe that spectra with very weak silicate 
emission are at least 3 times more frequent among M stars than among earlier spectral types, 
which may be an evidence of inner disk evolution. Finally, we find that 5 of the high- and 
intermediate-mass stars have SEDs and IRS spectra consistent with debris disk models 
involving planet formation, which could indicate debris disk formation at ages as early as 
4 Myr. 
\end{abstract}

\keywords{accretion disks --- planetary systems: protoplanetary disks --- 
stars: pre-main sequence }

\section{Introduction \label{intro}}

The dissipation of protoplanetary disks around low- and intermediate-mass
stars and the processes resulting in disk evolution and eventually planet 
formation are essential for the understanding of the formation of our own 
Solar System. More than two decades of observations have allowed to pin 
down the main timescales for the presence of protoplanetary disks. Optically 
thick, accreting disks are nearly universal at ages around 1 Myr, become very 
rare at or after 10 Myr, and are absent in objects older than 30 Myr (Strom et al. 
1989; Skrutskie et al. 1990; Lada et al. 2000; Carpenter 2000; Haisch et al. 2001; 
Armitage et al. 1999, 2003). The Spitzer Space Telescope has started to probe the 
processes of disk dissipation and evolution for large samples of targets (Porras 
et al. 2003; Meyer et al. 2004; Megeath et al. 2004) and to target the rare 
``transitional'' disks, which are presumably evolving from optically thick disks 
to debris-like disk via dust settling and grain growth in the inner disk, and 
probably via the opening of physical inner gaps (see the examples of CoKu Tau4 in 
Taurus, Forrest et al. 2004; D'Alessio et al. 2005; TW Hya, Calvet et al. 2002; 
Uchida et al. 2004; among others). 

Observations suggest that disk evolution occurs faster in the innermost regions, 
as already predicted by Hayashi et al. (1985). The evidence of inner disk
evolution is stronger in older regions, which also contain smaller disk fractions. 
In order to study large samples of evolved disks, we targeted the Cep OB2 Association 
(Patel et al. 1995, 1998). This bubble-shaped region, located at a distance of 900 pc 
(Contreras et al. 2002), contains the clusters Tr 37 and NGC 7160 (Platais et al. 1998).
Both clusters have been extensively studied with optical photometry and spectroscopy, and 
near- and mid-IR photometry (Spitzer IRAC and MIPS) in order to determine the cluster
members and their characteristics (Sicilia-Aguilar et al. 2004, 2005,2006a,2006b, 
from now on Paper I, II, III and IV, respectively). 

Our previous work revealed more than 180 low-mass members in Tr 37 (spectral types G to 
M2) and over 60 in NGC 7160, with average ages 4 and 12 Myr and accretion disk fractions
of $\sim$45\% and $\sim$4\%, respectively (Paper II, III), belonging to the epoch when 
most disk dissipation and planet formation is believed to occur (Haisch et al. 2001).
Combining observations at multiple wavelengths, interesting connections 
between the gas and dust evolution appear: grain growth and/or settling in 
the innermost disk affects nearly 95\% of the objects, and the accretion 
rates fall in many cases well below the median rates in younger regions, to the point 
that some ``transition objects''  (objects with inner physical and/or opacity gaps
resulting in no near-IR excesses at $\lambda \leq$ 6$\mu$m, but optically thick disks
at longer wavelengths; Paper III, IV) are consistent with no accretion (Paper IV).
The accretion rates for the accreting transition objects are usually very low ($\leq$10$^{-9}$ 
M$_\odot$/yr) compared to younger stars and to the bulk of accreting stars in Tr 37. This 
confirms their ``transitional'' stage, between Class II and Class III objects, and suggest 
a connection between gas and dust evolution. The transition objects represent about 10\% of 
the total number of disks in Tr 37, and maybe a larger fraction 
at older stages, when disks are rare (Paper III, IV). The small 
fraction of ``transition objects'' around low-mass stars suggests a relatively rapid (up to 
$\sim 10^6$ yr) timescale for the dissipation of the outer disk after the inner disk is 
cleared. These short lifetimes may be related either to photoevaporation caused by the
central star (Clarke et al. 2001; Alexander et al. 2006a,b; McCabe et al. 2006) or
to (giant) planet formation at few AU (Lin \& Papaloizou 1986, Bryden et al. 1999,
Quillen et al. 2004). However, the presence of accretion and, therefore, an inner gaseous
disk in roughly half of these objects, suggests that photoevaporation alone cannot explain 
this behavior, at least not in most low-mass stars (Paper IV). 

Among the high- and intermediate-mass stars (spectral types BAF) identified in Tr 37 and 
NGC 7160 ($\sim$60 in each cluster), our IRAC and MIPS data revealed the 
presence of a Herbig Be star in Tr 37 and several candidates in both 
clusters, with 24 $\mu$m excess suggesting either debris disks (formed by
second-generation dust after planetesimal or planet formation) or early-type ``transitional 
disks'' with large gaps. These disks would be among the few disks around BAF stars with 
well-determined ages (based on the low-mass cluster members) and, if they were confirmed to 
be debris disks, they would be among the youngest debris disks known. 

 In order to study the characteristics of dust grains, the presence of grain growth and 
settling, and the nature of potential debris disks and ``transition objects'', we have 
targeted a sample of objects in Cep OB2 with the IRS spectrograph onboard Spitzer (see the 
list of targets and their characteristics in Table \ref{target-tab}). Our sample contains 20 
low-mass stars (spectral types GKM) with different properties, and all the 13 high- and 
intermediate-mass stars (spectral types BAF) with any IRAC/MIPS mid-IR excess. In Section 
\ref{obs} we describe the data taking and reduction. In Section \ref{analysis} we present 
the collection of IRS spectra and fit the silicate feature at 10$\mu$m in order to estimate 
the grain composition and properties. In Section \ref{results} we compare the IRS data with 
the available optical and IR information, and finally we summarize our conclusions.

\section{Observations and Data Reduction \label{obs}}

We observed the Cep OB2 objects as part of the GTO program ``Spectroscopy of protostellar, 
protoplanetary and debris disks'' in the IRS Campaigns 15, 16, and 22.  They were observed 
with the nodded IRS Staring Mode.  We utilized all three orders of the IRS Short-Low (SL) 
spectrograph (5.3-15 $\mu$m, $\lambda$/$\Delta \lambda \sim$ 90) and all three orders of the 
IRS Long-Low (LL) spectrograph (15-40 $\mu$m, $\lambda$/$\Delta \lambda \sim$90).
This produces spectra covering the region from $\sim$5.4 $\mu$m to $\sim$35 $\mu$m, which contains
the most relevant dust features and traces the disk from $\sim$0.1 to $\sim$30 AU, approximately. 
In both clusters, we targeted all the high- and intermediate-mass
stars with spectral types BAF and mid-IR excess (8 in Tr 37, 5 in NGC 7160),
plus 19 low-mass stars in Tr 37 (about 1/8 of the total number of known members)
and the only known accretor in the 12 Myr-old cluster NGC 7160. 

Given that parts of the two clusters do still show diffuse nebulosity that
may result in some IR emission, we used the PRCS optical pickup on guide stars from the
catalog for pointing. The exposure times were calculated individually based on the mid-IR 
fluxes measured with IRAC and MIPS. In order to be able to study the silicate feature at 
10 $\mu$m, we required enough integration time to achieve a signal-to-noise ratio (S/N) of 
at least 30-50 in the 10 $\mu$m region. At longer wavelengths, our S/N requirements were only 
about 20, in order to optimize the GTO program.
Except for a couple of stars, which had very low mid-IR fluxes, we
achieved or surpassed our required S/N, being able not only to
study the 10 $\mu$m feature, but to identify some other crystalline
features in the 20 $\mu$m region as well (see Section \ref{seds-low}).

We reduced the spectra using the Spectral Modeling, Analysis, and Reduction 
Tool (SMART; Higdon et al. 2004).  We began with S12.0.2 BCD data from the 
SSC data reduction pipeline, which were already processed with stray light 
and flatfield corrections, and we departed from normal practice at several steps 
in order to achieve the best sensitivity and calibration. We corrected the pixels 
identified as NaN and rogue by interpolating the nearest non-bad neighbors
(since, in case of bright and variable background, the correction of response in the rogue 
pixels by the SSC-recommended procedure of sky subtraction is not efficient).
For each object that had multiple DCEs per exposure ID, the DCEs were averaged 
together.  Next, we subtracted the sky by subtracting the off-nod position of the 
same order.  We continued by extracting the spectra as a point-source for each 
nod position with a variable column extraction.  We chose a width of 4.046 pixels 
at the center of each order and we let the widths vary from 3.2 to 4.9 pixels in 
SL2, 3.6 to 4.4 pixels in SL bonus order, 2.7 to 5.4 pixels in SL1, 3.2 to 4.9 
pixels in LL2, 3.8 to 4.3 pixels in LL bonus order, and 2.5 to 5.5 pixels in LL1.
This custom variable-width extraction window fits very tightly to the spectrograph
point-spread function, so we can achieve good sensitivity. This point-source spectral 
extraction and calibration method was developed and refined for the IRS GTO surveys of 
Taurus, Chamaeleon and Ophiuchus.

After extraction, the spectra of each nod were divided by the nodded spectra of 
our identically extracted photometric low resolution standard star $\alpha$ Lac, and 
then multiplied by the template spectrum of $\alpha$ Lac (Cohen et al. 2003).  After reducing 
the data by this method, we merged the modules of each nod together, and then 
found the mean of the two nods.  We determined the error in our spectra from the 
difference between nods.  In a few of the objects, there are some small mismatches 
in flux between modules that are within our absolute spectrophotometric 
accuracy of 10\%. In these cases, we scaled the modules to match each other
(see Table \ref{scaling-tab} for scaling factors and notes on the reduction).

\section{Analysis \label{analysis}}

\begin{figure}
\epsscale{1.1}\plottwo{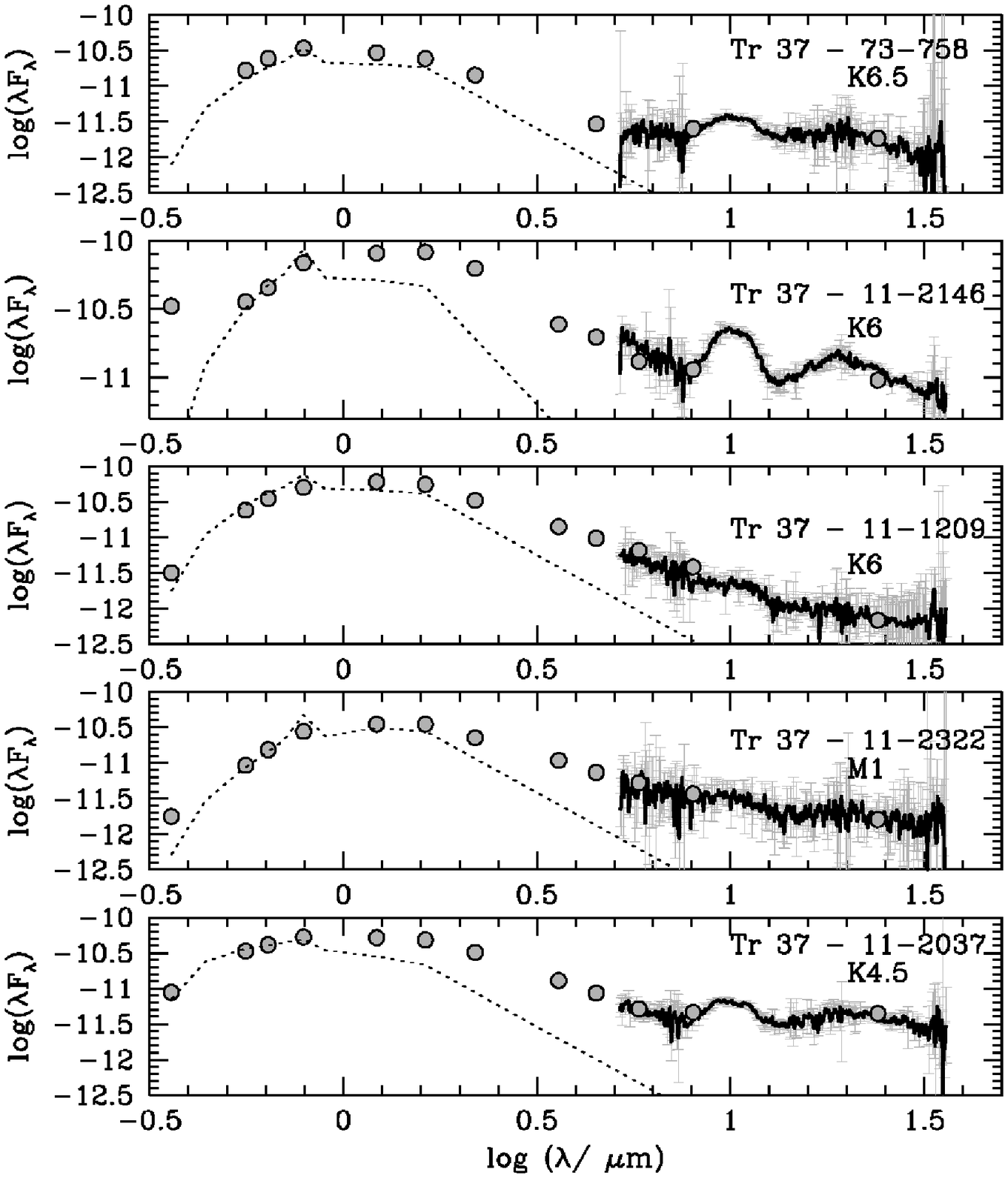}{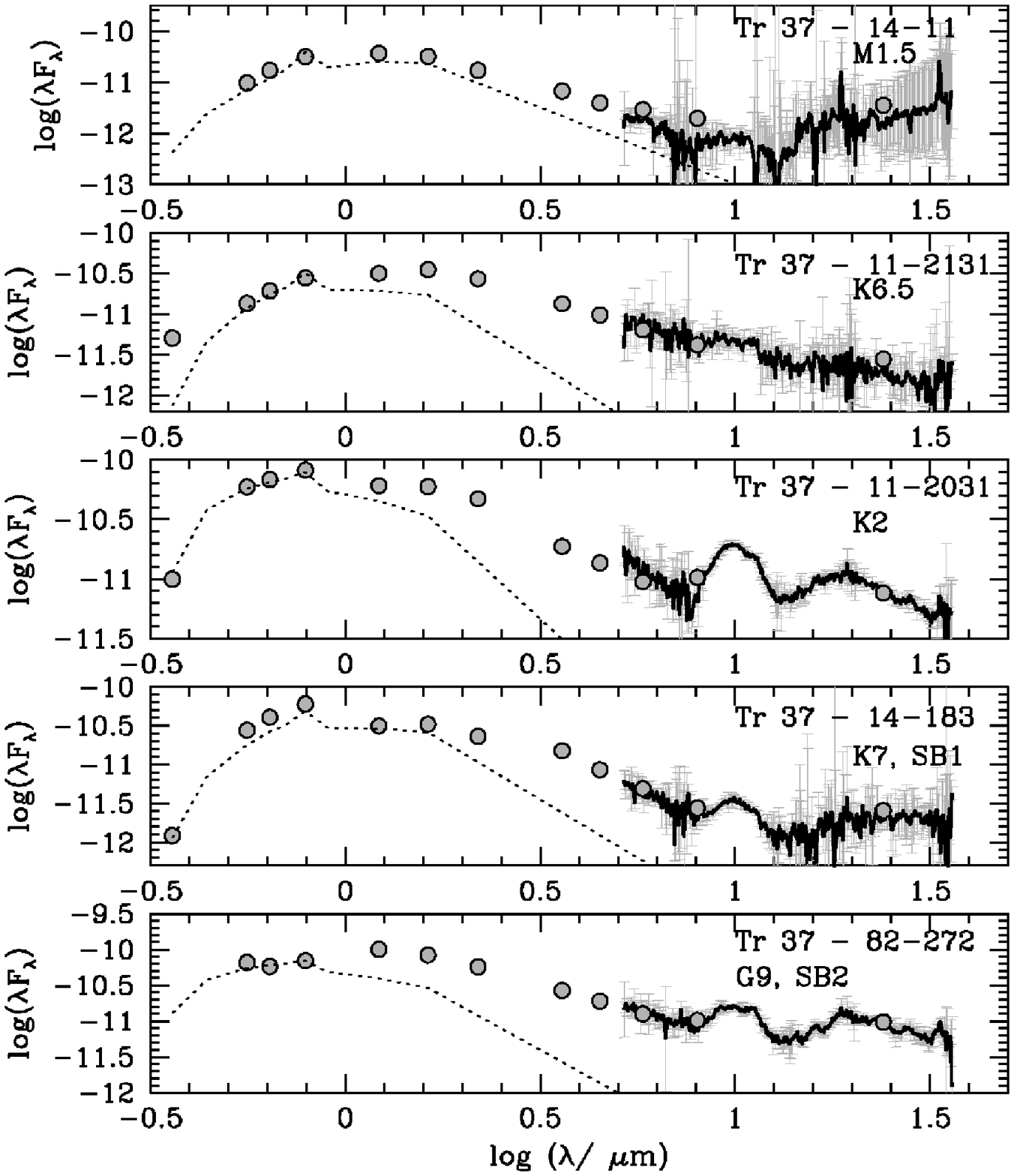}
\caption{Spectral Energy Distribution of low-mass stars in Cep OB2.
Optical (UVRI), 2MASS (JHK), IRAC (3.6-8.0 $\mu$m) and MIPS (24$\mu$m) data
is represented by dots. The IRS spectrum is displayed with its error.
A photosphere from Kenyon \& Hartmann (1995) is displayed for each
spectral type. The flux in log($\lambda$ F$_\lambda$) is displayed in
units of erg cm$^{-2}$ s$^{-1}$. 
 \label{sed1}}
\epsscale{1}
\end{figure}

\begin{figure}
\epsscale{1.1}\plottwo{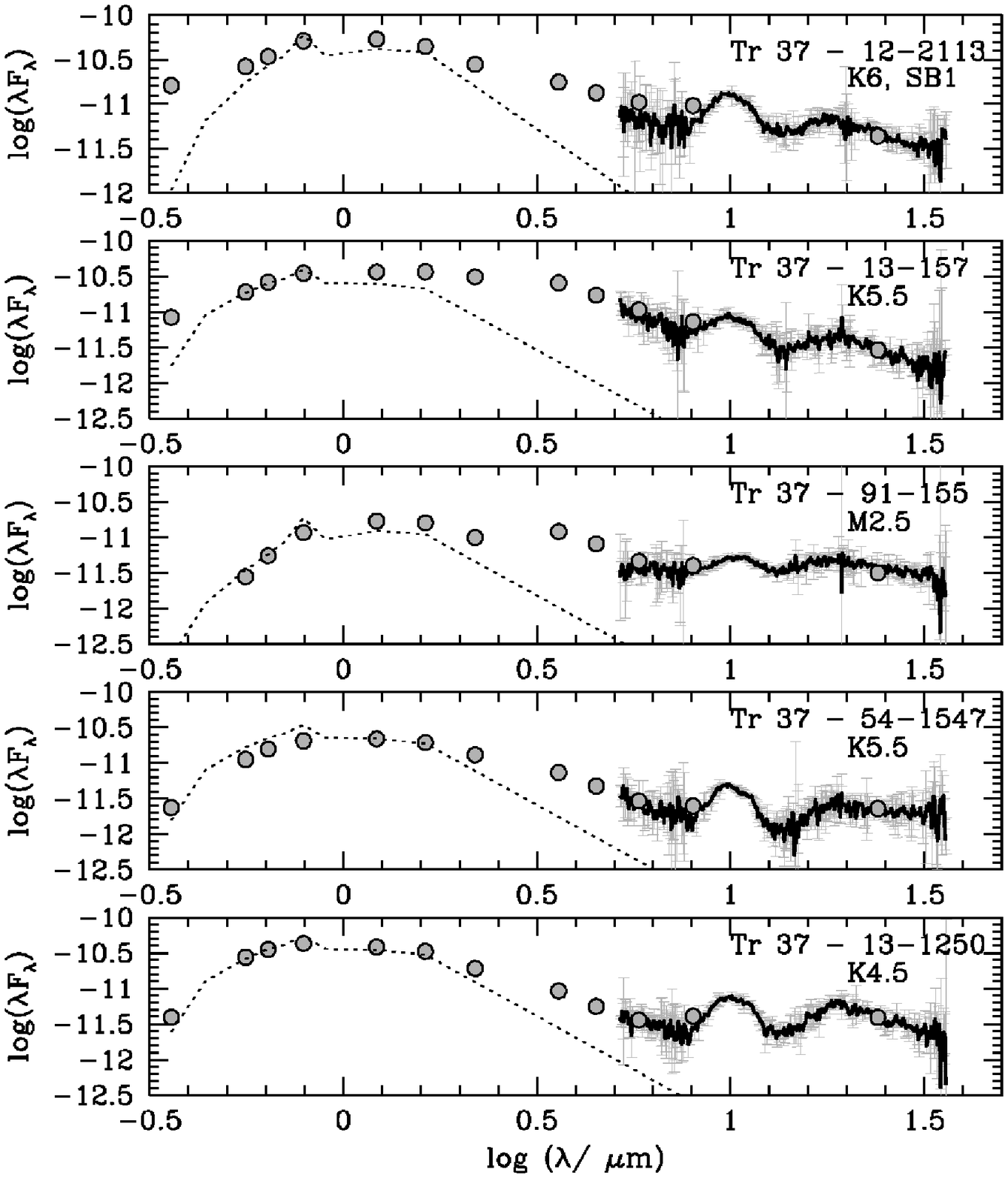}{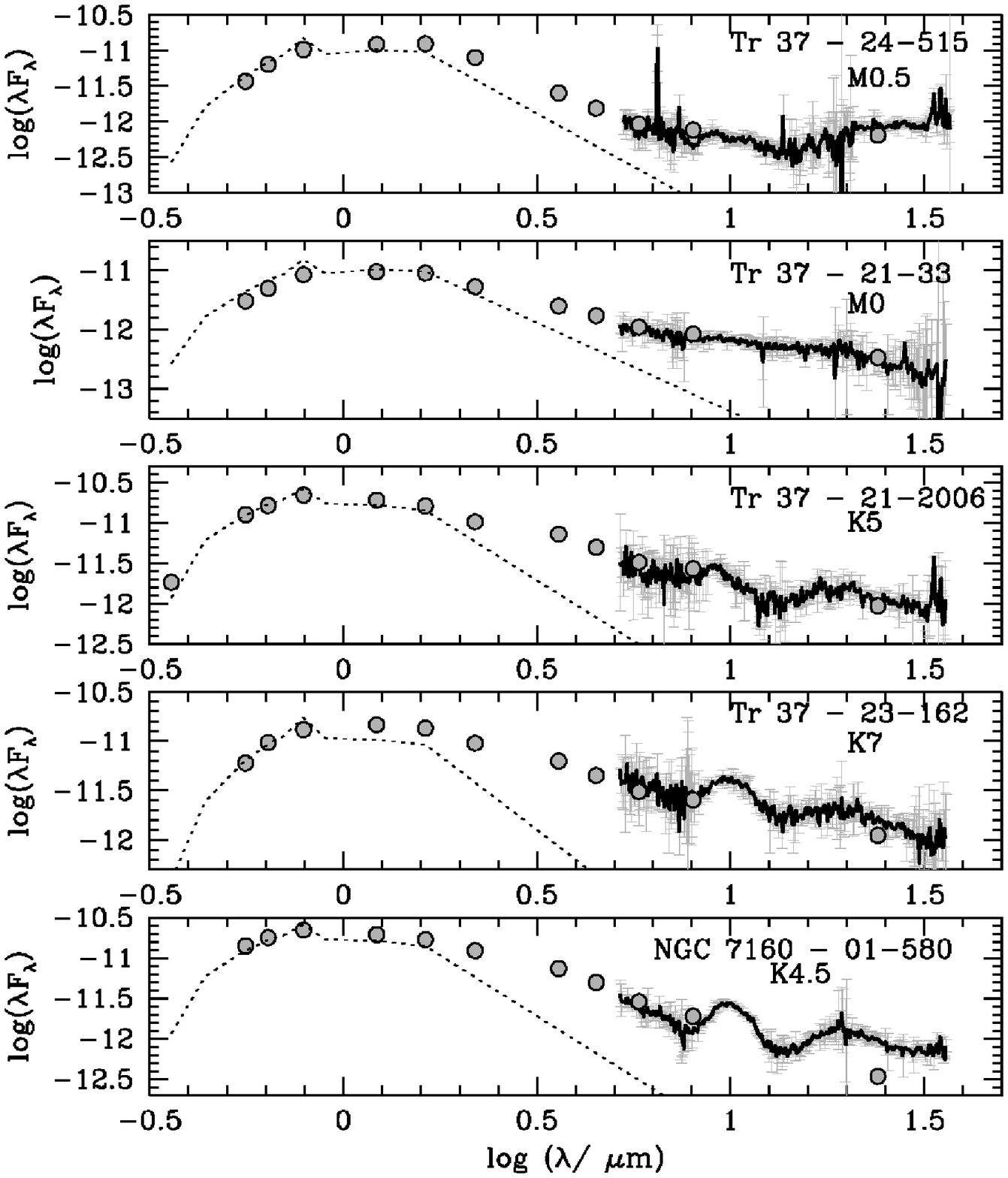}
\caption{Spectral Energy Distribution of low-mass stars in Cep OB2 (continued).
Caption as in Figure \ref{sed1}.
 \label{sed2}}
\epsscale{1}
\end{figure}

\begin{figure}
\epsscale{1.1}\plottwo{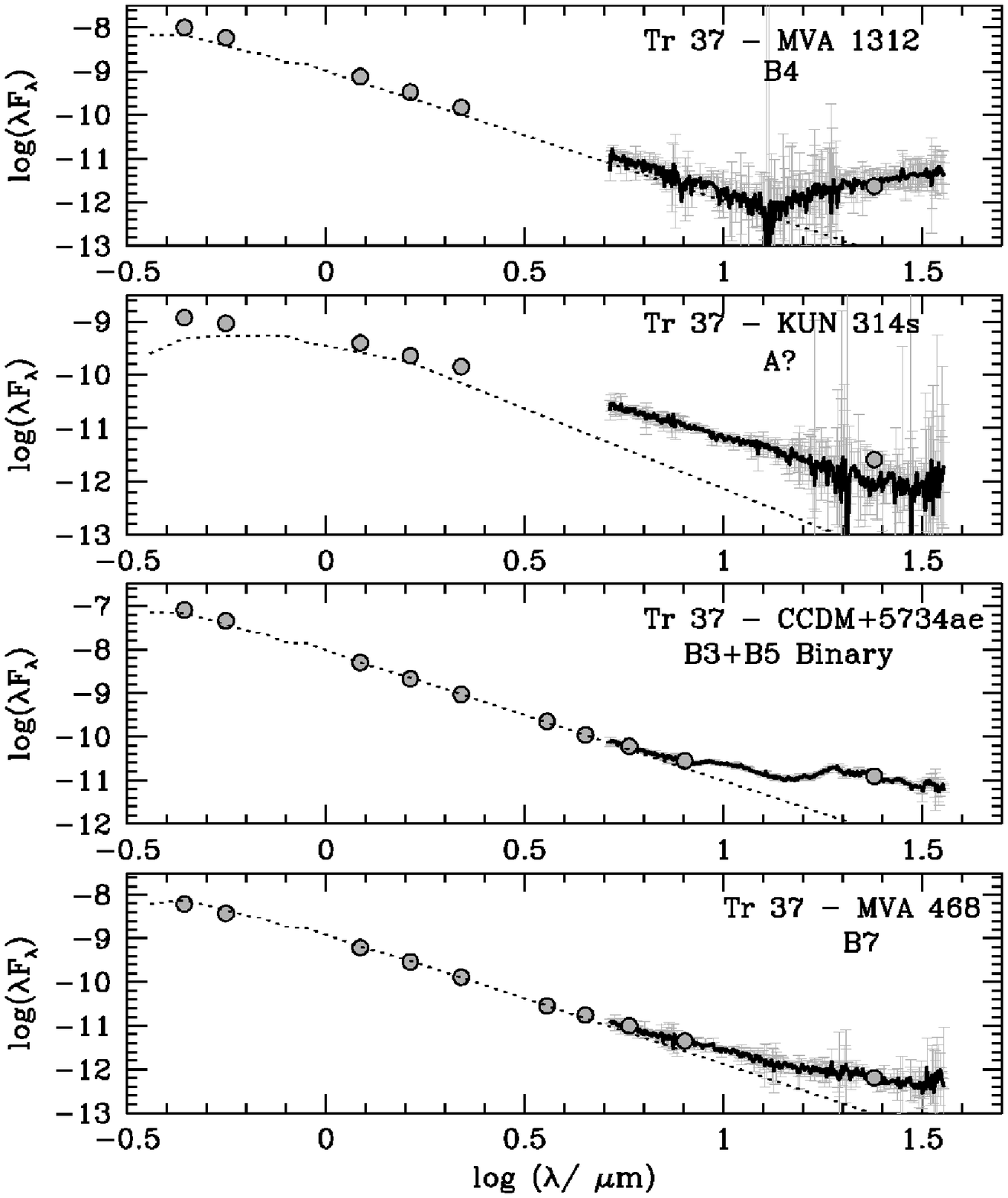}{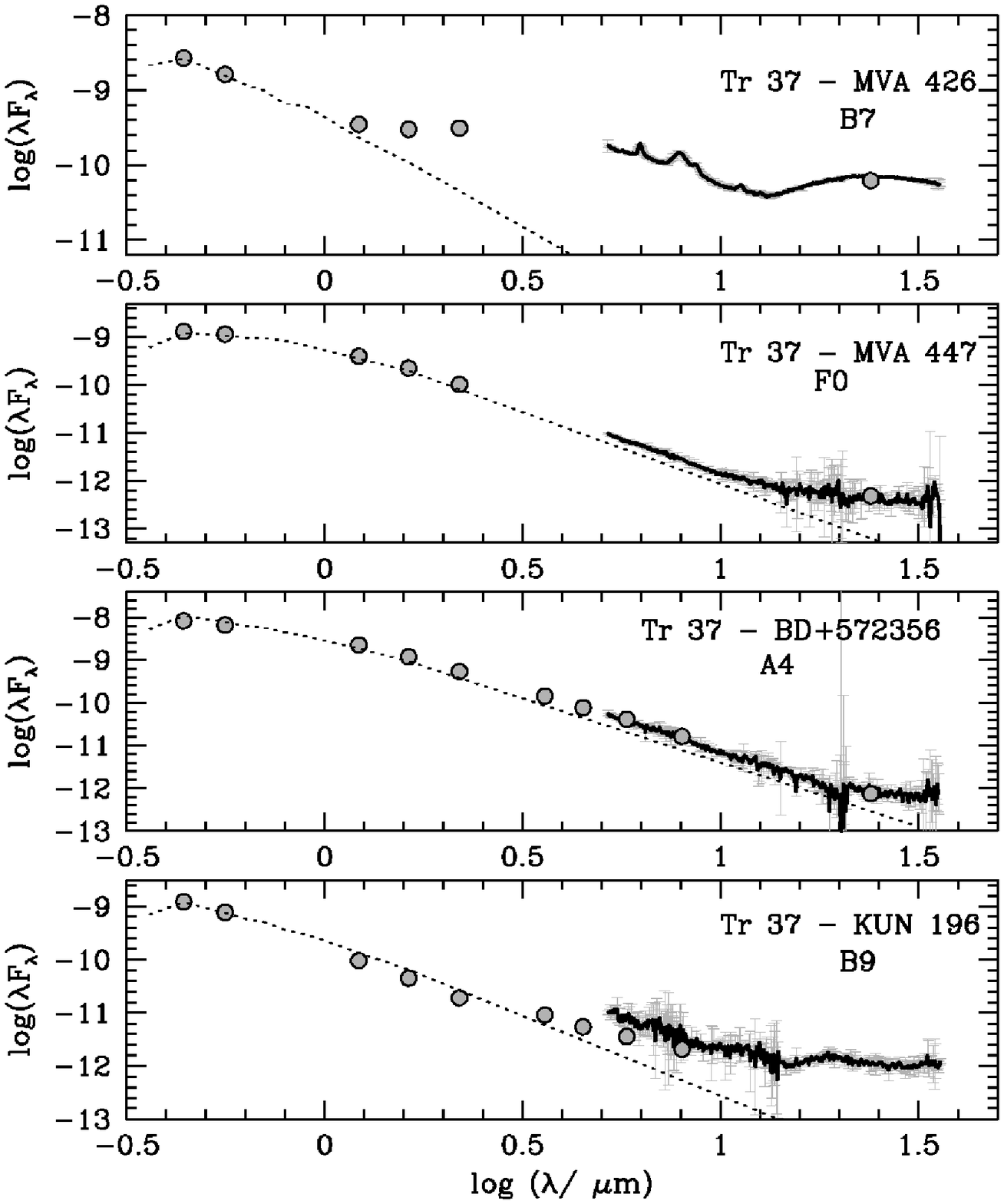}
\caption{Spectral Energy Distribution of intermediate- and high-mass stars in Tr 37.
Optical (UB), 2MASS (JHK), IRAC (3.6-8.0 $\mu$m) and MIPS (24$\mu$m) data
is represented by dots. The IRS spectrum is displayed with its error.
A photosphere from Kenyon \& Hartmann (1995) is displayed for each
spectral type. The flux in log($\lambda$ F$_\lambda$) is displayed in
units of erg cm$^{-2}$ s$^{-1}$. 
 \label{sed3}}
\epsscale{1}
\end{figure}

\begin{figure}
\epsscale{0.5}\plotone{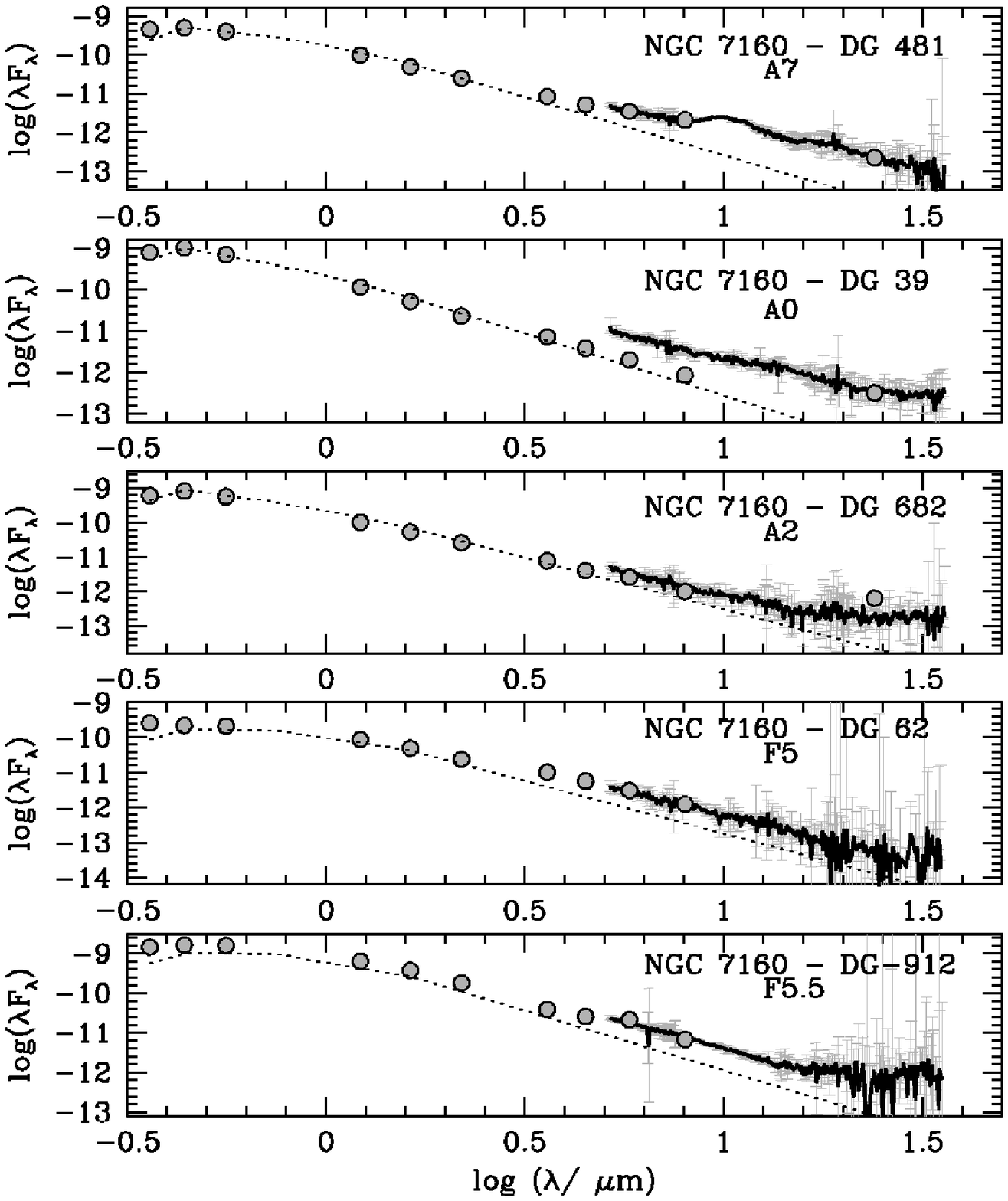}
\caption{Spectral Energy Distribution of intermediate- and high-mass stars in NGC 7160.
Optical (UBV), 2MASS (JHK), IRAC (3.6-8.0 $\mu$m) and MIPS (24$\mu$m) data
is represented by dots. The IRS spectrum is displayed with its error.
A photosphere from Kenyon \& Hartmann (1995) is displayed for each
spectral type. The flux in log($\lambda$ F$_\lambda$) is displayed in
units of erg cm$^{-2}$ s$^{-1}$. 
 \label{sed4}}
\epsscale{1}
\end{figure}

The IRS spectra of the Cep OB2 members are displayed in Figures \ref{sed1} to \ref{sed4}.
For comparison and completeness, we display the available fluxes in the optical and near-IR 
(UBVRI, JHK from 2MASS; Paper I, II), and IRAC/MIPS (from 3.6 to 24 $\mu$m; Paper III).
We also compare the resulting spectral energy distributions (SED) with the photospheric 
emission of a star with the same spectral type (Kenyon \& Hartmann 1995). 
It should be noted that all the IRS fluxes are in very good agreement with the IRAC/MIPS
ones, finding significant deviations (over the 3-sigma level) only for the 24 $\mu$m flux 
of the star 01-580 in NGC 7160, for KUN-196 at 4.5 $\mu$m, and for DG-39 at 4.5 and 8.0$\mu$m. 

\begin{figure}
\epsscale{1.1}\plottwo{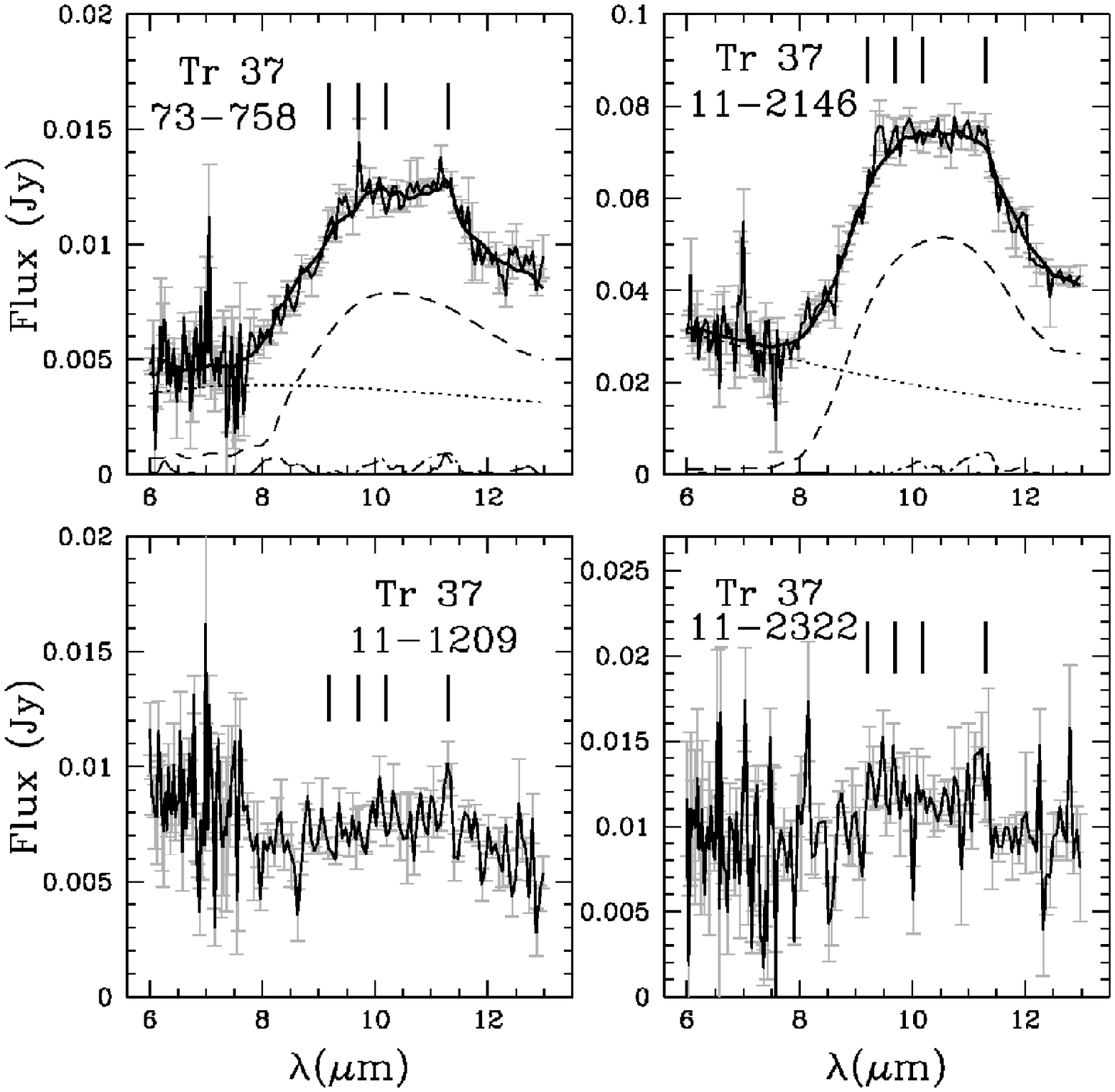}{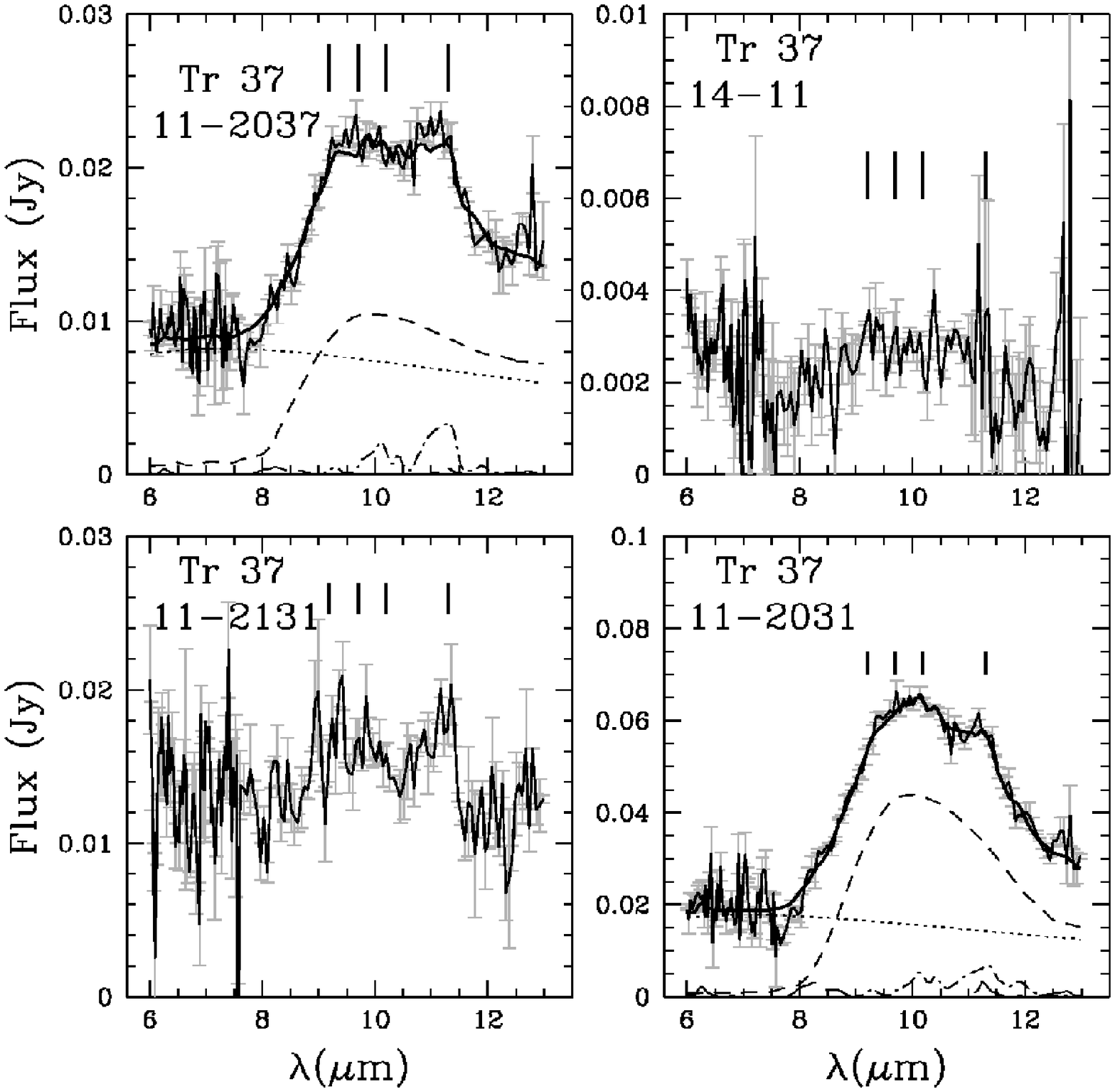}
\caption{Fit for the Silicate feature at 10$\mu$m for low-mass stars in Cep OB2. The observed spectrum is 
displayed with its errors, and the best fit is  drawn as a thick line.
The continuum level is displayed as a dotted line; the emission due to amorphous grains is
displayed as a  dashed line; the emission due to crystalline grains is
represented by a dotted-dashed line; the PAH contribution is a long-dashed
line. Vertical lines are displayed to mark the locations of the peaks of enstatite 
(9.2 $\mu$m),amorphous silicate of olivine and forsterite stoichiometry (9.7 and 10.2 $\mu$m, 
respectively), and crystalline forsterite (11.3 $\mu$m).
 \label{sil1}}
\epsscale{1}
\end{figure}

\begin{figure}
\epsscale{0.4}\plotone{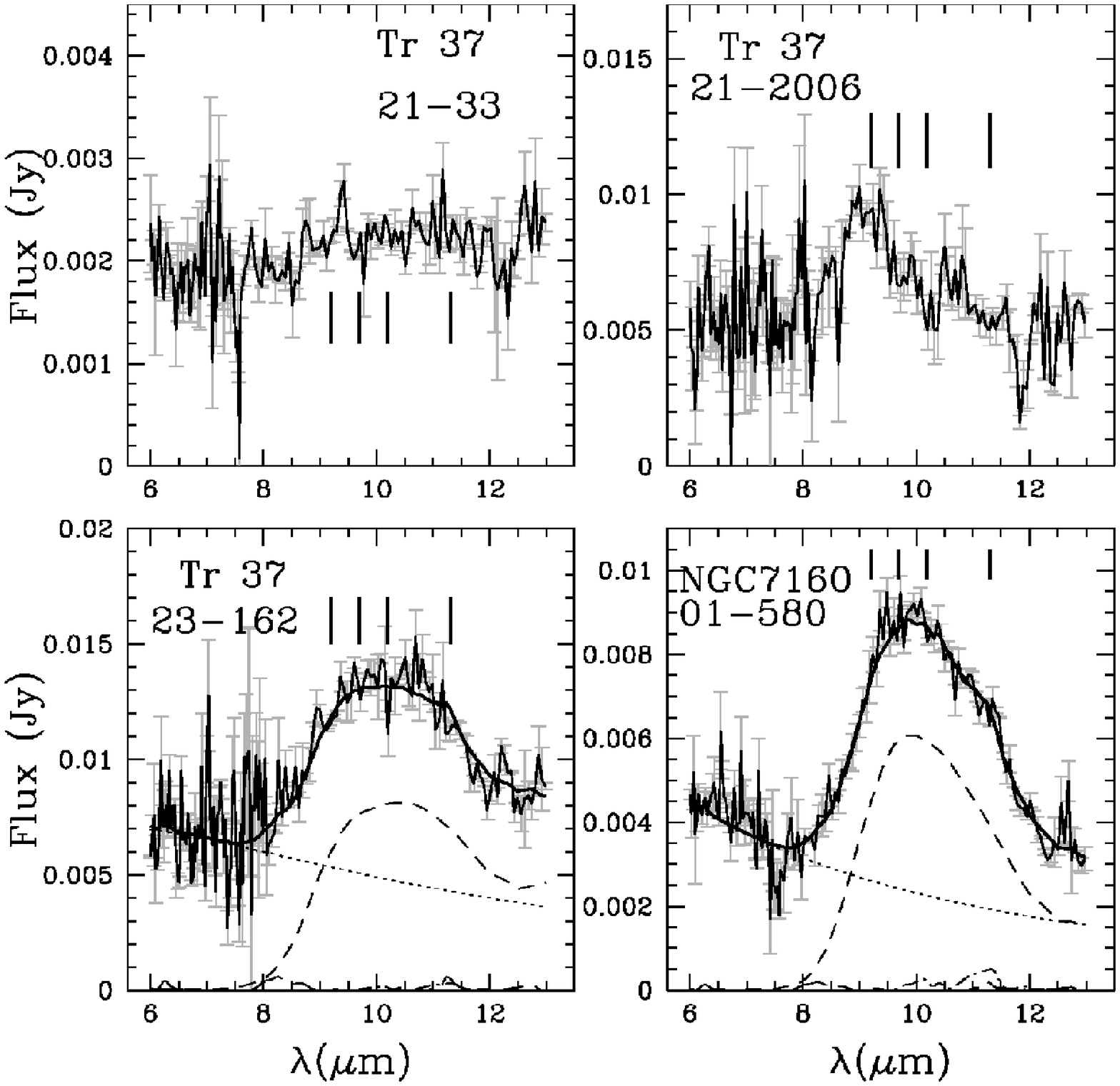}
\caption{Fit for the Silicate feature at 10$\mu$m for low-mass stars in Cep OB2 (continued).
See caption in Figure \ref{sil1}.
 \label{sil3}}
\epsscale{1}
\end{figure}

\begin{figure}
\epsscale{1.1}\plottwo{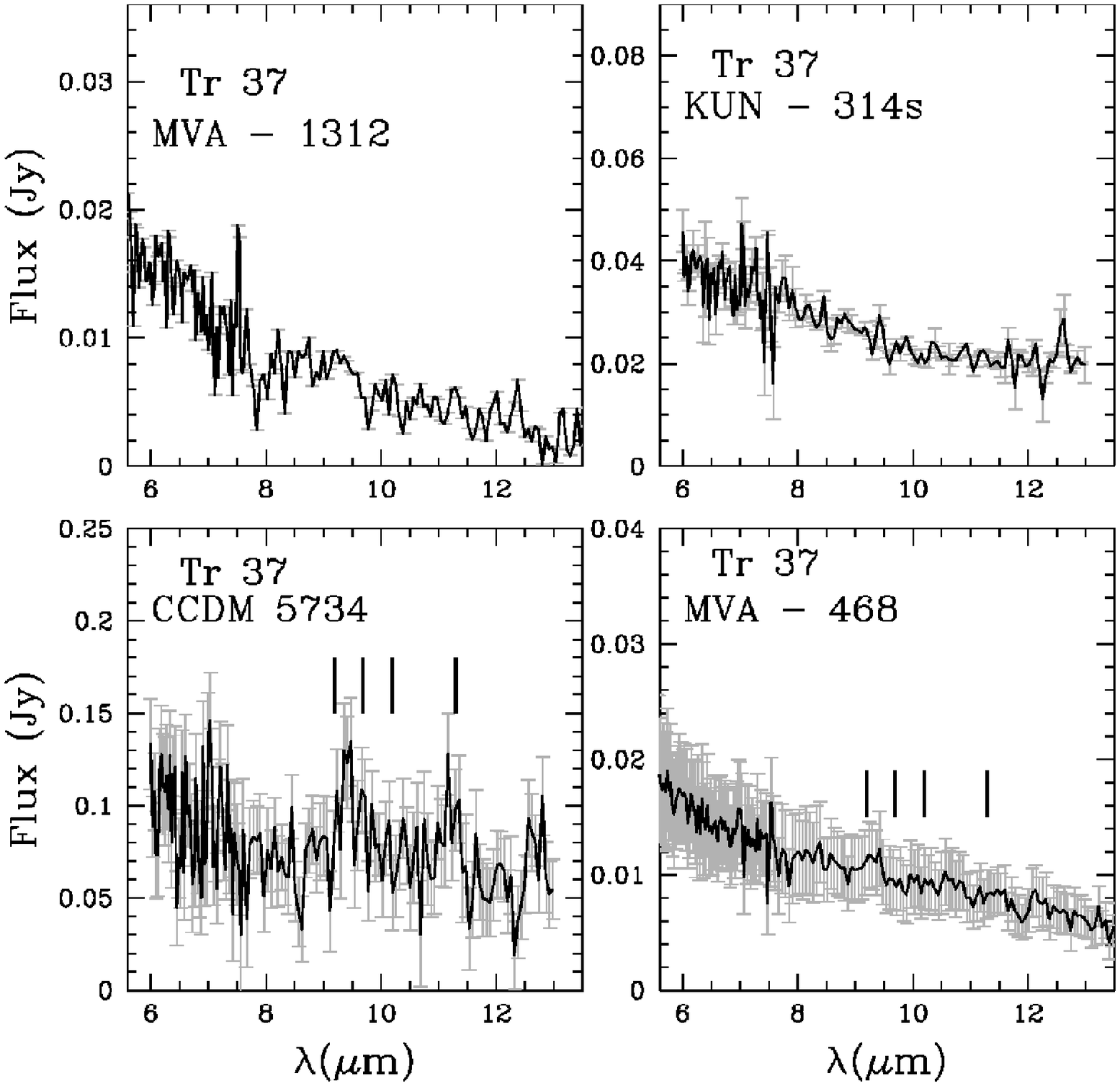}{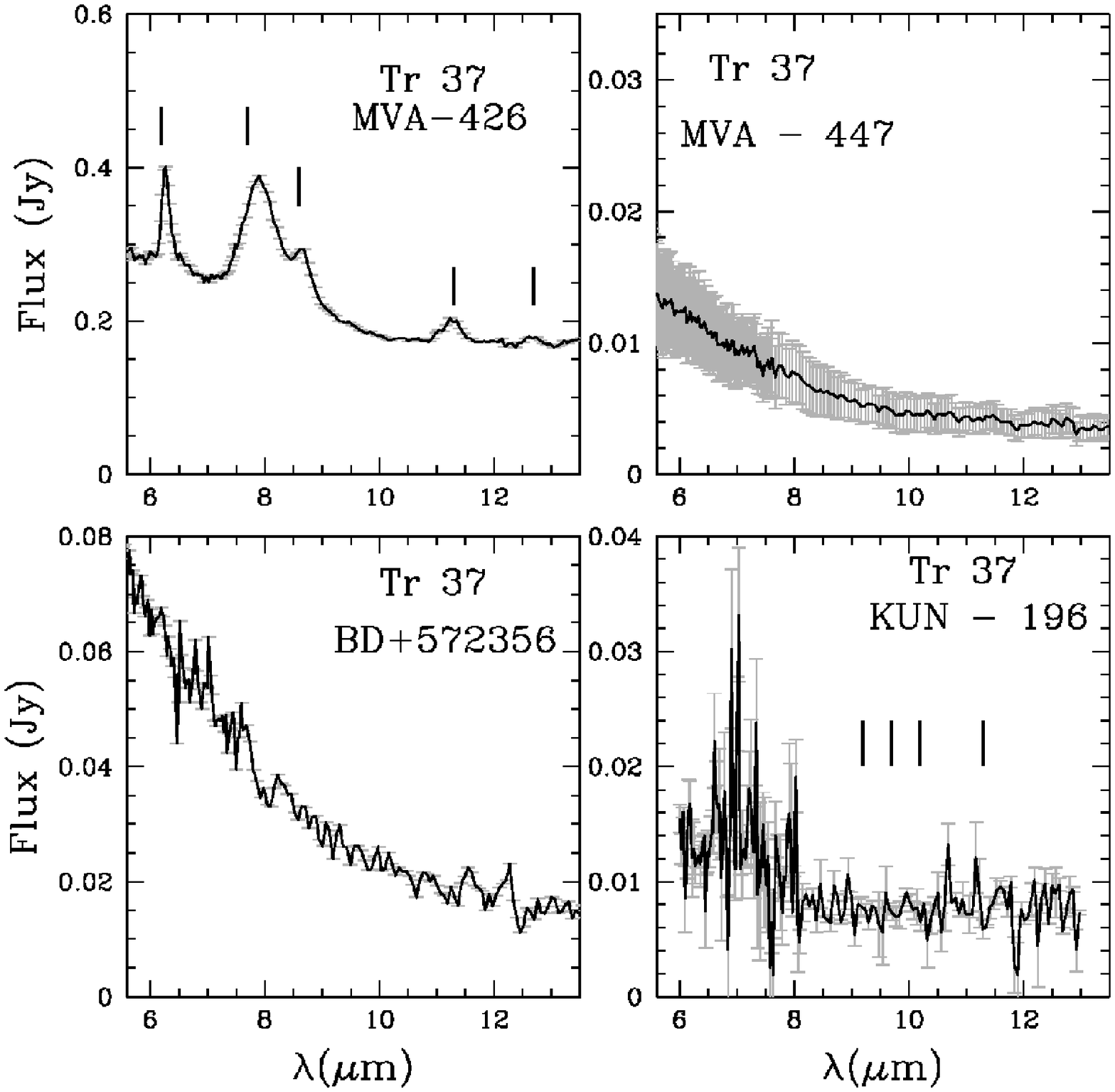}
\caption{Fit for the Silicate feature at 10$\mu$m for intermediate- and high-mass stars 
in Tr 37. The observed spectrum is displayed with its errors, and the 
best fit is  drawn as a thick grey line. The continuum level is
displayed as a dotted line; the emission due to amorphous grains is
displayed as a  dashed line; the emission due to crystalline grains is
represented by a dotted-dashed line; the PAH contribution is a long-dashed
line. Vertical lines are displayed to mark the locations of the peaks of enstatite 
(9.2 $\mu$m),amorphous silicate of olivine and forsterite stoichiometry (9.7 and 10.2 $\mu$m, 
respectively), and crystalline forsterite (11.3 $\mu$m). In the case of the Herbig B7 star MVA-426,
which lacks any silicate feature, we mark the strong PAH features at 6.2, 7.7, 
8.6, 11.3 and 12.7 $\mu$m.
 \label{sil4}}
\epsscale{1}
\end{figure}

\begin{figure}
\epsscale{0.9}\plotone{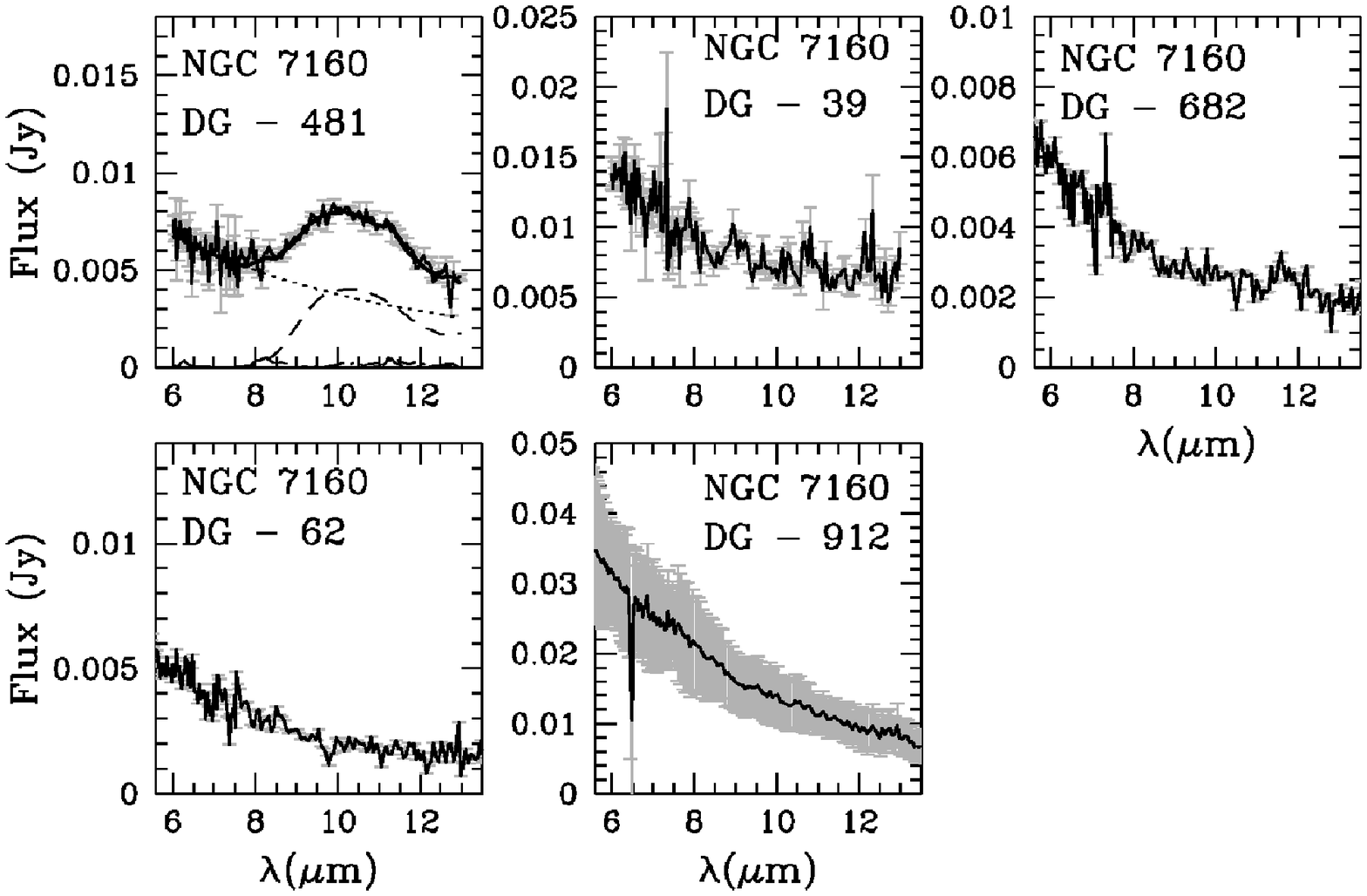}
\caption{Fit for the Silicate feature at 10$\mu$m for intermediate- and high-mass stars 
in NGC7160. The observed spectrum is displayed with its errors, and the best 
fit is  drawn as a thick grey line. The continuum level is
displayed as a dotted line; the emission due to amorphous grains is
displayed as a  dashed line; the emission due to crystalline grains is
represented by a dotted-dashed line; the PAH contribution is a long-dashed
line. 
 \label{sil5}}
\epsscale{1}
\end{figure}

Figures \ref{sil1} to \ref{sil5} display a zoom in the 10$\mu$m region of the spectrum. 
The silicate feature in emission is produced by warm ($\sim$300-500 K) dust grains of small 
sizes ($\sim$0.1-2 $\mu$m) in the optically thin upper layer (disk atmosphere) of
the disks (see Calvet et al. 1992; Chiang \& Goldreich 1997; Menshchikov \& Henning 1997,
among others). The atmosphere of flared disks is heated by the stellar radiation, resulting 
in a higher temperature in the outer layers with respect to the midplane of the disk, which 
is responsible for the optically thin silicate emission in disks that are optically thick
in the near- and mid-IR. Therefore, the absence of silicate features
is associated to disks that have no warm small dust grains, either 
because the inner part of their disks (where irradiation can heat the disk atmosphere
to the temperatures required) is evacuated from dust, and/or because there are no
small dust grains in the disk atmosphere (due to grain growth and/or dust
settling to the midplane). All the objects have spectra with good enough S/N at 10 $\mu$m to 
determine whether the silicate emission is present, although in a few cases the S/N and/or the 
strength of the silicate feature was not enough to allow proper fitting (see discussion 
in Section \ref{silfit}). The possibilities of finding silicate emission features 
at longer wavelengths (18 $\mu$m dust resonance, or the crystalline features at 27.5 and 
33.5 $\mu$m; see J\"{a}ger et al. 1994; Henning et al. 1995; Waelkens et al. 2000; among others) 
is limited by our S/N in most of the stars.

\subsection{Disks, Dust, and SEDs in Low-Mass Stars\label{seds-low}}

Protoplanetary (or accretion) disks are optically thick, flared disks
composed of gas and dust grains, generally extending from the innermost
regions ($\sim$0.1 AU from the star) out to up to hundreds of AU. 
These are the disks typically observed around low-mass or classical T-Tauri stars
(CTTS; Kenyon \& Hartmann 1987; Chiang \& Goldreich 1997, among others), 
and they can be characterized by excess at near- and mid-IR
wavelengths and ongoing accretion. Nearly all the disks found around low-mass stars
belong to this class, probably due to both relative numbers (timescale-dependent) and
sensitivity limitations. As disks evolve, the IR emission from the
inner disk tends to decrease (Paper III) and inner gaps may open
and extend outward (Forrest et al. 2004; Paper III), changing the
disk structure so that the SED and the silicate emission are affected.
Based on the presence of the silicate feature and the SED shape at 
different wavelengths, the disks around low-mass stars are usually classified
as Class II protoplanetary disks (flared, with excesses at all wavelengths from near- 
to mid-IR, and silicate features in emission), transitional disks or ``transition 
objects'' (with no excesses at $\lambda \leq$6$\mu$m, but CTTS-looking,
optically thick disks at longer wavelengths, may have silicate emission), and
debris disks (typically optically thin disks or rings, presumably formed 
by second-generation dust produced by collisions 
after planet formation has removed and/or conglomerated most of the disk 
material \footnote{Note that our limited sensitivity
in our 24$\mu$m MIPS survey did not allow to detect such disks around
stars later than F.}).

\begin{figure}
\epsscale{1.1}\plottwo{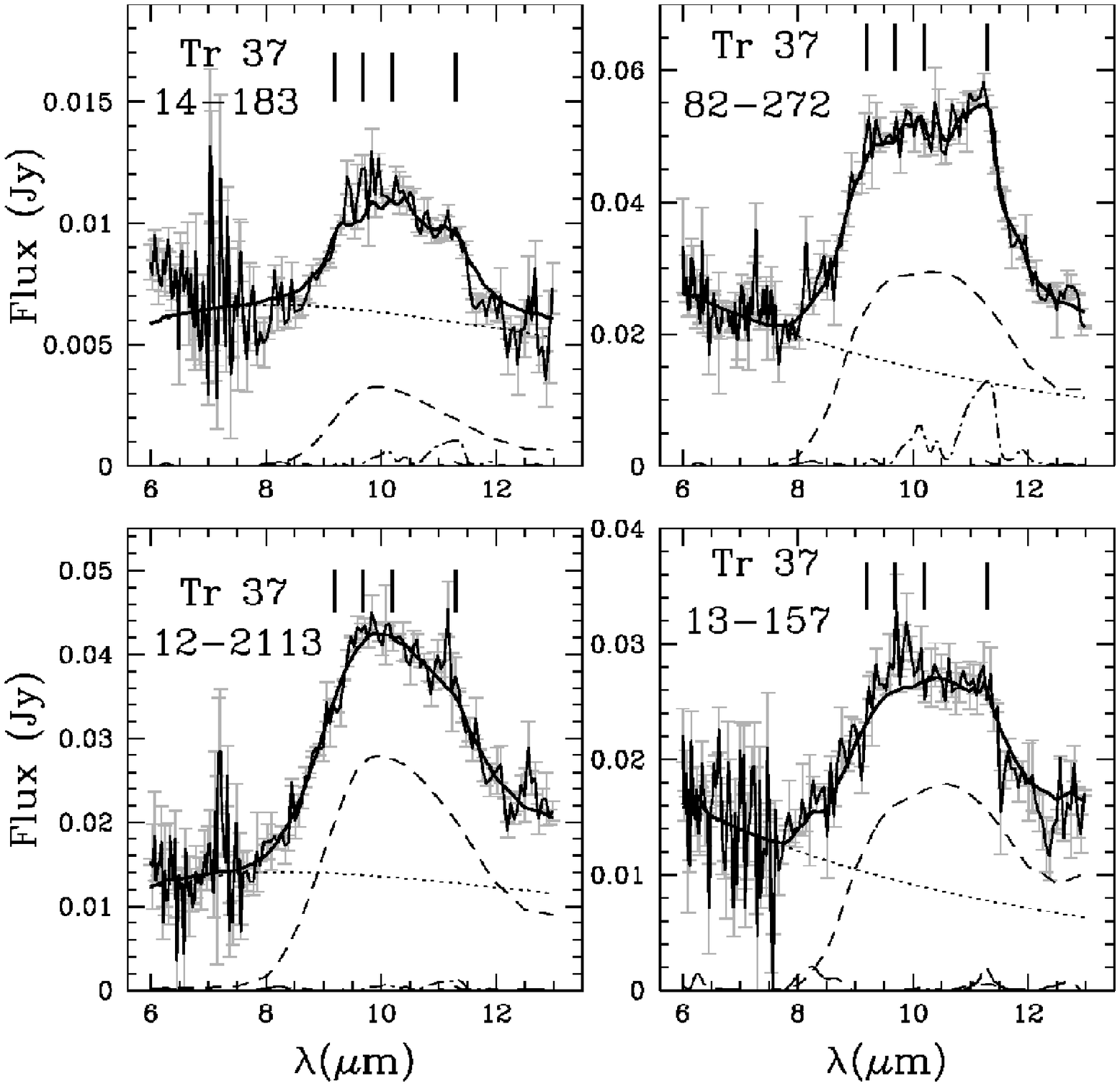}{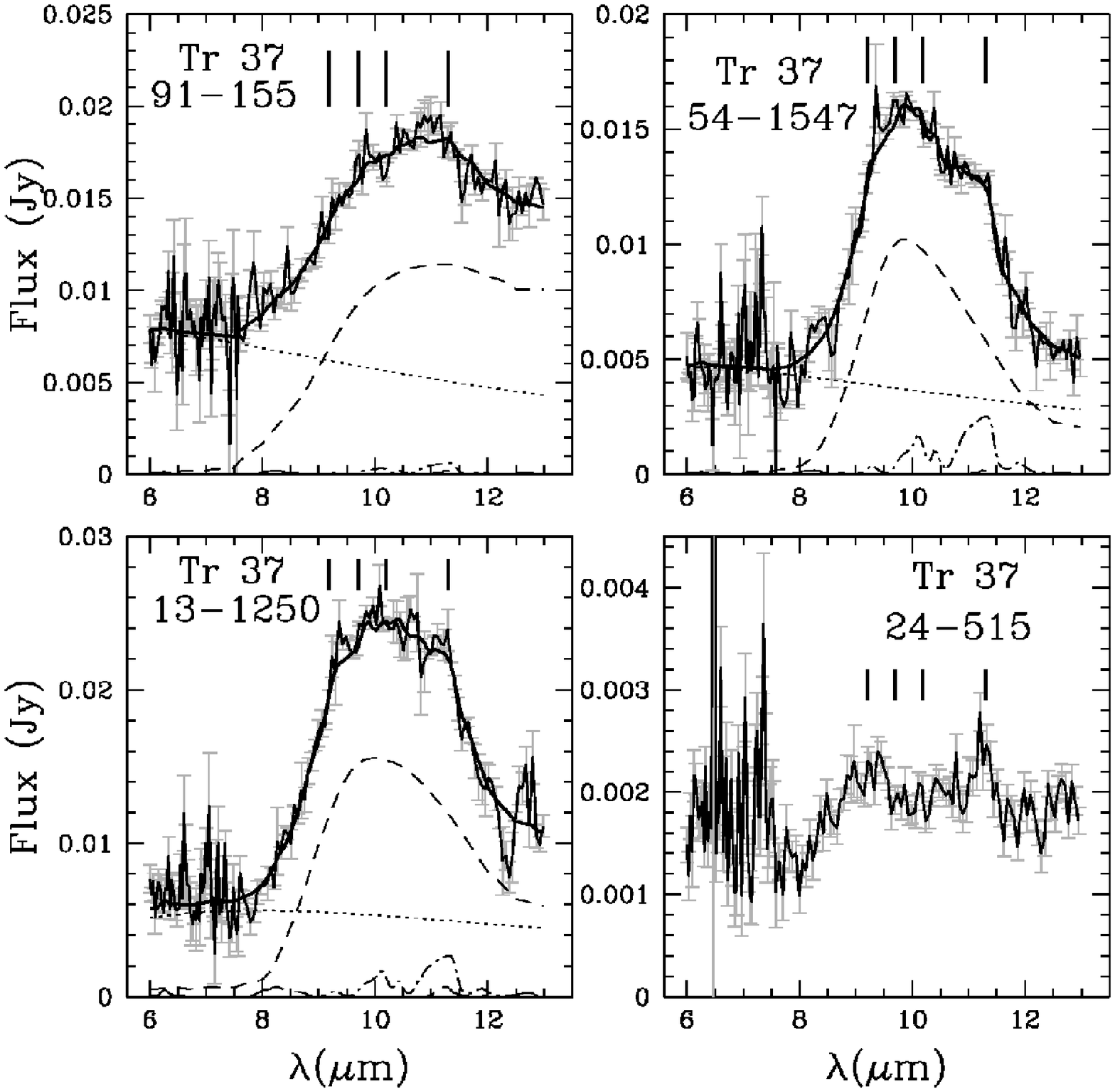}
\caption{Fit for the Silicate feature at 10$\mu$m for low-mass stars in Cep OB2
(continued). See caption in Figure \ref{sil1}.
 \label{sil2}}
\epsscale{1}
\end{figure}

A large excess over the photosphere at IRS wavelengths, together with excesses 
in the IRAS and 2MASS colors (see Paper III), characterize the CTTS-like disks. 
In Figures \ref{sed1}-\ref{sed4} and \ref{sil1}-\ref{sil3} we see that the shape and strength of
the silicate feature varies among stars with similar spectral types and 
similar SED slopes. This large variety of silicate emission had been noted in other 
studies (see Meeus et al. 2003; Kessler-Silacci et al. 2005). The fact that young, 
low-mass stars display such a large variety of disk morphology, dust size and dust composition, 
suggests that several factors, including probably age, spectral type, disk flaring, 
initial conditions, environment, and the presence of companions, affect 
the evolution of the disk. As examples, stars such as 11-2146, 11-2031, 
and the 9 Myr-old 01-580 show very strong silicate features, characteristic of small 
amorphous grains; stars such as 11-2037, 23-162 and 14-183 show a weaker
silicate emission, with some peaks associated to crystalline grains; and stars like 11-1209, 
11-2322 and 11-2131 show nearly no silicate emission at 10 $\mu$m over 3$\sigma$ (see 
Section \ref{silfit} for more details). In addition, some stars show amorphous and 
crystalline features at longer wavelengths, like 11-2131 and 11-2037 (both showing the 28$\mu$m
forsterite signature), 54-1547, 13-1250, and 21-2006. Note that the stars with 
detectable silicate features in the 18-30 $\mu$m spectral range are among the higher-mass 
(G-early K) stars in our sample, which is probably due to the S/N limitations at 
$\lambda >$18 $\mu$m for the faintest stars.

Among the CTTS-like systems, our sample contains a double-lined spectroscopic binary (SB2), 82-272, and two 
single-lined spectroscopic binaries (SB1), 14-183 and 12-2113 (Paper IV). The star 82-272 shows remarkable
emission features from crystalline and amorphous silicates, including the amorphous
silicate feature at 18$\mu$m and the crystalline forsterite bands at 23 and 28 $\mu$m.
Meeus et al. (2003) suggested that the presence of companions may accelerate the evolution 
of the disk and the formation of crystalline silicates by inducing vertical mixing and stirring up the disk.
Crystalline silicate emission at longer wavelengths is not detectable in 14-183 (maybe because of
S/N limitations), but the 11.3$\mu$m forsterite shoulder is marked (see Figure \ref{sil2}).
The third binary, 12-2113, is similar to the rest of the stars in the sample. 

\begin{figure}
\plotone{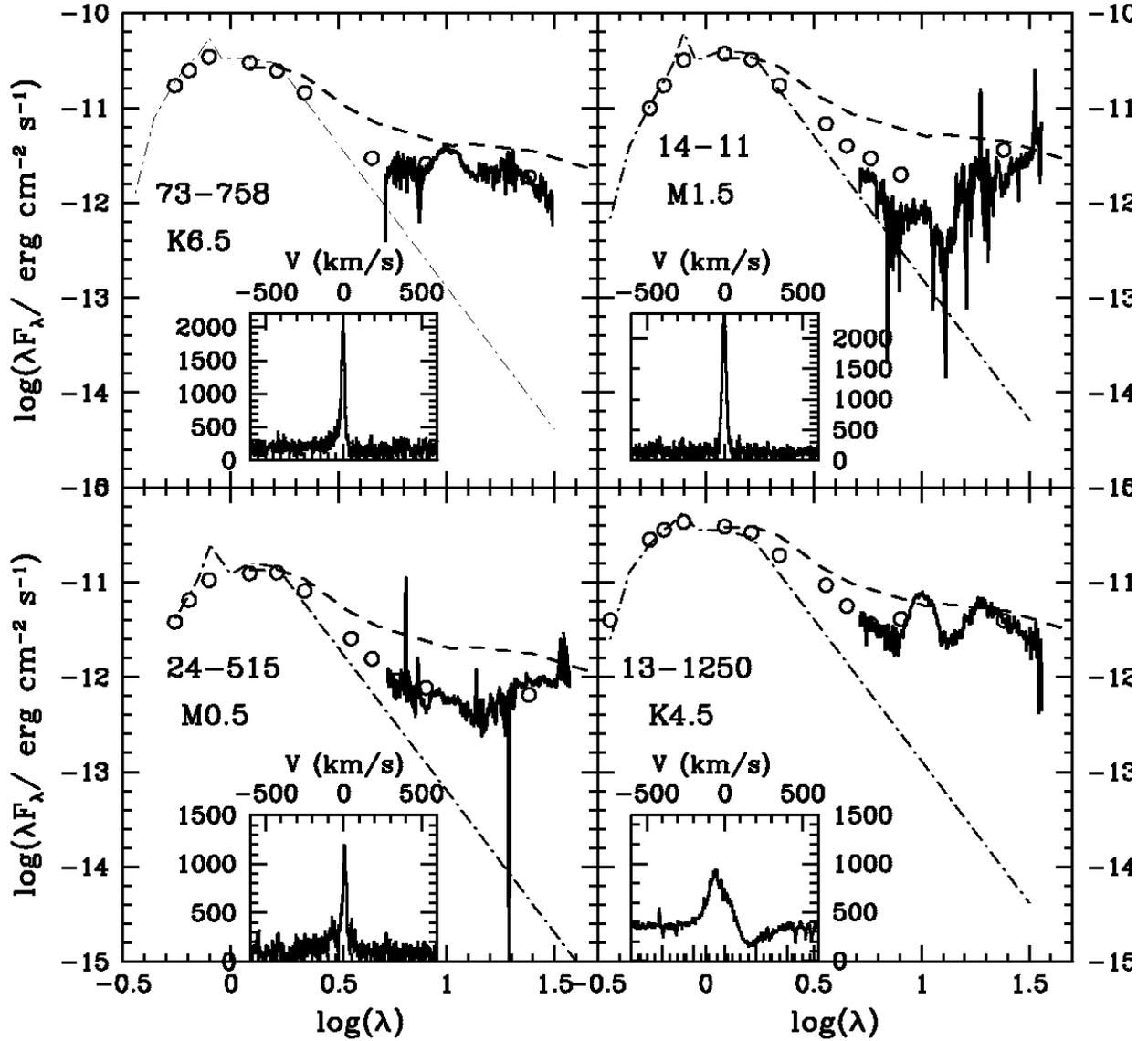}
\caption{SEDs of the 3 ``transition disks'' 73-758, 14-11, and 24-515,
and the very evolved disk 13-1250. The optical (UVRI), JHK (from 2MASS),
IRAC and MIPS (24 $\mu$m) photometry points are displayed together with the
IRS spectrum. The stellar photosphere and the median CTTS SED from Taurus 
(Hartmann et al. 2005) are shown for comparison as the dotted-dashed line
and the dashed line, respectively. The H$\alpha$  line (observed with Hectochelle,
R$\sim$ 34,000) is shown in the figure inset. Note that 73-758 and 14-11 
do not show any velocity wings, and that the velocity wings of 24-515
are only marginal.
 \label{transition}}
\end{figure}

The IRS sample contains 3 of the transitional disks detected with IRAC/MIPS in Tr 37 (see
Figure \ref{transition}). The star 14-11 is the most representative case of a non-accreting 
transition object in the IRS sample. Its IR excess starts in the region
of the 10$\mu$m silicate feature, very similar to CoKu Tau/4 (Forrest et al. 2004). 
Its age is about 1 Myr, again similar to the age of CoKu Tau/4 , as 14-11 
belongs to the young 1-Myr-old population associated to the Tr 37 globule 
(see Paper III for a discussion of the two populations in Tr 37). The silicate
feature in 14-11 is very weak, being only visible due to the lack of
continuum excess at shorter wavelengths. The IRS flux from $\sim$10-30 $\mu$m
is roughly consistent with a black body with T$\sim$70 K, which would approximately
correspond to a ``wall'' at a distance of $\sim$10 AU for this M1.5 star.
Given the lack of accretion signatures (so \.{M}$\leq$ 10$^{-12}$M$_\odot$/yr,
Paper IV), it is a good candidate for hosting one or more large bodies or even planets
(Quillen et al. 2004). 

The stars 73-758 and 24-515 are also transition objects, being the first a non-accretor, 
whereas the second is accreting at a very low rate (\.{M} $\sim$ 10$^{-9}$-10$^{-12}$ 
M$_\odot$/yr, marginally detectable in the H$\alpha$ high-resolution
spectra, Paper IV). The weak silicate feature in the transitional disks
is consistent with a few-AU opacity gap, that would result in very little warm dust 
in small grains in case of low-mass stars. The K4.5 star 13-1250 had been classified
originally as a potential accreting transition object (Paper III), although
some small excess may be present at shorter wavelengths, suggesting rather
strong dust settling and/or grain growth as a mechanism to decrease 
the near-IR opacity rather than a physical gap. The presence of a relatively
strong silicate feature in 13-1250 indicates that, despite the
inner disk grain growth and settling that we infer from its small near-IR excess, 
there are still enough small dust grains in the disk atmosphere at distances $\sim$1-3 AU.

As mentioned before and explained in Paper IV, possible explanations
of these gaps in transitional objects could be either the strong dust coagulation 
up to planetesimals and maybe 
planet sizes (see Lin \& Papaloizou 1986; Bryden et al. 2000; Quillen et al. 2004 
among others), or the photoevaporation of the innermost disk by the central 
star (Clarke et al. 2001). Nevertheless, the presence of accretion (and hence, gas) in about 50\% of
the transitional disks (Paper IV) seems to favor strong coagulation to create the opacity gaps.
The formation of one or more planet-sized bodies within the disk could lead to the
opening of gaps that could be deep enough to prevent the flow of gas throughout the 
disk and would result in the end of the accretion processes as it presumably
happens in CoKu Tau/4 (Quillen et al. 2004). 
Further IRS observations of the rest of transitional disks in Tr 37 (Paper IV)
would be useful in order to determine the relation between the presence silicate feature (and thus,
the presence of small grains in the inner disk) and other properties of the transition disks.

\subsection{Disks, Dust, and SEDs in High- and Intermediate-Mass Stars\label{seds-hi}}

The high- and intermediate-mass stars (spectral types BAF) with disks are usually
classified as Herbig Ae/Be stars (higher mass analogs of T Tauri disks;
Henning et al. 1998; Dullemond et al. 2001; Bouwman et al. 2001; Meeus et al. 2003;
van Boekel et al. 2003), transitional disks (with
inner holes, also analogs to the low-mass transition objects), or debris disks
(with inner gaps but basically depleted from gas and small dust particles after presumably planet
formation, containing reprocessed second-generation dust replenished by collisions; 
e.g. Sylvester \& Mannings 2000; Rieke et al. 2005; Kenyon \& Bromley 2005). 
In general, distinguishing optically thick, primordial, protoplanetary disks from 
optically thin, second-generation, debris disks at mid-IR wavelengths is not easy. 
Therefore, we will focus on the presence of silicate
features as a proof of small grain content, characteristic of protoplanetary disks
rather than of debris disks. Nevertheless, the weakness of the silicate feature in some 
of the optically thick disks around low-mass stars suggest that more detailed observations 
at longer wavelengths (including mm and submm), SED modeling and disk mass estimates will be 
required to probe the nature of the potential debris disks. Given the difficulties
in distinguishing primordial and second-generation dust from mid-IR spectra only,
we assign these disk the labels ``transitional'' or ``debris'' disks as the
most probable nature of the objects, that will be studied in more detail with longer
wavelength observations.

As in the low-mass case, we observe very different spectral signatures
and SED shapes in our sample of 13 high- and intermediate-mass stars
(Figures \ref{sil4} and \ref{sil5}). Here we list the characteristics of the most 
remarkable stars in Tr 37:

{\bf MVA-426}: It is a B7 star in Tr 37, found to have excess at all IR wavelengths
(from JHK to 70 $\mu$m) and strong emission lines in optical spectra 
(Contreras et al. 2002; Paper III). It shows the characteristic features of a Herbig Be 
star, with very strong continuum emission and remarkable PAH lines at 6.2, 7.7, 
8.6, 11.3 and 12.7 $\mu$m, being the only star in the whole sample with detectable PAH features.
It has no evidence of silicate emission at any wavelength, which could be either due to the
reduced contrast with the high continuum level, or to the shielding of part of the disk 
by an inner wall. This makes MVA-426 very similar to HD 100453 (Meeus et al. 2003). 
Its IR excess emission is dominated by two ``bumps'' peaking at near-IR and mid-IR 
wavelengths, as it has been observed in other Herbig Ae/Be stars (see Hillenbrand et al. 
1992; Meeus et al. 2003, among others).

{\bf KUN-314s}: It is the next star with the largest excess at 24 $\mu$m . This Ae
star displays a featureless, relatively bright continuum spectrum, which is consistent 
with a classical Ae star with IR excess produced by the free-free emission 
of a gaseous disk (Porter \& Rivinius 2003) or with a more evolved Herbig Ae star 
(Hillenbrand et al. 1992). It also shows an excess at all the IRS wavelengths and even a 
marginal 3-$\sigma$ excess in JHK (Hern\'{a}ndez et al. 2006), which could suggest that 
it is a protoplanetary disk with strong sedimentation affecting the silicate emission.

{\bf CCDM+5734Ae/w}: This B3+B5 binary displays a weak but
detectable silicate feature at 10$\mu$m, together with a relatively bright 18$\mu$m silicate emission.
The lack of an excess at wavelengths shorter than $\sim$8 $\mu$m suggests
a large inner gap in its non-accreting disk, and the presence of small silicate grains
is consistent with a transition disk rather than with a debris disk. 
The fact that the 18 $\mu$m feature is comparatively stronger than the 10 $\mu$m emission,
and that even the forsterite line at 33.5 $\mu$m is marginally detectable, 
may suggest that the bulk of small silicate grains is located at larger distances
from the star (in colder regions), probably due to the presence of the inner gap.

{\bf KUN-196}: This B9 star has excess at 5.8 and 8.0 $\mu$m IRAC channels, and a
relatively smooth spectrum with very small silicate emission at 10 $\mu$m.
This suggests the presence of a transitional protoplanetary disk with an inner gap 
and important dust settling and/or grain growth in its innermost region.  

{\bf MVA-1312}: It is a B4 star with a featureless spectrum showing emission similar to 
(but lower than) the young debris disk of HR 4796 A (Jayawardhana et al. 1998), with 
no excess over the photosphere up to $\sim$12 $\mu$m, and a smoothly increasing
excess at longer wavelengths. The shape of the emission is compatible with a black body
with T$\sim$100 K, that would mean a wall or ring at a distance of about 300 AU for a B4 star.
Therefore, this object is our best candidate for a debris disk, although observations at 
longer wavelengths (mm and submm) are required to reveal the extent of the disk, the
grain characteristics, and to find out if it is optically thin. 

The remaining intermediate-mass stars in Tr 37 (MVA-468, MVA-447 and BD+572356, spectral 
types B7, F0 and A4, respectively), do not show any silicate emission. The featureless
spectra with small excesses starting at wavelengths from 8 to 20 $\mu$m suggest that they
are most likely debris disks, analogous to the Vega-like systems described in
Sylvester \& Mannings (2000).

In the 12 Myr old cluster NGC 7160, no high- and intermediate-mass stars were found to have
optical emission lines (Paper II). The only star with a strong silicate emission, suggestive 
of a protoplanetary disk with a large inner gap, is the A7 star DG-481, which is the only one 
among the high- and intermediate-mass stars with a 10$\mu$m feature strong enough to be analyzed,
(see Section \ref{silfit}). The stars DG-39, DG-682 and DG-912 (spectral types A0, A2 and F5.5) 
are consistent with objects surrounded by debris disks similar to the observed in Tr 37. The F5 
star DG-62 does not seem to have a detectable excess larger than 3-$\sigma$ 
over the photospheric level.

\subsection{The Shape and Strength of the Silicate Emission: Connections to the Inner Disk \label{normsil}}

\begin{figure}
\plotone{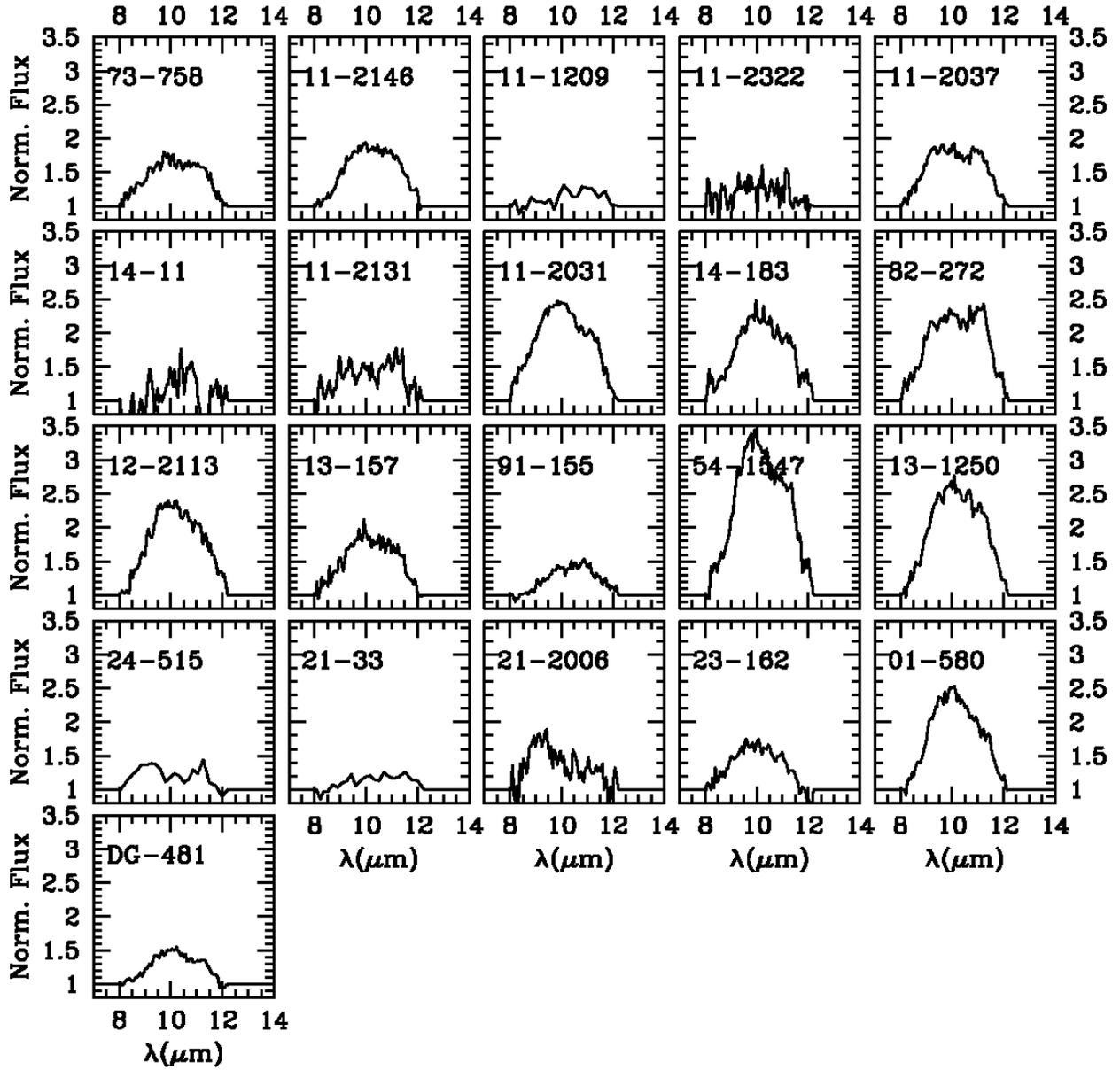}
\caption{Normalized flux in the 10 $\mu$m region for the stars showing
any silicate features in Tr 37 and NGC 7160.
 \label{fluxnorm}}
\end{figure}

In order to analyze the silicate emission in a simple and systematic way, we
 normalize the silicate feature to the continuum level for all the high- and low-mass stars
with significant excess at 10 $\mu$m. The continuum is approximated by a
straight line connecting the fluxes at 8 and 13 $\mu$m (Bouwman et al. 2001),
and the normalized silicate flux is defined as (F$_f$-F$_c$)/F$_c$, where ``f''
refers to the normalized flux in the feature, and ``c'' to the continuum
normalized flux. This normalization helps to extract and compare the information contained
in the silicate emission in a simple way, without assuming any particular model nor
grain properties, and independently of the continuum level observed, and of the shape of 
the SED. Figure \ref{fluxnorm} displays the extracted normalized flux in the 10 $\mu$m 
region for all the stars where the continuum can be successfully extracted (all the low-mass
stars plus the A7 star DG-481). In order to examine the characteristics of the silicate emission,
following the study of Herbig AeBe and T Tauri stars by Bouwman et al. (2001, 2006),
we measure the peak of the normalized flux at different wavelengths (8.6, 9.8, and 11.3 $\mu$m), 
corresponding to the broad shoulder noted by Bouwman et al. (2001), the peak of amorphous 
silicates with olivine stoichiometry, and the peak of crystalline forsterite. In addition, 
we measured the flux at the maximum of the feature. 

\begin{figure}
\plotone{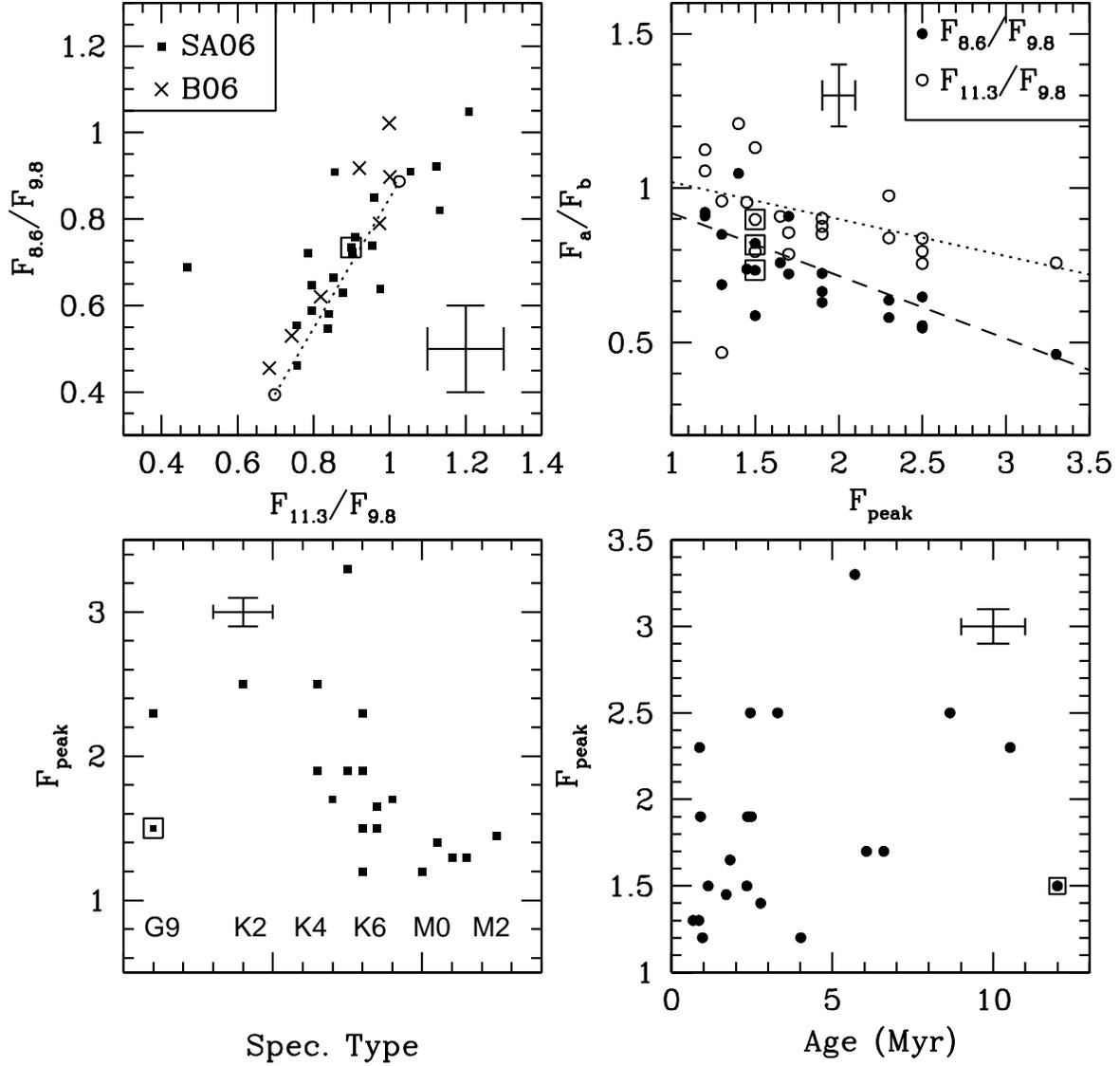}
\caption{Ratios of the normalized flux in the 10 $\mu$m region for the stars showing
silicate features in Tr 37 and NGC 7160, compared to the peak of the normalized flux 
and to two stellar properties (spectral type and age). The typical error bar is shown. The
A7 star DG-481 is marked with a large square to distinguish it
from the rest of the sample, containing only T Tauri stars. Note that DG-481
is displayed with a shift in the spectral type plot. The trend for amorphous grains
with sizes increasing from 1 to 10$\mu$m is shown in the top left panel as a dotted
line. The data points of Bouwman et al. (2006) are represented as well in the panel,
marked by crosses. In the top right figure, the trends of the data of Bouwman et al. 
(2006) for the F$_{11.3}$/F$_{9.8}$ and F$_{8.6}$/F$_{9.8}$ are shown as dotted and 
dashed lines, respectively.
 \label{normfluxratio}}
\end{figure}

As a first step, we examine the flux ratios following the scheme designed by Bouwman 
et al. (2001, 2006). As stated there, we find a strong correlation between
the flux ratios at 8.6, 9.8 and 11.3 $\mu$m (see Figure \ref{normfluxratio}), suggesting the 
same origin and similar composition for the silicate features. The values are
different from the Herbig stars in Bouwman et al. (2001), but comparable
to the ones observed for TTS in Bouwman et al. (2006) and in Kessler-Silacci et al. (2005). 
We do not see any difference in the behavior displayed by the A7 star DG-481,
compared to the observations of T Tauri stars. We also observe a correlation between the 
silicate feature strength and the different flux ratios (Figure \ref{normfluxratio}), similar 
to the correlation strength/shape described by van Boekel et al. (2003,2005), Przygodda et 
al. (2003), Bouwman et al. (2006), and Kessler-Silacci et al.(2005). The universal behavior 
of the relative fluxes, which do not depend on the spectral types nor other disk and stellar properties,
seems to indicate that the components of the silicate feature and the processes leading to grain 
growth and dust evolution are very similar in stars with different spectral types (from A7 to M2.5).

Since the properties of the stars in our sample are well-known (see Table \ref{target-tab}),
we have explored the correlations between the different stellar and disk properties and
the measured normalized fluxes. The significancy of these and other correlations investigated 
are listed in Table \ref{correlation-tab}. There is a very strong correlation between the 
peak flux and the spectral type (see Figure \ref{normfluxratio}), which shows that the strength of 
the silicate feature decreases for the later spectral types. This effect could be related to differences
in the region where the silicate feature is produced. Since late-type stars are less luminous,
the areas where the irradiation is enough to heat the disk atmosphere to the temperatures 
required to produce the silicate emission are smaller and closer to the star for M-type stars
compared to K-type and earlier. Therefore, the silicate emission in the lower-mass stars may be 
affected strongly by inner disk evolution (grain growth/dust settling/inner gap
opening), or by the changes of the slope in the innermost disk (due to disk evolution) 
where the silicate emission is produced. Moreover, the size of the region producing the optically 
thin silicate feature is smaller for late-type stars, resulting in general in weaker emission. 
The A7 star DG-481 seems to be out of the general trend for the peak flux versus spectral type, 
but this is not surprising taking into account the difference in spectral type and in disk 
morphology with the rest of the sample. The silicate emission in an A7 star should be produced 
roughly at distances of 2-6 AU from the star, so relatively large gaps can develop before 
the silicate feature disappears. 

\begin{figure}
\plotone{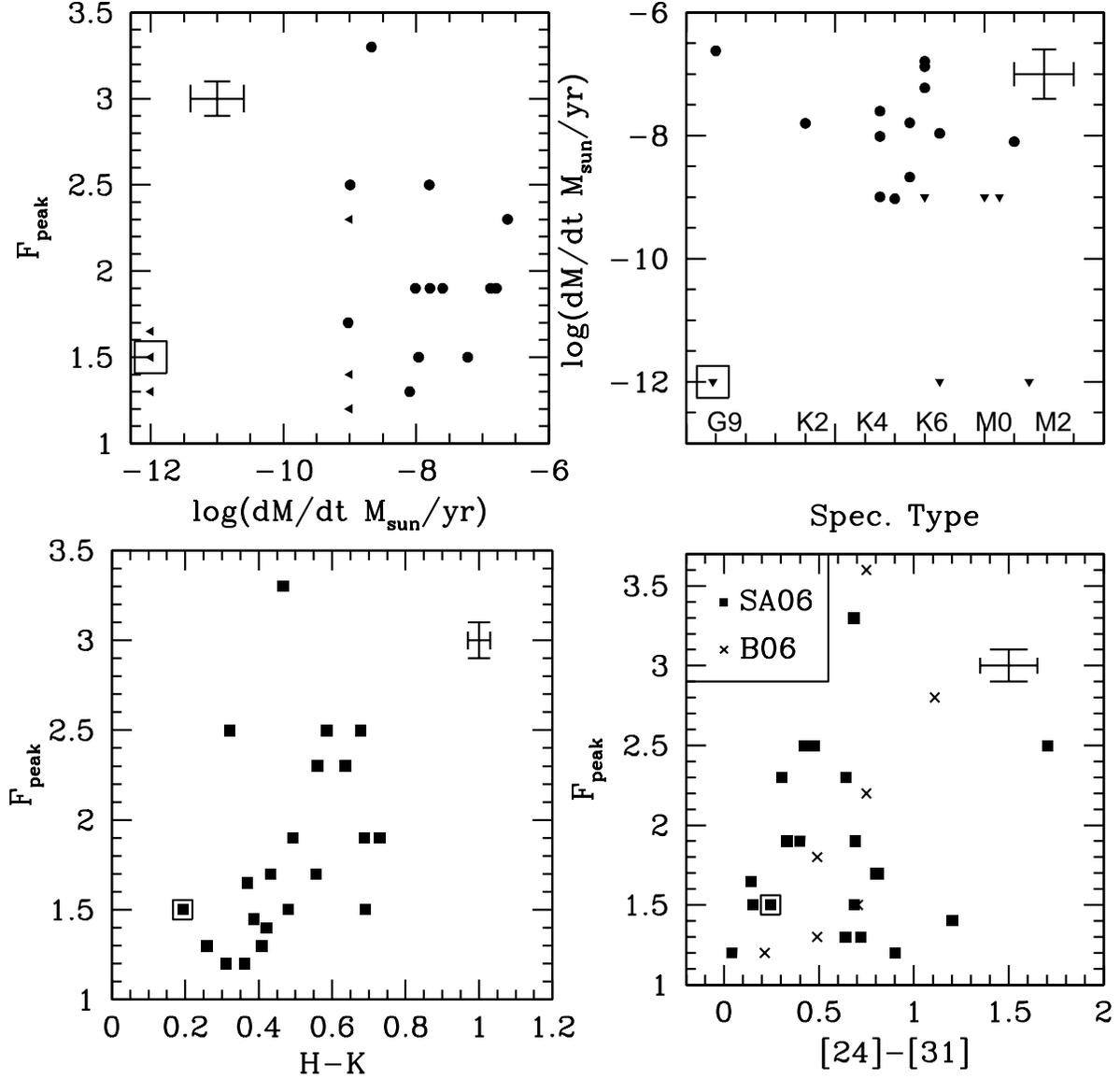}
\caption{Peak of the normalized flux in the 10 $\mu$m region for the stars showing
silicate features in Tr 37 and NGC 7160, versus different disk and
stellar properties (accretion rate and IR colors). The A7 star DG-481 is marked with a box.
Typical error bars are displayed in each diagram. The correlation between the
flux at the normalized peak and the [24]-[31] color is consistent with the data of Bouwman et al. (2006)
for low-mass stars, shown as crosses in the bottom right panel. The trend showing that
earlier spectral type stars tend to have higher accretion rates is shown in the top right
panel for comparison with the trends relating accretion rates to other silicate and stellar properties.
 \label{peakflux}}
\end{figure}

We also find a moderate correlation between the accretion rates and the silicate strength, 
showing that stars with strong features tend to have large accretion rates. The two stars 
with larger accretion rates in our sample (82-272 and 11-2146, with \.{M}$\sim$10$^{-7}$M$_\odot$/yr) 
have very strong silicate emission. This is in agreement with the observations of FU Ori objects 
(Green et al. 2006), which are found to have strong silicate emission (although none of the objects
examined here has accretion rates as high as FU Ori objects). The correlation
between silicate strength and accretion rate may be related to the correlation between
the spectral type and the accretion rate (Hartmann et al. 1998), given that stars
with earlier spectral types tend to have stronger accretion rates (see Figure \ref{peakflux}).
We will explore this behavior in more detail in Section \ref{lowmass}.

A possible dependence of the silicate feature on the SED shape, which affects the illumination
of the disk by the central star, is explored in Figure \ref{peakflux}. There is a general 
tendency for more flared (redder-colored) disks to have stronger silicate features, although 
we find strong scattering. This tendency can be understood as more flared disks are more efficient
capturing the light from the star, so they can be heated to the temperatures required for producing
silicate emission over larger areas. As noted in Paper III, we observe a general trend for a smaller
flux at $\lambda$ $\leq$8 $\mu$m in Cep OB2 compared to younger regions (i.e., Taurus)
that could be interpreted as the effect of differential dust settling/grain
growth  (more efficient in the innermost disk) increasing with age. Given that it is in the innermost
disk where most of the silicate emission for late-type stars is originated,
if grain growth/settling/disk flattening is more efficient there, the 
silicate emission will tend to be weaker in later spectral types.

\subsection{Fitting the 10 $\mu$m Silicate Feature: Grain Size and Mineralogy \label{silfit}}

In order to analyze the 10$\mu$m silicate feature in a more quantitative way,  we used the 
models developed for studying the objects in the FEPS sample (Bouwman et al. 2006), 
based on the previous work and observations of Bouwman et al. (2001) and van Boekel et al. (2003). 
These models, similar to the ones used by van Boekel et al. (2005), include 2 temperatures 
(for continuum and for grains), three different sizes for grains (0.1, 1.5 and 6.3 $\mu$m), 
and 6 different species (olivine, pyroxene, forsterite, enstatite, silica, and PAHs; see 
Table \ref{species-tab}). The models assume that the overall emission can be reproduced
by adding the emission from the individual constituents so that the total flux can
be written:

\begin{equation}
F_\nu = B_\nu(T_{cont}) C_0 + B_\nu(T_{dust}) \sum_{i=1}^3 \sum_{j=1}^5 C_{ij} \kappa_{\nu}^{ij} + 
C_{PAH} I_{\nu}^{PAH}
\end{equation}

We fit the spectral range from 6 to 13 $\mu$m, taking into account the
errors at each wavelength and using a least-square minimization for obtaining
the best fit. The fit could be extended to the silicate features in the 18-20 $\mu$m 
region, but in our case the S/N  in the LL part of the spectra is not high enough.
There are some limitations to this approach. One of the limitations is the single 
temperature for the continuum and the single temperature for all the grains of different 
sizes and compositions, problem that could be improved by using detail disk models. 
The addition of the contributions of the different constituents is
accurate if grains come in individual components or form very porous 
aggregates (see Bouwman et al. 2006 for details). In this study, we use
this approximation as a first approach to characterize the grain sizes and
composition,  leaving the detailed disk modeling for future work.

Spectra with very weak and/or noisy silicate features cannot be properly fitted. 
In our low-mass star sample, we find 4 stars with good S/N and non-detectable or 
marginally detectable silicate feature (11-1209, 11-2322, 11-2131, and 21-33),
and 3 more stars with weak and noisy features (14-11, 24-515, and 21-2006),
none of which could be properly fitted. Among the high- and intermediate-mass stars, 
we could only fit the silicate feature of DG-481, which is the only one showing a prominent
feature at 10 $\mu$m. Figures \ref{sil1} to \ref{sil5} show the
details of the silicate feature, including the fits whenever they were significant.
Table \ref{fit-table} summarizes the amount and size of the different materials 
needed to reproduce the silicate feature with this simplified models. 
It is worth mentioning that our fits reveal a very small amount of
crystalline silicates (typically few percent, and never larger than
20\% of the total mass, consistent with Meeus et al. 2003, Honda et al.
2003, 2006, and Sargent et al. 2006), and no significant evidence for enstatite (only marginal
detections within 1-3$\sigma$, consistent with the above mentioned references).

\section{Discussion \label{results}}

\subsection{Parallel Dust and Gas Evolution: Grain growth, Turbulence, and Settling \label{lowmass}}

\begin{figure}
\epsscale{1.1}\plotone{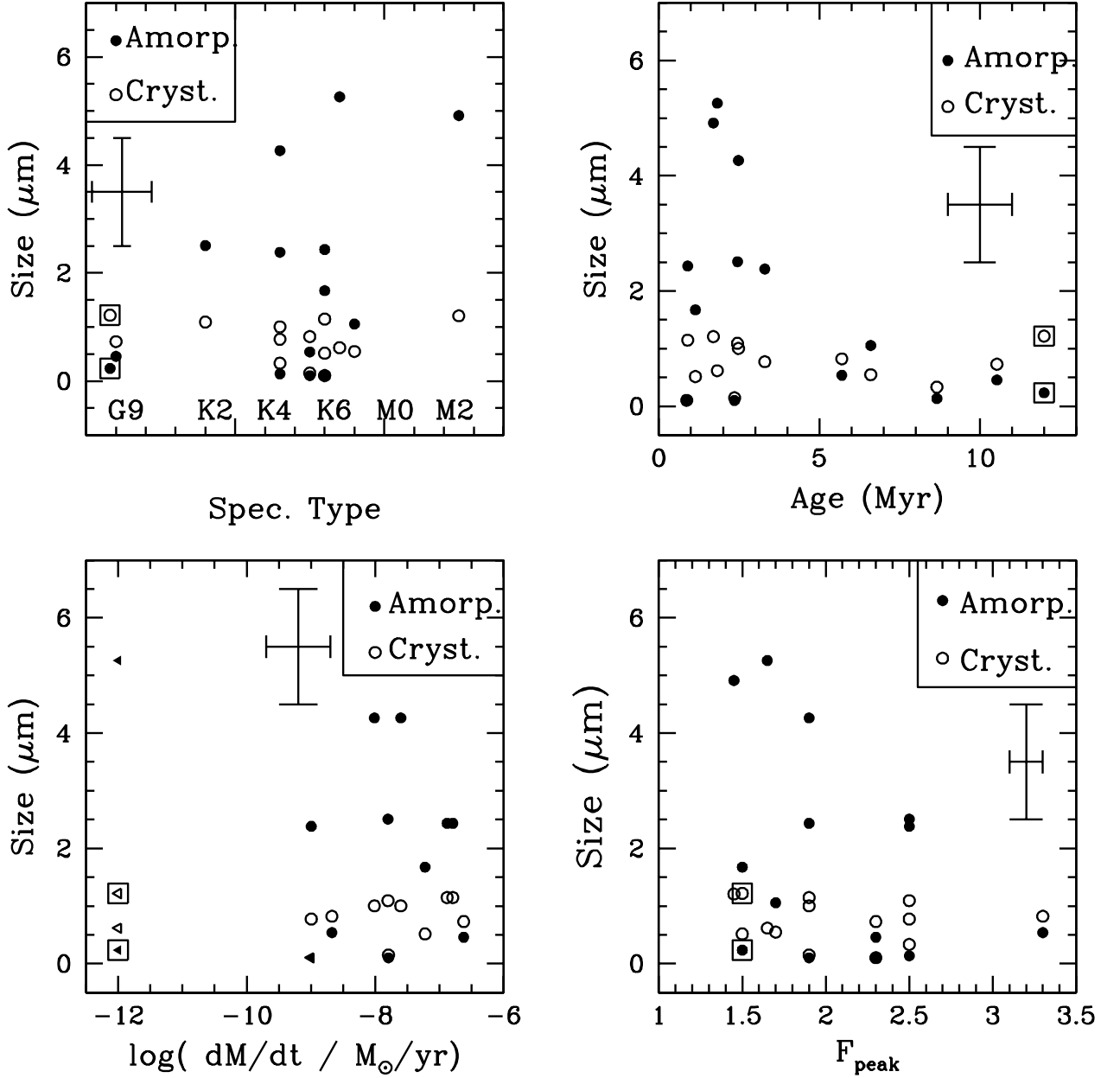}
\caption{Grain size versus some stellar and disk properties (spectral type, age, 
and accretion rate). The A7 star DG-481 is marked with a large square (note that it is
displaced in the spectral type figure for clarity), the
rest of stars are TTS. The fact that larger (amorphous) grains
are found preferentially in young stars suggest dust settling with age. Most 
of the large grains are found in stars with the higher
accretion rates, as it would be expected if the disk atmosphere is replenished
by stirring the grains that may be settling in more turbulent disks. 
The described behavior seems to affect only the amorphous grains (which constitute
most of the mass), the behavior of the (smaller) crystalline grains is not
found to correlate significantly with other properties than binarity.
The correlation between grain size and normalized flux strength (F$_{peak}$) is
also displayed. \label{Sizecorr}}
\epsscale{1}
\end{figure}

\begin{figure}
\epsscale{1.1}\plotone{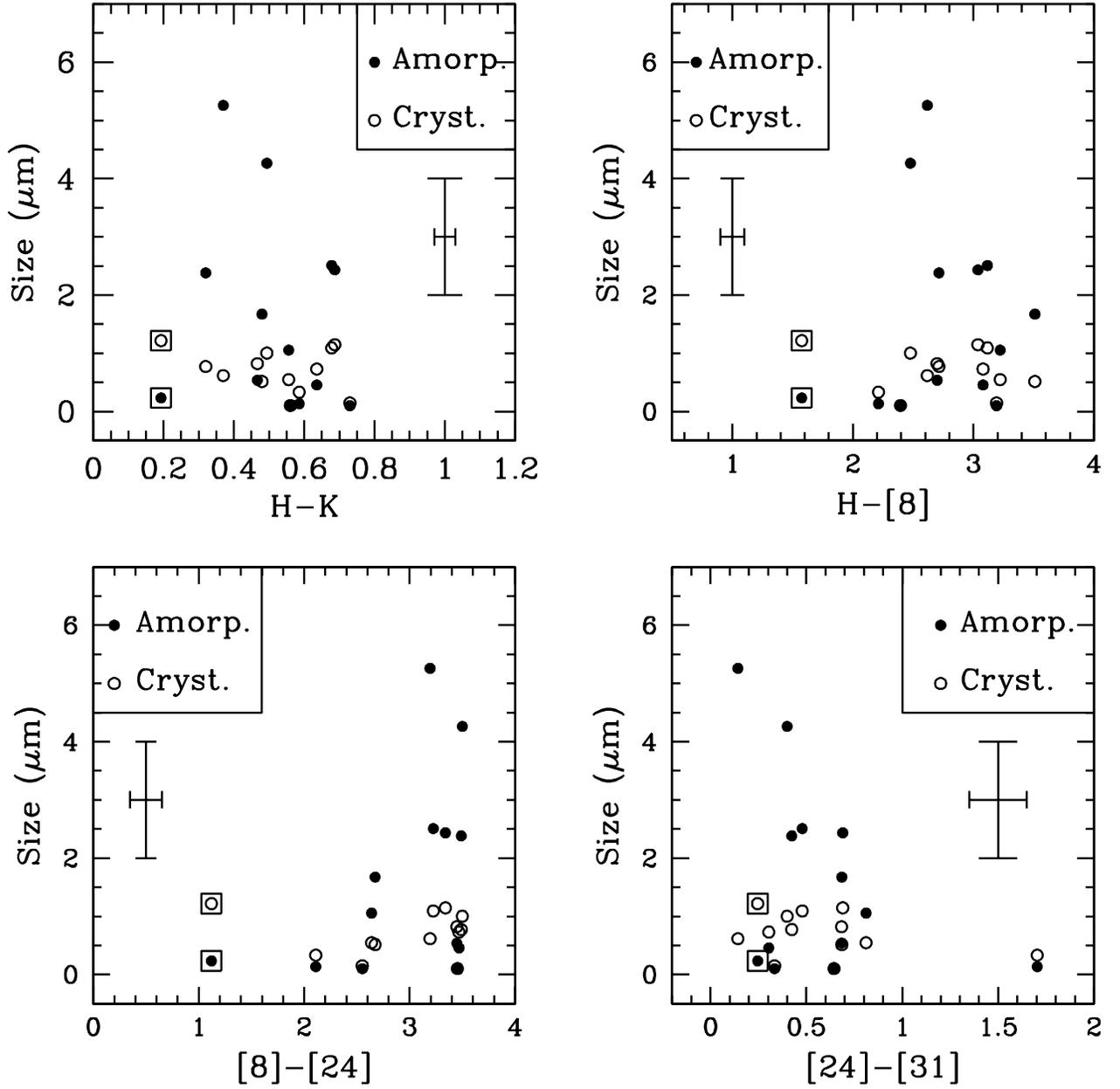}
\caption{Grain size versus the SED slope at different radii.
The A7 star DG-481 is marked with a large square, the
rest of stars are TTS.
 \label{SizeIRcorr}}
\epsscale{1}
\end{figure}

\begin{figure}
\epsscale{1.1}\plotone{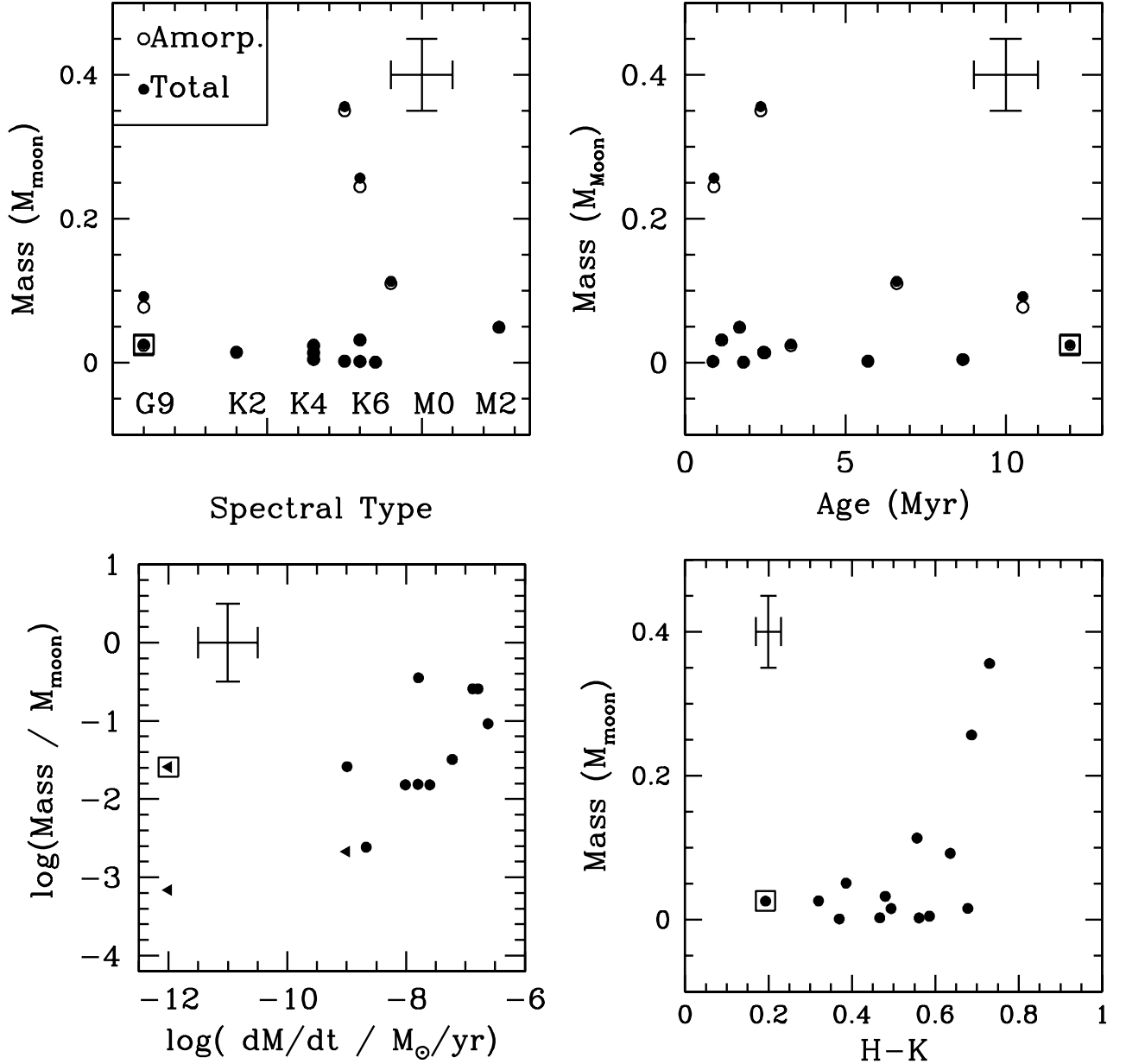}
\caption{Total mass from the silicate fit, and mass in amorphous grains, versus
different disk and stellar properties (spectral type, age, accretion rate, and near-IR color). 
There is a weak correlation between the silicate mass
and the H-K color (related to the flaring of the inner disk/inner wall), suggesting that
the amount of warm silicates depends on the irradiation (therefore, temperature) of 
the inner disk. No correlation is observed between mass and spectral type, as we would
expect if the amount of silicates depends rather on disk geometry and properties than on stellar
properties. The correlation between the mass in silicates and the accretion
rate could be related to the higher turbulence of the stronger accretors, but could be also
due to the differences in spectral type. The A7 star
DG-481 is marked by a large square. Note that its spectral type is shifted for better
viewing of the low-mass stars. 
 \label{Mcorr}}
\epsscale{1}
\end{figure}

\begin{figure}
\epsscale{1.1}\plotone{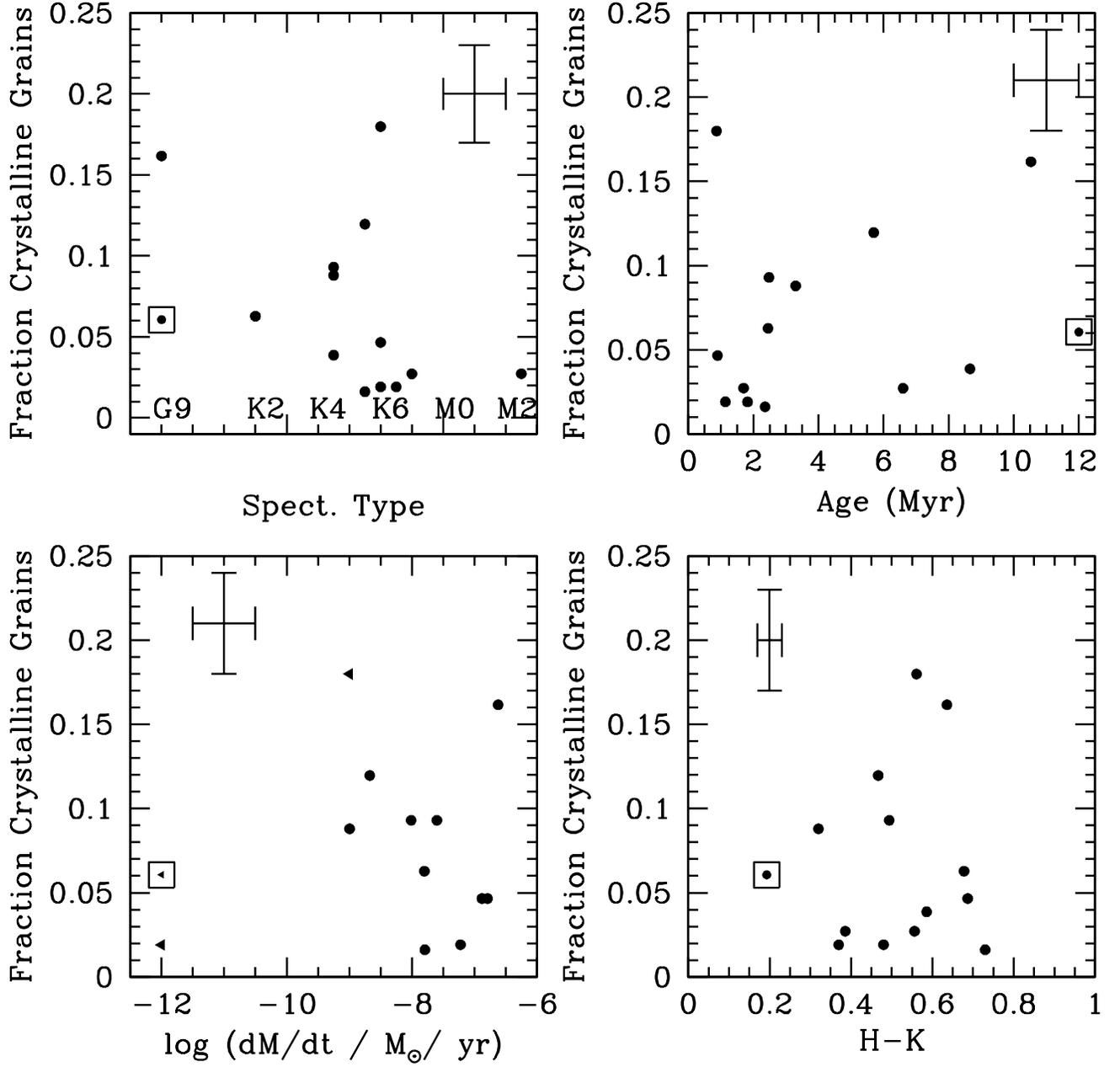}
\caption{Fraction of crystalline silicates versus different characteristics of
the disks and stars. The fraction of crystalline silicates does not seem to be correlated
with any stellar property (spectral type, SED shape, mass accretion rate, age), 
although it is remarkable that the two stars with larger crystalline fraction are
spectroscopic binaries. The A7 star
DG-481 is marked by a large square. Note that its spectral type is shifted for better
viewing of the low-mass stars.
 \label{Cryscorr}}
\epsscale{1}
\end{figure}

Using the grain properties obtained from model fitting (see Section \ref{silfit},
Table \ref{fit-table}), we examine the possible correlations
between the characteristics of silicates, the age and the accretion
rates. Figures \ref{Sizecorr}, \ref{SizeIRcorr}, \ref{Mcorr}, and \ref{Cryscorr} 
show some of the correlations explored between the silicate size (obtained as a mass weighted
average of the components required to reproduce the feature), the amount of mass in
silicates, and the crystalline fraction. Their statistical significancy is compiled in 
Table \ref{correlation-tab}. Even though the sample is small, since only 13 of the low-mass stars and 
1 intermediate-mass star have enough S/N in the 10$\mu$m region to allow size
and composition analysis, we can draw some conclusions from the Tr 37 data. As already mentioned
in Section \ref{normsil}, the behavior of the A7 star DG-481 is not significantly different 
from what we observe for the low-mass members, so we will analyze the whole sample together. 

There is a strong correlation that shows that the size of the grains tends to be 
smaller at older ages (Figure \ref{Sizecorr}). This observation seems counterintuitive, 
since we would expect the grain size to increase with time as dust coagulates (Dullemond
\& Dominik 2005). Nevertheless, the 10 $\mu$m silicate feature traces only a 
(presumably small compared to the total) population of micron-sized grains in the upper layers
of the disk. The population of grains that produces the silicate feature would be
representative of the whole disk content only in the case of very strong
turbulence (inconsistent with our moderate to low mass accretion rates) that would replenish
the upper disk layers with coagulated and evolved grains formed in the deep layers of the
disk. Therefore, this behavior is not inconsistent with grain growth to larger sizes occurring 
at later stages nor with inner disk removal (Paper III). Moreover, the lack of large grains (as 
traced by the silicate emission feature) at later evolutionary stages is consistent with 
dust settling increasing with time and affecting preferentially the larger grains.
A KS test comparing the sizes of grains in disks with ages below 5 Myr and
over 5 Myr gives a probability of 0.07 (7\%) that both distributions are identical.
Taking into account the small size of the sample, the decrease of grain size with age
is statistically significant. Since viscous disk evolutionary models (Hartmann et al. 1998)
and observations of regions with different ages (Muzerolle et al. 2000, Paper II)
show a decrease of the accretion rate with age, the decrease of silicate size with age
would be a natural consequence if the evolution of the inner dusty disk occurs somehow 
parallel to the evolution of the gaseous component and turbulence/accretion (Paper IV). 

Moreover, the fact that we observe relatively large (nearly 6 $\mu$m) grains at very early 
stages may be an indication that grain growth (at least, to the sizes that we can still 
detect at 10 $\mu$m) starts already at very young ages (Dullemond \& Dominik 2005).
This observation implies that we must be careful before considering that stars showing a very 
``pristine'' silicate feature (composed mostly of small grains) are very young or in an early
evolutionary stage, since evolution (dust settling) may cause the large grains to be 
removed from the disk atmosphere, leaving only the smaller ones. A good example of
this could be the only accreting star in NGC 7160, 01-580, which has a silicate feature 
due to very small amorphous grains, and an age near 9 Myr (5 Myr older than the average 
star in Tr 37). These observations seem in contradiction with the general idea that
small grains are removed first in the atmospheres of Herbig AeBe stars because of 
coagulation (Kessler-Silacci et al. 2005), but ages of intermediate-mass stars should be
regarded with care.  Additionally, the average size of silicates is weakly 
connected to the slope of the SED (Figure \ref{SizeIRcorr}).
This result is similar to the observations in HAeBe stars in Bouwman et al. (2001),
stating that the amount of silicate processing is not dominated by the SED shape,
even if the overall strength of the feature may be affected by the disk flaring.

Another strong correlation observed is that stars with higher accretion rates tend to have 
larger average silicate grains (Figure \ref{Sizecorr}) in the disk atmosphere, compared to 
stars with low accretion rates or even no accretion. Given the relationship between 
accretion and turbulence (Hartmann et al. 1998; Schr\"{a}pler \& Henning 2004), and that 
the silicate feature traces the grains in the warm disk atmosphere, these observations 
could be an indication that more turbulent disks can replenish the disk atmospheres with more 
(and larger) grains that might otherwise settle in the middle plane in absence of strong
turbulence. Nevertheless, the scattering is large, and, as we mentioned before,
some of the more massive stars (82-272, 11-2037) may have large accretion rates because 
of their more massive disks, and moreover, strong silicates due to heating of larger disk 
areas to the temperatures required for producing silicate emission. The fact that 
we observe more mass in warm silicates in stars with higher accretion rates
(Figure \ref{Mcorr}) is likely to be due mostly to the higher accretion rates around 
earlier spectral type stars, since the total mass in silicates is mostly correlated to the
size of the area producing the silicate feature. As we will discuss later (Section \ref{mstarsevol}), 
the stars with very weak silicate emission are preferentially M-type, so effects such as 
illumination or inner disk evolution may be more important for them. But, in any case, 
the correlation also holds for K and earlier spectral types, so for a given spectral type, 
the silicate emission may reflect the influence of turbulence in the grain 
content of the disk atmosphere.

We do not observe any correlation between the crystalline grains or 
the degree of crystallinity, and any other property (age, accretion rate, SED slope, 
spectral type), except for the fact that the two objects with more than 15\% of the
mass in crystalline grains (Figure \ref{Cryscorr}) are the spectroscopic binaries 
82-272 (SB2) and 14-183 (SB1). The double-lined spectroscopic binary 82-272 (Paper IV)
has two spatially unresolved accreting components, and a very strong silicate feature
at 10 $\mu$m, plus relatively strong amorphous and crystalline features at
18-28 $\mu$m. The single-lined binary 14-183 has strong crystalline features at 10 $\mu$m as well,
but the third SB1, 12-2113, has a crystalline fraction similar to the rest of the stars,
which have all less than 10\% of mass in crystalline grains. As mentioned by Meeus et al. (2003),
the higher crystallinity fraction in close binaries could be a consequence of their
stronger activity and more turbulent disks. Nevertheless, the other effect of enhanced dust 
processing mentioned by Meeus et al. (2003), that binaries would have larger grains, is not 
observed here, although we must keep in mind that the silicate feature traces only the upper 
layers of the disk, and large grains can be removed from the disk atmosphere via dust settling. 

The lack of correlation between crystalline silicates and other disk and stellar
properties has been noted by Meeus et al. (2003) and Sargent et al. (2006), among others.
Additionally, we do not observe crystalline grains with sizes larger than 1.5 $\mu$m, in 
agreement with Bouwman et al. (2006), who suggest that this fact may be due to agglomeration 
of the (less abundant) crystalline silicates with the bulk of amorphous grains, producing 
mixed crystalline and amorphous aggregates, rather than large, fully crystalline grains. 
The lack of correlation between age and crystallinity suggests that crystallinization may 
occur already at the very early stages of star formation and disk evolution (Dullemond et al. 2006).

We have IRS spectra for two ``transition objects'' with accretion rates below 10$^{-12}$ M$_\odot$/yr 
(consistent with no accretion, confirmed via both the lack of an U band excess and
the narrow H$\alpha$ profiles at R$\sim$34,000 resolution, Paper IV), 
named 73-758 and 14-11, for a transitional disk with ongoing accretion (24-515),
and for nearly-transitional disk (a disk with nearly photospheric near-IR colors), 13-1250. 
The silicate feature is very weak in 14-11, and absent or extremely faint in 24-515, so only 
73-758 could be fitted. Although we cannot fit the silicate feature of some ``transition objects'', 
its weakness is consistent with the general trend showing that stars with lower 
accretion rates tend to have weaker silicate features, and supports the idea of dust evolution
and gas evolution happening somehow in parallel in accretion disks.

\subsection{Age, Inner Disk Evolution, and the Silicate Feature \label{mstarsevol}}

\begin{figure}
\epsscale{1.1}\plottwo{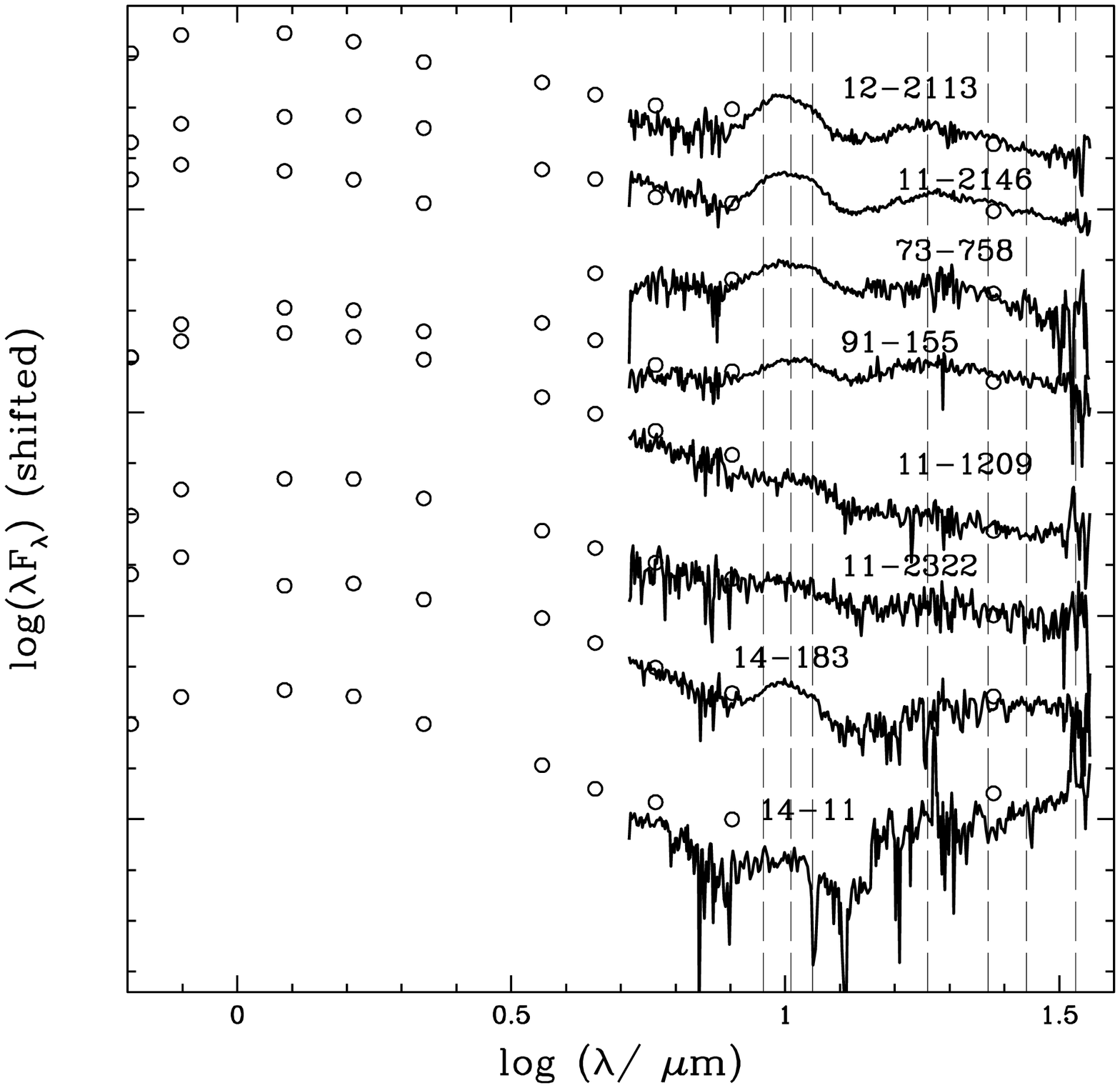}{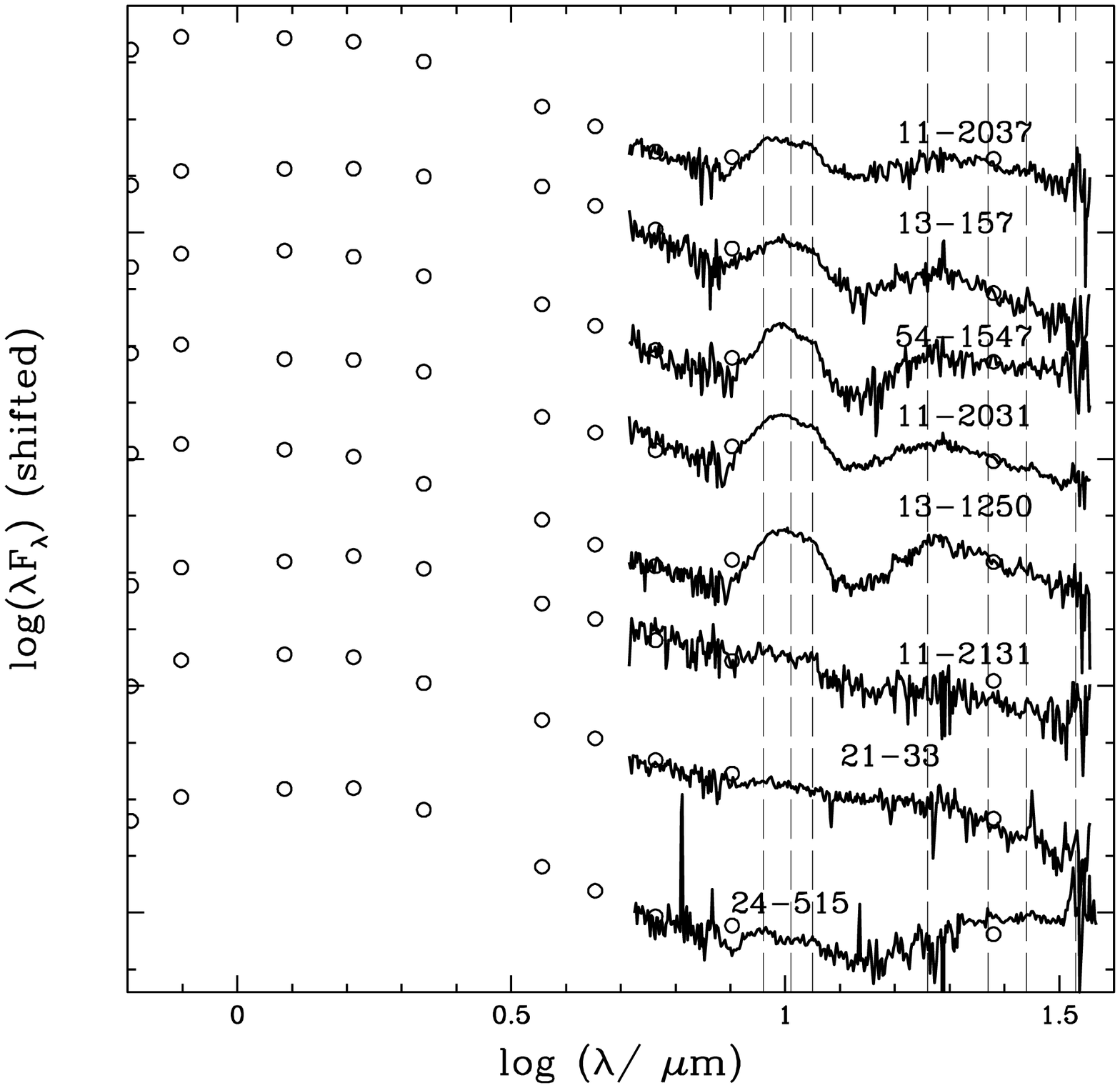}
\caption{IRS spectra for stars with ages 0-2 Myr (left) and 2-6 Myr (right).
The location of the main silicate peaks (9.7 and 18 $\mu$m peaks of the amorphous 
silicates of olivine stoichiometry, 10.2 and 23.5 $\mu$m peaks of the amorphous
silicates of forsterite stoichiometry, 11.3 $\mu$m crystalline forsterite, and the 28.5 and
33.5 $\mu$m crystalline bands) is marked by dashed lines; the SED from our photometry
studies are depicted as well. Spectra are shifted in flux to allow
the visualization of them all, as absolute fluxes do not matter for our discussion.
 \label{age1}}
\epsscale{1}
\end{figure}

\begin{figure}
\epsscale{0.5}\plotone{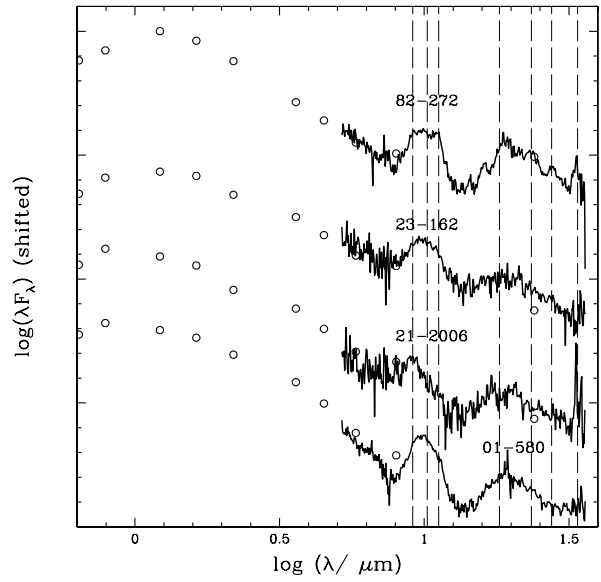}
\caption{IRS spectra for stars with ages over 6 Myr. Note that the age
for the G9 star 82-272 is uncertain. As in Figure \ref{age1},
we have shifted the spectra for display and we mark
the location of the main silicate peaks by dashed lines.\label{age2}}
\epsscale{1}
\end{figure}

\begin{figure}
\epsscale{1.1}\plottwo{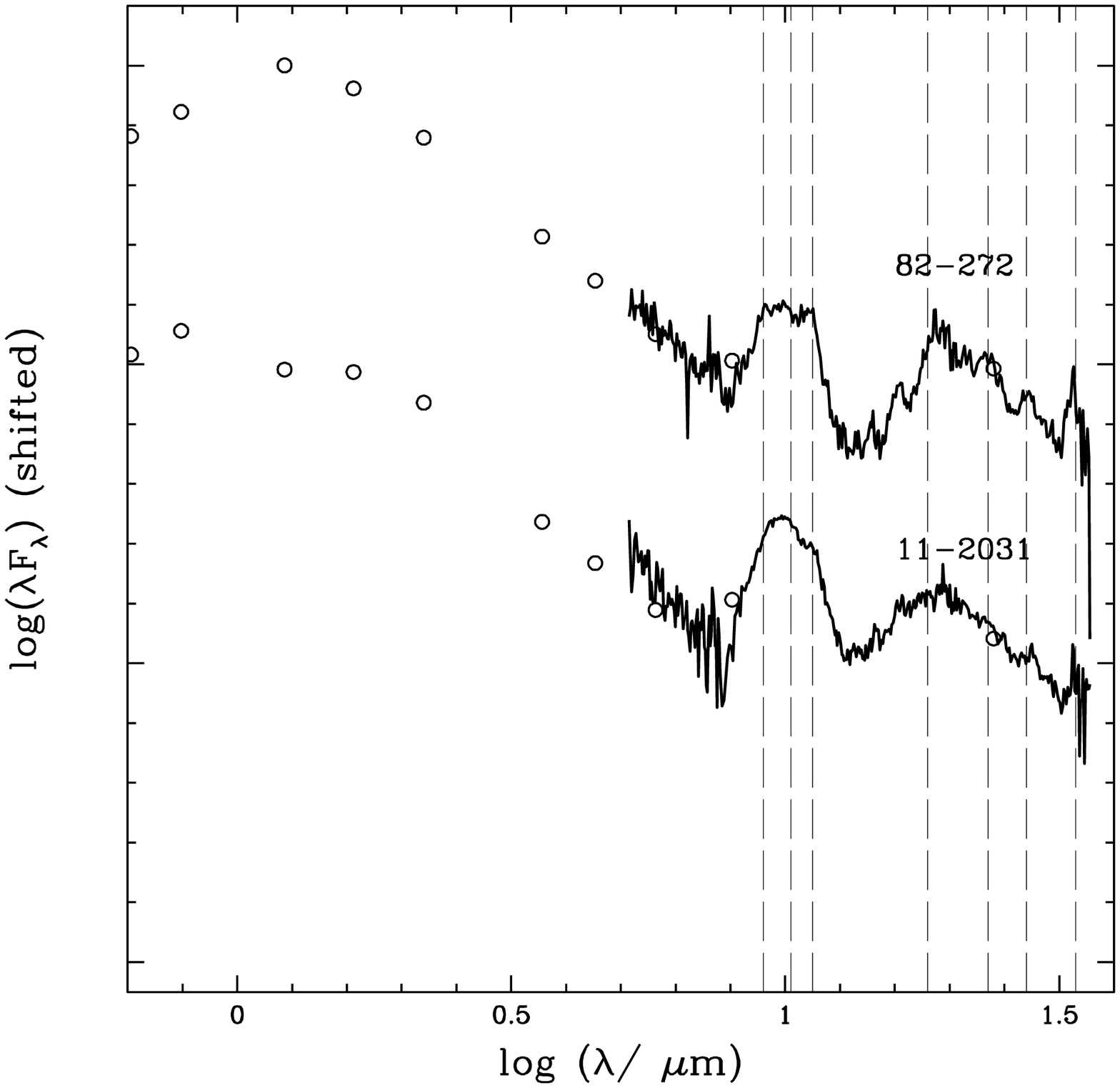}{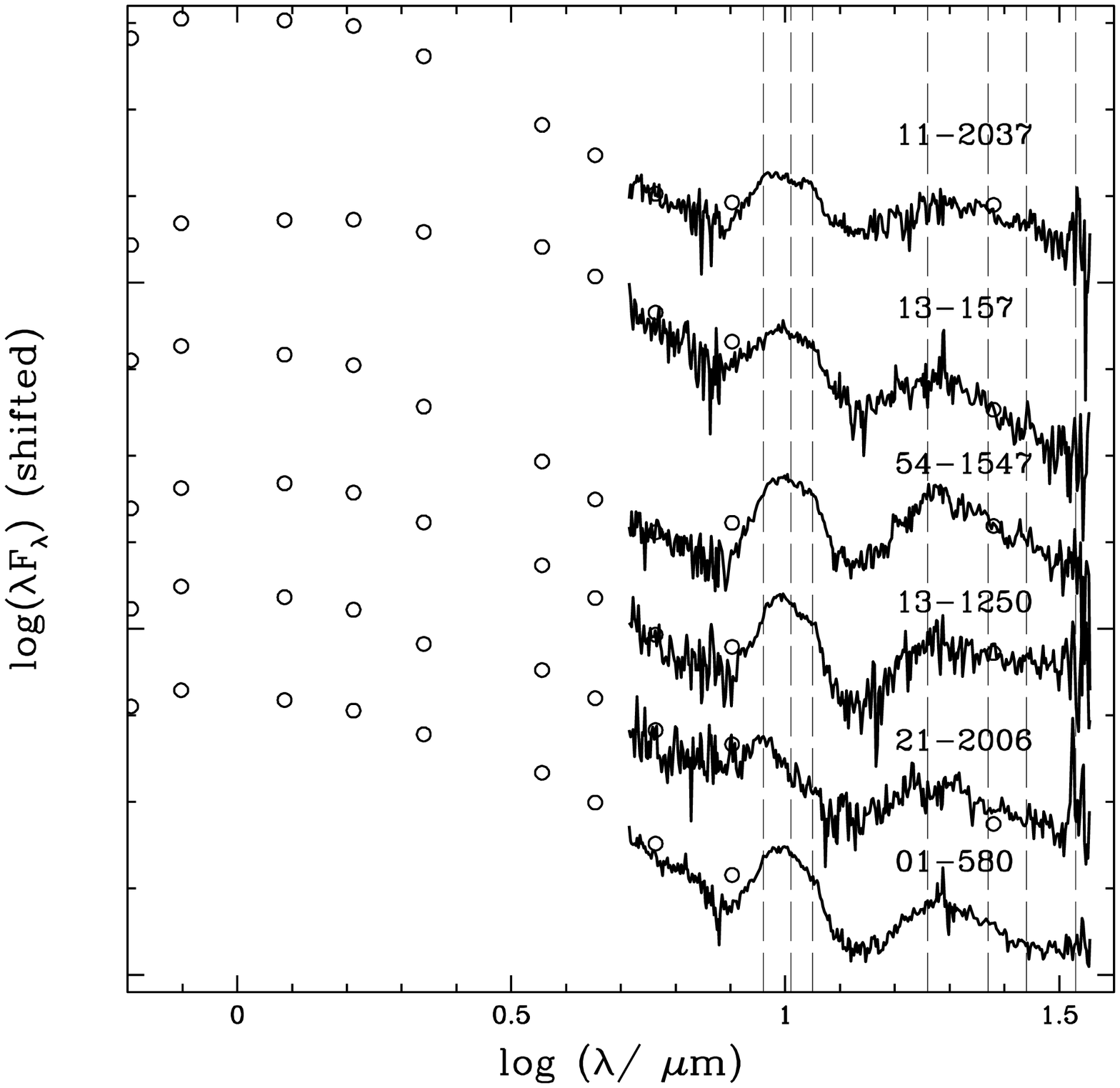}
\caption{IRS spectra for stars with spectral types G9-K2 (left) and K4-K5 (right).
 As in Figure \ref{age1}, we have shifted the spectra for display and we mark
the location of the main silicate peaks by dashed lines.
Among the earlier spectral types, the lack of silicate emission seems to be
highly infrequent.
 \label{G9K2K4K5}}
\epsscale{1}
\end{figure}

\begin{figure}
\epsscale{1.1}\plottwo{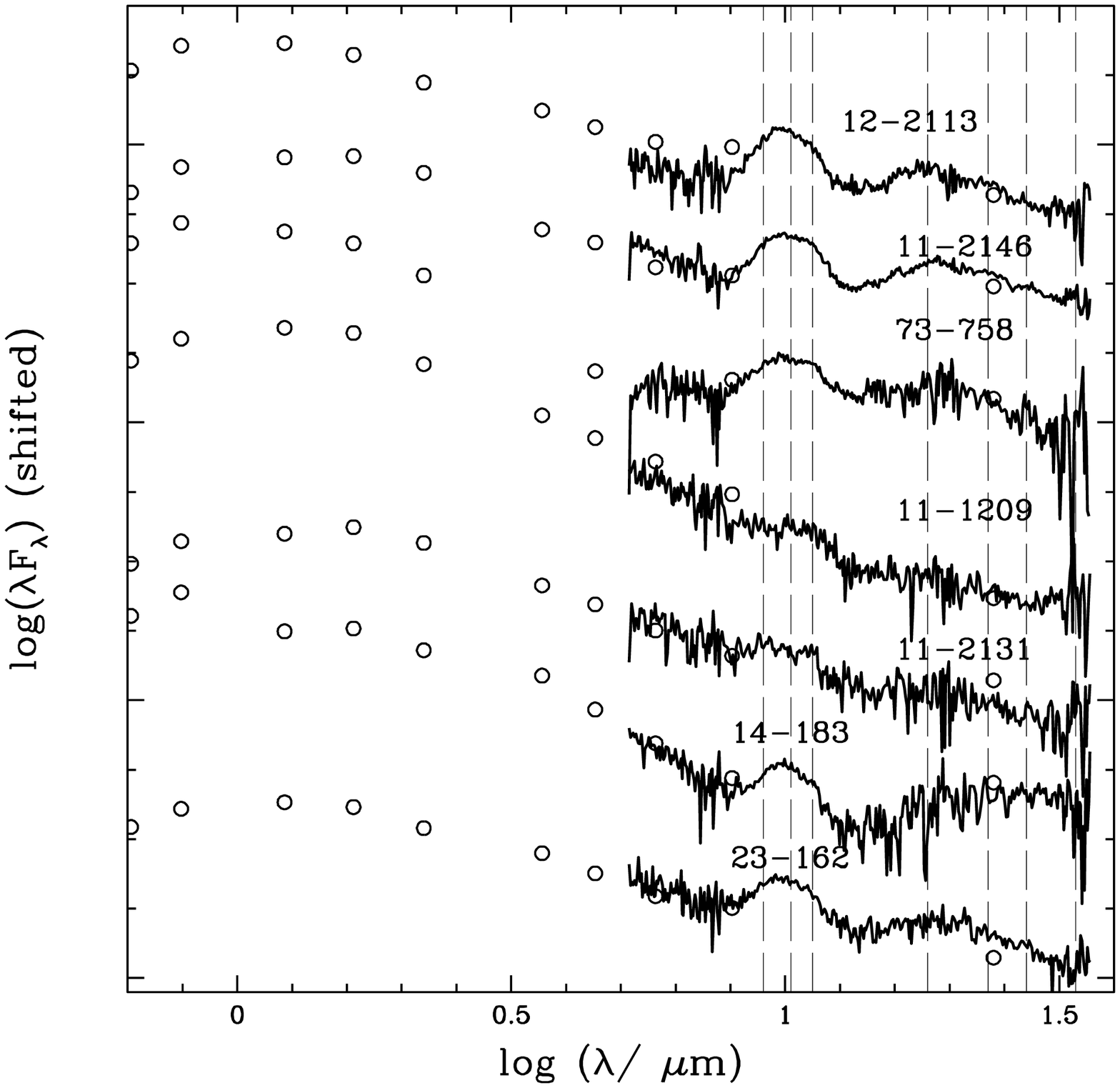}{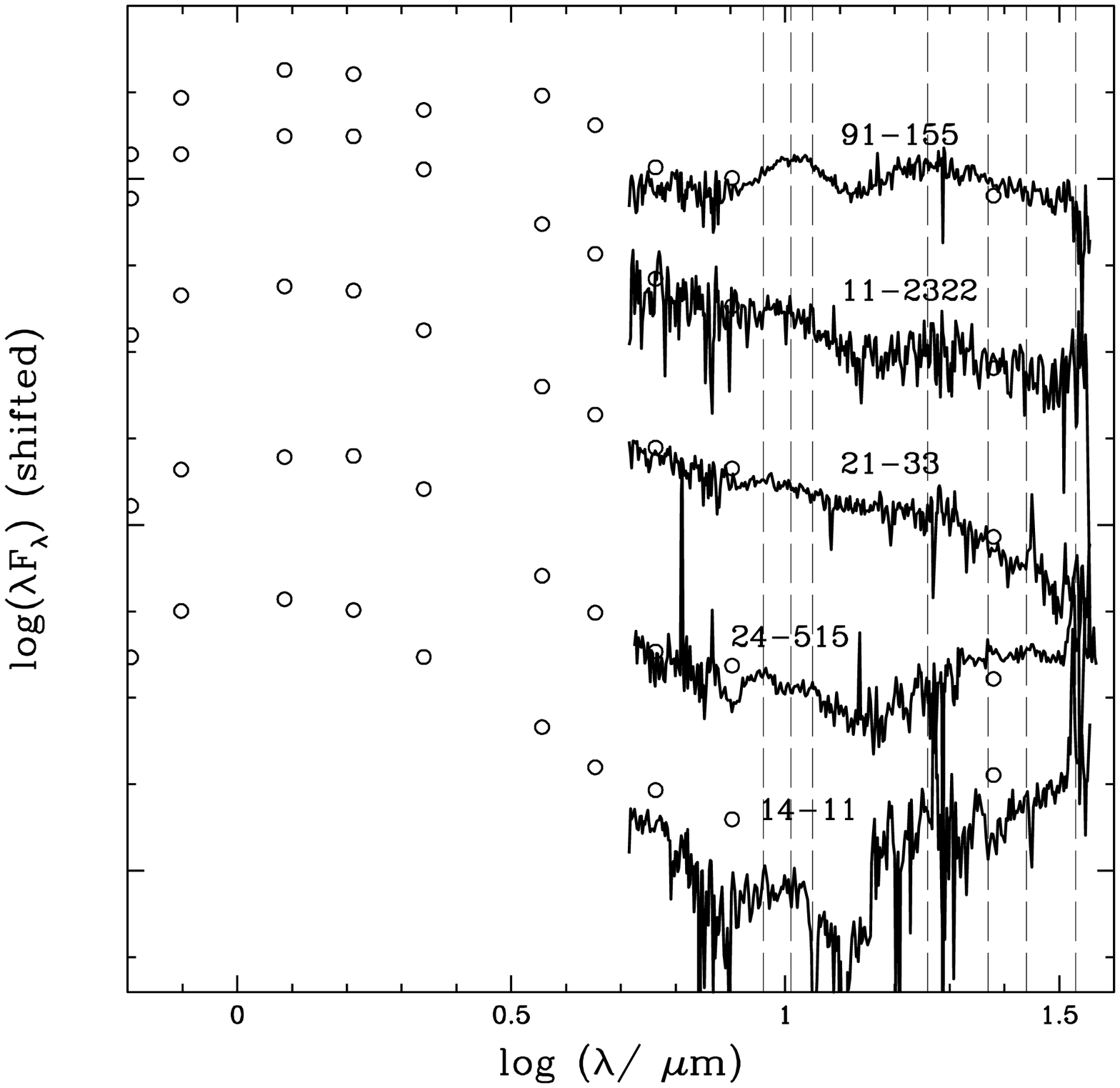}
\caption{IRS spectra for stars with spectral types K6-K7 (left) and M0-M2.5 (right). 
As in Figure \ref{age1}, we have shifted the spectra for display and we mark
the location of the main silicate peaks by dashed lines.
Stars with later spectral types seem to display a lack of silicate emission more
frequently (see discussion in text).
 \label{K6K7M0M2}}
\epsscale{1}
\end{figure}

In order to complete our understanding of the differences in silicate emission due to age
and spectral type in these stars, we have analyzed the IRS spectra grouping them in classes
according to similar ages and similar spectral types.  Figures \ref{age1} and \ref{age2}
display the younger population (0-2 Myr, all but 91-155 associated with the young
globule population, see Paper II), the average age in Tr 37 (2-6 Myr), and the older group 
members (over 6 Myr, including the  only accretor in NGC 7160, 01-580, aged 9 Myr).
Figures \ref{G9K2K4K5} and \ref{K6K7M0M2} display the spectra in groups with similar spectral
types (G9-K2, K4-K5, K6-K7, and M0-M2.5), containing (except the first one)
about the same number of stars in each group.

As for the age classification, we observe a large variety in the silicate emission morphology
and strength. Individual ages of young stars must be handled with care, as some errors may occur 
due to the presence of variability and the problems defining the birthline (affecting mostly 
G stars, see Hartmann 2003). Nevertheless, studies including variability effects (Burningham 
et al. 2005) have revealed that ages derived from the V vs. V-I diagram
are not much affected by variability, so that the main differences observed in these
color-magnitude plots are caused by real age differences. Therefore, as discussed in
Paper II, our ages are likely to be accurate up to $\sim$1 Myr, except for the G stars (the star 
82-272 in our sample), for which the uncertainties are larger. As noted in Section 
\ref{lowmass}, we observe that half of the stars with atmospheric silicates dominated by 
amorphous small grains are among the older group with ages over 6 Myr. This is the case of the 
NGC 7160 star 01-580, and of the 7 Myr-old 23-162. There are only 2 younger stars (among 16 
with ages $<$6 Myr), 12-2113 (a SB1, aged 1 Myr), and 13-157 (aged 2 Myr), which contain
mostly unprocessed small amorphous silicates (with average sizes $<$2$\mu$m) similar to 01-580 and
23-162. Despite the low-number statistics, the proportion of stars with pristine silicate features 
seems higher for older stars, which could be an additional indication of dust settling removing 
the large grains from the upper layers of the disks with time, in agreement with the anti-correlation 
between size and age described in Section \ref{lowmass}.

The classification according to the spectral types shows also interesting
results that escaped our analysis from the silicate fitting, given that 7 of
the low-mass spectra with very weak silicate features could not be properly fitted. 
Observing Figures \ref{G9K2K4K5} and \ref{K6K7M0M2}, we can see that, despite the large variety 
of silicate features present in all the spectral type groups, all the spectra with 
absent or nearly absent silicate features belong to the later spectral types, being
2 out of the 7 spectra with K6-K7 types, and 4 of the 5 spectra with
M0-M2.5 types. Among the 8 spectra with spectral types earlier than K6, we
find only one example of small but certainly well-detectable silicate feature, 21-2006.
The peak of the flux over the normalized continuum of the K5 star 21-2006 is 1.7$\pm$0.1,
larger the peak fluxes of stars like 12-1209, 11-2322, 14-11, 11-2131, 24-515, and 21-33, ranging from  
1.2 to 1.5$\pm$0.1. A KS test comparing the normalized peak flux of stars with spectral 
types earlier than K6, and K6 and later, gives a probability of 0.03 (3\%) that both samples come
from a similar distribution. Taking into account that we have 8 spectra earlier than K6, 
and 11 spectra with types K6 or later, and that they were drawn randomly from our
member sample in Cep OB2, this difference has a statistical significance.
About half of the late-type stars have very weak silicate features,
compared to roughly 10\% of the earlier-type stars, or said in a different way,
nearly 90\% of the stars with very weak or absent silicate feature in our 
sample are K6 or later. 

If we examine the sample of TTS studied by Honda et al. (2006), we find the same behavior:
there is only 1 star with very weak silicate feature among their 8 stars with spectral types
G2 to K5, but there are 6 among their 22 stars with spectral types K6 to M4 (all of them M-type, 
except for one K6 star), so $\sim$85\% of their stars with weak silicates are K6 or later. 
Although our sample seems richer in stars with weak silicate features (about 50\% of the later 
spectral types, compared to 30\% in Honda et al. 2006), the trend is the same: Stars with very 
weak or undetectable silicate feature are 3-5 times more frequent among the spectral types 
later than K6. Honda et al. (2006) relates the lack of silicate to strong accretion, which 
is obviously not the case in our sample (we moreover observe the opposite, as stars with stronger
accretion rates tend to have stronger silicate features). Nevertheless, all but one of the 
stars in their sample have H$\alpha$ normal for low-mass young stars accreting at typical rates, 
taking into account that M stars have larger H$\alpha$ EW than K stars for similar accretion rates 
(White \& Basri 2003). The star DN Tau (that they mention as having a large accretion rate causing 
the reduced silicate emission) was found to have an accretion rate of 3.5$\times$10$^{-9}$M$_\odot$/yr 
(Gullbring et al. 1998), so the presence of systematically higher accretion rates is not clear. 
Moreover, Green et al. (2006) has shown that even FU Ori objects have strong silicate features,
despite their extremely high accretion rates. Most of the stars in the sample of Honda et al. (2006) 
are Taurus stars, younger than ours, although they include a few from the TW Hya association (10 Myr age).
If inner disk evolution increases with time, as suggested in Paper III, we would then expect that the
number of stars with weak silicates increases with time. But even considering our 
sample, it is not easy at this point to derive a correlation between age and the proportion 
of stars with no silicates due to the few observations of disks aged 10 Myr. In the sample 
of stars investigated by Kessler-Silacci et al. (2005), which contains mostly younger stars as well, the 
M-type stars tend to have smaller peak fluxes than the higher-mass ones, although their sample has very 
few stars with weak silicate features, and some spectral types are largely uncertain. 

During some time, there was the open question of whether late M stars and brown 
dwarfs (BD) would show any silicate emission, precisely because of these requirements 
for having warm silicates. Nevertheless, the work of Apai et al. (2004, 2005) and 
Mohanty et al. (2004), among others, revealed that BD can have strong silicate features,
including strongly processed grains, and there are no doubts that early M-type stars as those in our sample
can have strong silicate emission (Bouwman et al. 2001, 2006; Sargent et al. 2006; among others).
But most of the systems studied contain younger stars (typically, 1-2 Myr),
and this may explain part of the difference, taking into account the changes in stellar luminosity
and the evolution of the inner disk with age. 

There are several factors that could be responsible for the very weak 
silicate features in M-type stars, and one of them could be the region where 
the silicate feature is produced, the warm disk atmosphere.
Due to the lower effective temperature of M-type stars, the silicate feature is produced in a
region much closer to the star (and thus smaller) in M stars than in early K and G stars. 
In order to achieve a temperature around 300 K, required for the silicate emission at 10 $\mu$m,
the small grains in the atmosphere need to be at distances closer than 0.7 to 1.5 AU for an M0 star, 
compared to 1.2 to 3 AU for a K5 star, approximately, and the area of the disk capable of
producing silicate emission is therefore about 4-5 times larger in a K5 star than in a M0.
As the luminosity of the star decreases with age, the warm silicate region moves inwards. 
The change of luminosity is remarkable for M-type stars, for which it may drop by an 
order of magnitude from the age of 1 to 10 Myr. For a typical M2 star, 
the warm disk region would change from being located at $\sim$0.4 to 1.2 AU, to
being reduced to $\sim$0.1 to 0.4 AU. Moreover, we observed that important
inner disk evolution has occurred by the ages of $\sim$ 4 Myr (Paper III), which can be 
inferred from the weaker near-IR emission, suggesting significant grain growth and/or 
dust settling. Therefore, we may not only be probing a different area of the disk, but 
a more evolved one as well, if dust evolution (dust coagulation/settling/opacity 
changes/gap opening) is faster and more efficient in the innermost disk.
Although we do not observe any significant difference between the inner disk emission,
including the presence of inner gaps, for the different spectral types from late G to 
early M (Paper IV), grain growth and/or dust settling would affect the silicate emission 
of M stars more than for K stars. K and G stars could have part of their inner disks depleted 
of small grains (as it may happen to the K4.5 star 13-1250, which has a very reduced near-IR 
excess that makes it very close to a ``transition object''), but the higher stellar luminosity
will be still enough to produce the temperatures required for silicate emission at larger 
distances from the disk, unlike in M-type stars. Therefore, the silicate feature could be a good 
instrument to probe the evolutionary stage of the innermost disk in late type stars, even before 
other effects like SED flattening in the near-IR become evident.

Although our study in Papers III and IV did not reveal any correlation between the
presence of inner gaps and the spectral type, we noted in Paper IV that 
other works have suggested that near-IR gaps (detectable at IRAC wavelengths)
tend to appear with a higher frequency around M-type stars (McCabe et al. 2006),
which could be a proof of photoevaporation removing the inner disk (the gaps around
earlier-type stars would be larger, and not detectable with IRAC). Cep OB2 is again 
the older region where a statistically significant study of transitional disks has been
made. Therefore, the fact that at 4 Myr we do find inner gaps in K stars similar to the
gaps found for M star by McCabe et al. 2006 at 1-2 Myr may as well indicate
that the evolution of M-type stars differs in some aspects from the evolution of
later spectral types. Nevertheless, since we find that half of the transition 
disks are still accreting, we need probably other effects instead of or in addition
to photoevaporation, in order to produce the inner gaps. The idea of M stars suffering 
evolutionary processes different from the processes affecting their 
higher mass counterparts could result in the 
differences observed here in the strength and shape of the silicate emission.
Lower silicate emission might as well suggest less turbulence in the 
lower-mass disks, and maybe different disk structure and evolution. If disks around
M-type stars differ in some aspects from more massive T Tauri disks, it could explain
the difficulties for M-type stars to form giant planets 
suggested by the models of Laughlin et al. (2004) and the lack of giant planets around
the less massive stars in the observations of Endl et al. (2006).
Despite the historical considerations that K and early M stars evolve similarly,
recent studies suggest differences in disk evolution around M stars, detectable as
lower disks fractions and higher percentages of ``flattened'' disks at a given age,
compared to K-type stars (Lada et al. 2006). In addition, accretion mechanisms operating
in M-type stars and very low-mass objects may also need to be reconsidered, as the depth of the
active layer of the disk and the presence of dead zones may be very different in
case of the very low-mass disks around late-type objects (Hartmann 2006). Further 
observations of very low-mass stars at different ages are strongly required in order 
to investigate all these hypotheses.

\subsection{Transitional and debris disks around intermediate- and high-mass stars \label{himass}}

The IRS data on high- and intermediate-mass stars in Tr 37 reveal that
two of these stars (CCDM+5734Ae/w and KUN-196) are more likely non-accreting
transitional protoplanetary disks with inner gaps, as the silicate 
feature at 10$\mu$m is present.
The 10$\mu$m feature suggests that, despite of the inner gaps, some small grains are
present at distances of the order of 10-50 AU for the B3+B5 star CCDM+5734Ae/w,
and from $\sim$5 to 15 AU for the B9 star KUN-196. KUN-314s may be either an evolved
Herbig Ae star (with a disk flatter than a typical Herbig Ae star), or a classical Ae star.
The rest of the stars (MVA-1312, MVA-447, MVA-468, BD+752356) do not show any silicate 
emission, so they are consistent with typical ``debris disks'' as stated in Paper III. In NGC 7160 
we do not find any Herbig Ae/Be star. The A7 star DG-481 shows a very clear
silicate feature, produced mainly by small amorphous grains that would be located at distances
between $\sim$2 and $\sim$6 AU, so the system is probably a non-accreting transitional protoplanetary
disk. The rest of stars (DG-39, DG-682 and DG-912\footnote{ The excess in DG-62 is
probably not significant, since it is consistent with photospheric levels within
the errors, so we cannot confirm the detection of a disk for this star.})
do not show any evidence of silicate emission (Figure \ref{sed4}), being consistent
with ``debris disks'' (Sylvester \& Mannings 2000). 

Analyzing the high- and intermediate-mass disks, we must mention the
difference in protoplanetary disk frequencies compared to low-mass stars, 
given that only 3-4 stars in Tr 37 (out of 58 members) and 1 star
in NGC 7160 (out of 66 members) present protoplanetary disks (with or 
without inner gaps). The lack of classical Herbig Ae/Be stars in the 
older cluster and the fact that the total number of stars with excesses is double in Tr 37
compared to NGC 7160, due to the presence of more potential protoplanetary
disks with large gaps in Tr 37, seems to point to disk evolution among these 
higher-mass stars (given that they contain 58 and 66 ABF members, respectively). 

The fraction of potential debris disks is similar in
both clusters, which would be also the expected result if the
signatures we observe correspond to the disks left over and stirred
by recently-formed planetary systems (Kenyon \& Bromley 2005).
In Table \ref{debris-tab} we list the excesses over the photospheric levels 
of all the high- and intermediate-mass stars. In agreement with Kenyon \& Bromley (2005), 
the observed excess luminosity over the photosphere of the potential debris disks 
is larger than the luminosity seen in older systems. This is consistent with the
predictions of the models of debris disk evolution, which include 
large bodies ($>$1000 km planets/planetesimals) in the inner disk
stirring the dust and replenishing the debris disk via collision events. The excess flux is
lower than the extremely bright HR 4796A disk, which is brighter than the predictions 
of typical young debris disk models.  

According to
Rieke et al. (2005), we would expect an excess at 24 $\mu$m of the order of 1.25 to 5 at 
ages under 24 Myr, although the fraction of stars with a strong excess ($\sim$5) would
be larger (11\% compared to 5\%) for objects under 10 Myr age. The decay of the luminosity
would be exponential, with a decay time of the order of hundreds of years.
According to Kenyon \& Bromley (2005), the predicted typical excesses over the photosphere
at 24$\mu$m for an A star, after planet formation in the innermost disk ($<$3 AU), 
would be of the order of log(F$_{24}$/F$_{24p}$)$\sim$1.25-0.8 at 1 Myr after the formation of the planets, 
0.8-0.5 after 4 Myr, and 0.2-0.6 after 12 Myr. Nevertheless, if the planets were to be 
formed at 3-20 AU, the excess could be of the order of $\sim$1.8 after 1 Myr, to decay to 
$\sim$1 after 10 Myr, and even 1.2 after 10 Myr if the planets are formed at 30-150 AU. 
These predictions are consistent with our observations, so all the disks but those of 
MVA-426 and KUN-314S are consistent with debris disks luminosities, as stated in Paper III.
Nevertheless, the presence of silicates in CCDM+5734A, KUN-196 and DG-481 suggest that they are
rather protoplanetary disks with large gaps. Although our sample is limited by the small 
number of stars, debris disks in Tr 37 tend to be brighter than their counterparts in NGC 7160 
(as we would expect from debris disk evolution), but here we should take into 
account that the stars in NGC 7160 have later spectral types.

However, it would be necessary to examine these disks at longer wavelengths (mm and submm) in
order to determine if their opacity is mostly due to second-generation dust
from collisions, and if they are optically thin, as the weak excess seem to suggest in all or
part of the cases (see Table \ref{debris-tab}). Longer wavelength observations
are specially important, given that in our low-mass
sample we have observed several examples of stars with optically thick protoplanetary
disks at 24 $\mu$m and extremely weak near-IR excess and silicate features. The higher temperatures
of the central stars and the larger areas subtended by the inner disks here 
would require much larger inner gaps in order to avoid silicate emission, making
it more plausible that they are at least partially optically thin. In that case, the stars 
detected in Tr 37 and NGC 7160 would be among the youngest debris disks detected, and some 
of the few ones with well determined ages (derived from the average ages of the low-mass 
clusters members), showing that debris disks are present already at ages as young as 4 Myr. 
The number of debris disks observed at 4 and 12 Myr is roughly similar (4 plus maybe 1 
versus 3 plus 1 uncertain, respectively). The emission over the photosphere of the presumed 
debris disks is variable in both clusters, as we would expect if initial and/or
environmental conditions (and maybe the characteristics of a planetary system
that it could have formed there) play an important role in disk evolution,
in addition to differences in age and mass (Kenyon \& Bromley 2005; Rieke et al. 2005).

\section{Conclusions \label{conclu}}

We presented IRS spectra of all the 13 high- and intermediate-mass stars with
IR excess, as well as a total of 20 low-mass stars selected among the stars with 
disks in the clusters Tr 37 and NGC 7160. Our sample contains stars that are on 
average older than the members of the best known regions, so we can examine with
more detail the effects of disk evolution in the shape and strength of the
silicate emission. The IRS data reveal a large variety of spectra and silicate 
features in both clusters, suggesting different stages in dust evolution, coagulation 
and presence of small-grains in the warm disk atmosphere, and probable dependence on
initial conditions and/or environment (Przygodda et al. 2003; Kessler-Silacci et al. 
2005; Dullemond \& Dominik 2005). We do not find any significant differences in grain 
size nor composition between the low-mass (G-M) stars and the only early-type star with 
strong 10$\mu$m feature, revealing that disks, silicate grains, and dust processing can 
be similar in systems with central stars of very different masses. Nevertheless, despite 
of the large variety of cases, there are significant trends that relate the disk structure, 
turbulence, the amount of settling, and the spectral type to the strength and 
characteristics of the silicate feature.

We observe clear evidence of the dependence of silicate emission with
spectral type. Objects with very weak silicate features are 3-5 times more
frequent among the M-type stars than among K-type. The differences
may be larger if we consider early K and G-type stars, as we do not observe 
any G or early K stars with very weak silicate features, although the total 
numbers are quite similar. One possibility to explain this could be 
inner disk evolution, given that the silicate feature is produced
closer to the star and in a smaller region in M-type star than in K 
and earlier types. Therefore, we would be probing different disk regions in K and M
stars, which may have suffered differential disk evolution (grain growth/dust settling).
The probed regions would change with time as the stellar luminosity decreases.
In that case, the silicate feature may be a very good indicator of the evolutionary
stage of the inner disk, even before than the effects of the grain growth and the
sedimentation produce detectable disk flattening. It could also indicate 
that evolutionary processes and timescales are somehow  different for the later 
spectral types, as some authors have suggested (based on 
disk fractions and accretion; Lada et al. 2006; Hartmann 2006), and
this could have effects in their capabilities to form planetary systems.

We find an apparently counter-intuitive correlation that
large average grain sizes ($\sim$5-6 $\mu$m) are found only among the 
youngest stars (aged 0-2 Myr) and the stars in our sample
that show very pristine silicate features (containing only small
amorphous grains) tend to be among the oldest ones ($>$6 Myr). 
This trend is consistent with dust settling
increasing with time, as large grains would drift faster towards the
middle plane, leaving only very small and unprocessed 
grains in the disk atmosphere at the later stages of disk evolution. 
Additionally, the fact that we observe already large grains
at very early ages suggests that considerable grain growth occurs
already within the very first years of disk formation and evolution, 
as suggested by Dullemond et al. (2006). Therefore,
one must be cautious before assigning an earlier evolutionary stage
to objects with silicate features due to smaller amorphous grains.

We observe that stars with low accretion rates have in general smaller
grains than the strong accretors, which could suggest that significant dust
settling has occurred in the more evolved, slow accreting, stars, and/or that
the presence of turbulence in disks would play an
important role replenishing the disk atmosphere from silicate grains that may otherwise
settle into the middle plane with time. This could be an additional evidence
of dust evolution occurring (or requiring) parallel to gas and accretion evolution (as
we mentioned in Paper IV), as it may be required that gas accretion decreases in
order to allow the dust settling to take place.

In the case of transitional disks with inner gaps, we find very weak features
among the 3 objects examined here, consistent with the trend of weak silicate
emission related to stars with very small or zero accretion. We find that
one of the transition objects, the M1.5 star 14-11, is very similar to CoKu Tau/4,
having no accretion and emission at $\lambda >$8 $\mu$m consistent with
a wall or rim surrounding an inner gap approximately 10 AU in size.

Regarding crystalline silicates, we confirm the observations of Meeus et al. (2003),
Bouwman et al. (2006), and Honda et al. (2006), finding  that there is no clear trend in the
degree of crystallinity versus spectral type, age, accretion rate nor
SED slope. We find that crystalline grains tend to be smaller than
amorphous ones, and that they are about the same size at any age or evolutionary
stage, maybe because of the difficulties to form entirely crystalline large grains 
(Bouwman et al. 2006). Crystalline grains may be therefore part of aggregates with 
the more abundant amorphous silicates. The degree of crystallinity is in general low, 
being always well below 20\%. The two objects with higher ($>$15\%) crystallinity fractions are two
spectroscopic binaries, which would be in agreement with Meeus et al. (2003),
who reported that binaries tend to have more crystalline disks, probably because
their disks are more turbulent. 

Among the high-mass stars, we can confirm a total of 7 potential debris disks objects 
with average ages 4 and 12 Myr. These potential debris disk lack the silicate emission and 
show mid-IR excess comparable to other typical debris disks, and consistent with the
excess predicted a few Myr after planet formation has occurred (Kenyon \& Bromley 2005;
Rieke et al. 2005). Further research at longer wavelengths (mm and submm) would be required 
to check whether these disks are debris disks, or they are in a more evolved phase within 
the transitional stage from Herbig AeBe disks to Vega-like systems.

The correlations between accretion rates/turbulence, spectral type and the silicate feature
observed here can fit within the same picture: With age, both the
gaseous and the dusty components of the disk evolve in a parallel way. The decrease
of the accretion rate (consistent with viscous disk evolution) is accompanied by the settling
of dust grains to the mid-plane, and the lower turbulence levels are not enough to
replenish the disk atmosphere with more small grains. The settling is stronger
in the innermost disk, and would extend outward with time, until strong dust coagulation 
results in the opening of opacity gaps and eventually, maybe in planet formation.

We want to acknowledge B. Mer\'{\i}n and C. Dullemond for interesting discussion.
We also want to thank the referee for the comments and suggestions to
clarify this paper.

\clearpage

\clearpage

\clearpage

\clearpage

\clearpage

\clearpage

\clearpage

\clearpage

\clearpage

\clearpage

\clearpage

\clearpage

\clearpage

\clearpage

\clearpage

\clearpage

\clearpage

\clearpage

\clearpage

\clearpage

\clearpage

\begin{deluxetable}{llcccccll}
\tabletypesize{\footnotesize}
\rotate
\tablenum{1}
\tablecolumns{9} 
\tablewidth{0pc} 
\tablecaption{Summary of Disk and Stellar Properties of the Observed Stars\label{target-tab}} 
\tablehead{
\colhead{Cluster} & \colhead{ID} & \colhead{RA(2000)} & \colhead{DEC(2000)} & \colhead{Sp.Type} &\colhead{Age (Myr)}  
&\colhead{\.{M} (10$^{-8}$M$_\odot$/yr)} & \colhead{Class} & \colhead{Comments}}
\startdata
Tr 37 & 73-758  & 21:35:08.39 & 57:36:03.2  & K6.5 & 1.8 &   0 & TO  &    \\
Tr 37 & 11-2146 & 21:36:57.69 & 57:27:33.3  & K6 & 0.9 &   16.2-13.2 & II   &     \\
Tr 37 & 11-1209 & 21:36:58.55 & 57:23:26.1  & K6 & 1.0 &   $<$0.1 & II    &  10$\mu$m feature marginal  \\
Tr 37 & 11-2322 & 21:37:01.96 & 57:28:22.3  & M1 & 0.8 &  0.80 & II       &  10$\mu$m feature marginal \\
Tr 37 & 11-2037 & 21:37:07.07 & 57:27:01.1  & K4.5 & 2.5   & 0.97-2.5 & II  & 28$\mu$m Forsterite     \\
Tr 37 & 14-11   & 21:37:10.30 & 57:30:19.1  & M1.5 & 0.7 &   0 & TO   &  $\sim$10 AU gap, BB$\sim$70 K, 10$\mu$m feature marginal \\
Tr 37 & 11-2131 & 21:37:12.19 & 57:27:26.6  & K6.5 & 2.3 &   1.1 & II   &  28$\mu$m Forsterite, 10$\mu$m feature marginal   \\
Tr 37 & 11-2031 & 21:37:15.95 & 57:26:59.5  & K2 & 2.5 &   1.6 & II   &     \\
Tr 37 & 14-183  & 21:37:38.53 & 57:31:41.7  & K6 & 0.9 &   $<$0.1 & II,SB1:  &      \\
Tr 37 & 82-272  & 21:38:03.53 & 57:41:35.4  & G9 & 10.5 &   23.9 & II+II,SB2  & 18$\mu$m Silica, 23-28$\mu$m Forsterite     \\
Tr 37 & 12-2113 & 21:38:27.47 & 57:27:21.3  & K6 & 1.1 &  6 & II,SB1  & \\
Tr 37 & 13-157  & 21:38:28.12 & 57:30:47.1  & K5.5 & 2.4 &   1.6 & II  &     \\
Tr 37 & 91-155  & 21:38:34.75 & 57:41:28.0  & M2.5 & 1.7 &   $\dots$ & II &       \\
Tr 37 & 54-1547 & 21:38:44.48 & 57:18:09.6  & K5.5 & 5.7 &   0.212 & II    &   \\
Tr 37 & 13-1250 & 21:39:12.17 & 57:36:16.9  & K4.5 & 3.3 &   0.10 & TO/II &  Small excess in near-IR, strong settling/growth    \\
Tr 37 & 24-515  & 21:39:34.10 & 57:33:32.1  & M0.5 & 2.8 &   $<$0.1 & TO   & 10$\mu$m feature marginal    \\
Tr 37 & 21-33   & 21:39:35.64 & 57:18:22.3  & M0 & 4.0 &   $<$0.1 & II   &  10$\mu$m feature marginal  \\
Tr 37 & 21-2006 & 21:40:13.91 & 57:28:48.5  & K5 & 6.1 &  0.09 & II & \\
Tr 37 & 23-162  & 21:40:44.52 & 57:31:31.9  & K7 & 6.6 &  $<$0.1 & II &       \\
NGC 7160 & 01-580  & 21:53:37.07 & 62:28:47.2  & K4.5 & 8.7 &  --- & II & \\
\\
Tr 37 & MVA-1312 & 21:35:56.82 & 57:20:53.1   & B4 & 4(*) &  0 & D & $\sim$300 AU gap, BB$\sim$100 K\\
Tr 37 & KUN-314s & 21:36:14.45 & 57:21:32.5   & A: & 4(*) &   --- & TO :/cBe: & Featureless\\
Tr 37 & CCDM+5734A & 21:37:40.93 & 57:33:37.5 & B3+B5 &   4(*) & 0 & TO, VB & 18$\mu$m Silica, Gap with small grains within 10-50 AU\\
Tr 37 & MVA-468 & 21:38:08.18  & 57:31:26.7   & B7 & 4(*) & 0 & D & \\
Tr 37 & MVA-426 & 21:38:08.71 & 57:26:48.0    & B7 & 4(*) &   --- & HAeBe & PAH \\
Tr 37 & MVA-447 & 21:38:26.64 & 57:28:40.8    & F0 & 4(*) &  0 & D & \\
Tr 37 & BD+572356 & 21:38:30.31 & 57:46:26.5   & A4 & 4(*) &  0 & D & \\
Tr 37 & KUN-196   & 21:40:15.14 &  57:40:51.2  & B9 & 4(*) &   0 & TO: & 10$\mu$ feature marginal, small grains at $\sim$5-15 AU  \\ 
NGC 7160 & DG-481 & 21:52:21.10 & 62:45:03.7 & A7 & 12(*) &   0 & TO  & 10$\mu$ feature strong, small grains at $\sim$2-6 AU\\
NGC 7160 & DG-39  & 21:53:27.73 & 62:35:19.0  & A0 & 12(*) &  0 & D &\\
NGC 7160 & DG-682 & 21:53:45.03 & 62:36:55.2  & A2 & 12(*) &  0 & D & \\
NGC 7160 & DG-62  & 21:54:30.23 & 62:31:16.1  & F5 & 12(*) &   0 & D: & Excess not significant\\
NGC 7160 & DG-912 & 21:55:47.50 & 62:35:43.7  & F5.5 &  12(*) & 0 & D & \\
\enddata
\tablecomments{Summary of the properties of the stars observed with IRS. 
For the low-mass stars, ages are derived from the V vs. V-I diagram,
using the isochrones by Siess et al. 2000 (see Sicilia-Aguilar et al. 2005 for a detailed
description of spectral typing and age/mass calculation). 
Accretion rates are based on U band emission (see Sicilia-Aguilar et al. 2005,2006a,2006b)
or are upper limits from high-resolution H$\alpha$ spectra. We denote
with \.{M}=0 those stars with no U excess and H$\alpha$ profiles ( \.{M}$<$10$^{-12}$ M$_\odot$/yr,
Paper IV). The ``Class'' can be class II (II), class III (III), transition object (TO),
Herbig Ae/Be star (HAeBe), classical Be star (cBe) or debris disk (D). We also include notes about binarity (SB1= single-lined
spectroscopic binary; SB2=double-lined spectroscopic binary; VB=visual binary). 
The ``Comments'' column include notes about the IRS spectra, among others. 
(*) For the high- and intermediate-mass
stars, no individual ages nor masses are estimated, and we do not have any evidence of accretion. 
Uncertain values are denoted by ``:''.  }
\end{deluxetable}

\begin{deluxetable}{ll}
\tabletypesize{\scriptsize}
%\rotate
\tablenum{2}
\tablecolumns{2} 
\tablewidth{0pc} 
\tablecaption{Scale Factors for the Combination of Orders\label{scaling-tab}} 
\tablehead{ \colhead{ID} &  }
\startdata
73-758  &   No scaling \\
11-2146 &    SL x 1.20  \\
11-1209 & No scaling \\
11-2232 & No scaling \\
11-2037 &    No scaling, used LL bonus order from 19.40 to 21.00 $\mu$m \\
14-11  &   No scaling \\
11-2131  &   No scaling \\
11-2031  &   No scaling \\
14-183  &   No scaling, used LL bonus order from 19.40 to 21.00 $\mu$m \\
82-272  &   No scaling \\
12-2113 &   No scaling \\
13-157  &   No scaling \\
91-155 &   No scaling \\
54-1547  &   No scaling, used LL bonus order from 19.40 to 21.00 $\mu$m \\
13-1250 &    No scaling \\
24-515 & No scaling \\
21-33 & No scaling \\
21-2006  &   No scaling, used bonus order from 7.53 to 8.10 $\mu$m (SL), from 19.40 to 21.00 $\mu$m (LL) \\
23-162  &   No scaling, used bonus order from 7.53 to 8.10 $\mu$m (SL), from 19.40 to 21.00 $\mu$m (LL) \\
\\
01-580 &  No scaling  \\
\\
MVA1312   &  No scaling, LL bonus order from 19.40 to 21.00 $\mu$m \\
KUN-314S  &   No scaling \\
CCDM+5734Ae &    SL x 1.1 \\
MVA-468 &  No scaling \\
MVA-426   &  No scaling \\
MVA-447 & No scaling \\
BD+572356 &  No scaling  \\
KUN-196 &    SL x 6.50, used bonus order from 7.53 to 8.10 $\mu$m (SL), from 19.40 to 21.00 $\mu$m (LL) \\
\\
DG-481 &    No scaling, used bonus order from 7.53 to 8.10 $\mu$m (SL), from 19.40 to 21.00 $\mu$m (LL) \\
DG-39 &    SL x 4.74, used bonus order from 7.53 to 8.10 $\mu$m (SL), from 19.40 to 21.00 $\mu$m (LL) \\
DG-682 &   SL x 2.00, used bonus order from 7.53 to 8.10 $\mu$m (SL), from 19.40 to 21.00 $\mu$m (LL)\\
DG-62  &   SL x 1.717    \\
DG-912 &   SL2 x 1.29,  SL1 x 1.79,  LL1 x 2.48 \\
\enddata
\tablecomments{ Scale factors in the data reduccion process. Where indicated, the SL or LL orders
were multiplied by a certain scaling factor in order to match the other orders. }
\end{deluxetable}

\begin{deluxetable}{lcccl}
\tabletypesize{\scriptsize}
%\rotate
\tablenum{3}
\tablecolumns{5} 
\tablewidth{0pc} 
\tablecaption{Dust species used to model the silicate feature\label{species-tab}} 
\tablehead{ \colhead{Species} & \colhead{State} & \colhead{Shape} & \colhead{Size ($\mu$m)} & \colhead{Reference}  }
\startdata
Silicate (Olivine stoichiometry) & Amorphous & Homogeneous & 0.1, 1.5, 6.0 & Dorschner et al. (1995) \\
Silicate (Pyroxene stoichiomtry) & Amorphous & Homogeneous & 0.1, 1.5, 6.0 & Dorschner et al. (1995) \\
Forsterite  & Crystalline & Inhomogeneous & 0.1, 1.5, 6.0 & Servoin \& Piriou (1973) \\
Enstatite   & Crystalline & Inhomogeneous & 0.1, 1.5, 6.0 & J\"{a}ger et al. (1998) \\
Silica      & Amorphous   & Inhomogeneous & 0.1, 1.5, 6.0 & Spitzer \& Kleinman (1960) \\
\enddata
\tablecomments{ Dust species used to reproduce the 10 $\mu$m silicate feature. See Bouwman et al. (2006)
for more details. All components were included in three different sizes. The shape refers to homogeneous 
spheres (with opacities derived from Mie theory) and to inhomogeneous spheres (using the distribution of hollow spheres
from Min et al. 2005).}
\end{deluxetable}

\begin{deluxetable}{lccccccccc}
\tabletypesize{\large}
\rotate
\tablenum{4}
\tablecolumns{10}
\tablewidth{0pc} 
\tablecaption{Summary of Dust Properties from Model Fitting\label{fit-table}} 
\tablehead{
 \colhead{ID} &\colhead{M$_{oli}$} &\colhead{M$_{pyr}$} &\colhead{M$_{fos}$} &\colhead{M$_{enst}$} &\colhead{M$_{sil}$} &\colhead{M$_{amo}$} &\colhead{M$_t$ (M$_{Moon}$)} & \colhead{T$_{cont}$} & \colhead{T$_{grain}$}\\
             &\colhead{S$_{oli}$} &\colhead{S$_{pyr}$} &\colhead{S$_{fos}$} &\colhead{S$_{enst}$} &\colhead{S$_{sil}$} &\colhead{S$_{amo}$} &\colhead{S$_{cry}$}          & \colhead{Comments}   }
\startdata
73-758  & 0.94$^{-0.24}_{+0.04}$  & 0.04$^{-0.04}_{+0.23}$  & 0.009$^{-0.004}_{+0.004}$  & 0.01$^{-0.01}_{+0.01}$  & --- & 0.98$^{-0.01}_{+0.01}$  & 0.0007$^{-0.0001}_{+0.00002}$ &  721$^{-139}_{+552}$  & 491$^{-37}_{+5}$   \\
	& 5.23$^{-0.21}_{+0.20}$  & ---                     & 0.13$^{-0.03}_{+0.40}$     & 1.26$^{-0.95}_{+0.24}$  & ---  	            & 5.21$^{-0.22}_{+0.23}$  & 0.62$^{-0.45}_{+0.38}$        &   \\
11-2146 & 0.35$^{-0.29}_{+0.55}$  & 0.60$^{-0.54}_{+0.29}$  & 0.02$^{-0.01}_{+0.01}$  & 0.02$^{-0.02}_{+0.02}$  & 0.01$^{-0.01}_{+0.02}$  & 0.95$^{-0.02}_{+0.02}$    & 0.26$^{-0.22}_{+0.49}$ &  1286$^{-496}_{+324}$  & 275$^{-92}_{+186}$    \\
	& 1.87$^{-1.78}_{+2.81}$  & 2.62$^{-2.08}_{+2.50}$  & 0.43$^{-0.31}_{+0.48}$  & 1.43$^{-0.87}_{+0.06}$  & 1.41$^{-0.68}_{+0.09}$  & 2.43$^{-1.87}_{+1.90}$    & 1.15$^{-0.16}_{+0.15}$ & 	  \\
11-2037  & 0.72$^{-0.36}_{+0.18}$  & 0.19$^{-0.17}_{+0.34}$  & 0.02$^{-0.005}_{+0.009}$  & 0.05$^{-0.01}_{+0.02}$  & 0.02$^{-0.01}_{+0.03}$  & 0.91$^{-0.04}_{+0.02}$  & 0.02$^{-0.01}_{+0.06}$ &  731$^{-93}_{+217}$  & 352$^{-121}_{+134}$  \\
	 & 5.28$^{-1.44}_{+0.41}$  & 0.38$^{-0.27}_{+3.63}$  & ---                       & 1.44$^{-0.15}_{+0.06}$  & 0.89$^{-0.55}_{+0.45}$  & 4.26$^{-2.22}_{+1.18}$  & 1.00$^{-0.16}_{+0.09}$ &	 \\
11-2031 & 0.73$^{-0.18}_{+0.11}$  & 0.20$^{-0.10}_{+0.15}$  & 0.05$^{-0.02}_{+0.02}$  & 0.01$^{-0.01}_{+0.01}$  & 0.01$^{-0.01}_{+0.01}$  & 0.94$^{-0.02}_{+0.02}$  & 0.02$^{-0.01}_{+0.03}$ &  723$^{-48}_{+56}$  & 334$^{-70}_{+99}$   \\
	& 3.08$^{-1.13}_{+0.53}$  & 0.11$^{-0.01}_{+0.51}$  & 1.07$^{-0.18}_{+0.10}$  & 0.75$^{-0.64}_{+0.63}$  & 1.40$^{-0.68}_{+0.10}$  & 2.51$^{-1.02}_{+0.64}$  & 1.09$^{-0.25}_{+0.12}$ & 	 \\
14-183  & 0.79$^{-0.29}_{+0.07}$  & 0.03$^{-0.03}_{+0.87}$  & 0.06$^{-0.02}_{+0.02}$   & 0.12$^{-0.05}_{+0.06}$  & ---    & 0.82$^{-0.07}_{+0.05}$  & 0.002$^{-0.001}_{+0.03}$ &  630$^{-29}_{+72}$  & 285$^{-37}_{+41}$   \\
	& 0.10$^{-0.00}_{+0.00}$  & 0.10$^{-0.00}_{+0.00}$  & ---                      & 0.11$^{-0.01}_{+0.11}$  & ---                       & 0.10$^{-0.00}_{+0.00}$  & 0.11$^{-0.01}_{+0.11}$   &  SB1	 \\
82-272  & 0.03$^{-0.03}_{+0.13}$  & 0.81$^{-0.13}_{+0.05}$  & 0.06$^{-0.01}_{+0.01}$  & 0.07$^{-0.02}_{+0.02}$  & 0.03$^{-0.02}_{+0.02}$  & 0.84$^{-0.04}_{+0.03}$  & 0.09$^{-0.05}_{+0.06}$ &  1332$^{-221}_{+185}$  & 198$^{-16}_{+21}$   \\
	& 1.13$^{-1.04}_{+4.14}$  & 0.46$^{-0.30}_{+0.45}$  & 0.14$^{-0.04}_{+0.25}$  & 1.31$^{-0.40}_{+0.17}$  & 0.63$^{-0.48}_{+0.45}$  & 0.46$^{-0.30}_{+0.45}$  & 0.73$^{-0.16}_{+0.12}$ &	SB2  \\
12-2113 & 0.61$^{-0.30}_{+0.14}$  & 0.37$^{-0.14}_{+0.31}$  & 0.01$^{-0.01}_{+0.01}$  & 0.01$^{-0.01}_{+0.03}$  & 0.004$^{-0.003}_{+0.01}$  & 0.98$^{-0.03}_{+0.01}$  & 0.03$^{-0.02}_{+0.21}$  &  628$^{-61}_{+119}$  & 291$^{-81}_{+55}$  \\
 	& 1.88$^{-1.43}_{+1.25}$  & 1.14$^{-0.97}_{+0.90}$  & 0.14$^{-0.04}_{+1.32}$  & 0.95$^{-0.85}_{+0.53}$  & 1.00$^{-0.80}_{+0.50}$  & 1.67$^{-1.36}_{+1.07}$  & 0.52$^{-0.39}_{+0.67}$  &	 SB1	 \\
13-157  & 0.00$^{-0.00}_{+0.00}$  & 0.98$^{-0.02}_{+0.01}$  & 0.004$^{-0.004}_{+0.006}$  & 0.01$^{-0.01}_{+0.02}$  & 0.00$^{-0.00}_{+0.01}$  & 0.98$^{-0.02}_{+0.01}$  & 0.36$^{-0.04}_{+0.06}$ &  1291$^{-138}_{+153}$  & 151$^{-3}_{+2}$   \\
	& ---                     & 0.10$^{-0.00}_{+0.00}$  & 0.10$^{-0.003}_{+0.10}$     & 0.10$^{-0.00}_{+0.00}$  & 0.28$^{-0.18}_{+0.99}$  & 0.10$^{-0.00}_{+0.00}$  & 0.15$^{-0.05}_{+0.80}$ &	 \\
91-155  & 0.16$^{-0.07}_{+0.18}$  & 0.81$^{-0.18}_{+0.07}$  & 0.00$^{-0.00}_{+0.00}$  & 0.02$^{-0.01}_{+0.01}$  & 0.002$^{-0.002}_{+0.002}$  & 0.97$^{-0.01}_{+0.01}$  & 0.05$^{-0.04}_{+2.75}$ & 944$^{-180}_{+282}$  & 262$^{-39}_{+25}$   \\
	& 0.97$^{-0.71}_{+1.26}$  & 5.54$^{-1.85}_{+0.42}$  & ---                     & 1.42$^{-0.56}_{+0.07}$  & 0.57$^{-0.66}_{+0.57}$  & 4.91$^{-1.26}_{+0.29}$  & 1.21$^{-0.41}_{+0.17}$ &		 \\
54-1547 & 0.74$^{-0.10}_{+0.08}$  & 0.14$^{-0.10}_{+0.14}$  & 0.05$^{-0.02}_{+0.02}$  & 0.05$^{-0.03}_{+0.04}$  & 0.02$^{-0.01}_{+0.02}$  & 0.88$^{-0.05}_{+0.04}$  & 0.0002$^{-0.001}_{+0.0003}$ &  921$^{-220}_{+350}$  & 340$^{-67}_{+110}$   \\
	& 0.49$^{-0.38}_{+1.91}$  & 0.48$^{-0.37}_{+3.02}$  & 0.28$^{-0.17}_{+0.34}$  & 1.12$^{-0.67}_{+0.33}$  & 1.47$^{-1.93}_{+0.03}$  & 0.54$^{-0.42}_{+2.01}$  & 0.82$^{-0.27}_{+0.21}$ &		 \\
13-1250 & 0.58$^{-0.48}_{+0.33}$  & 0.33$^{-0.31}_{+0.46}$  & 0.02$^{-0.01}_{+0.01}$  & 0.06$^{-0.02}_{+0.02}$  & 0.006$^{-0.001}_{+0.006}$  & 0.91$^{-0.03}_{+0.03}$  & 0.03$^{-0.02}_{+0.05}$ &  728$^{-144}_{+395}$  & 365$^{-180}_{+130}$   \\
	& 2.53$^{-2.47}_{+1.55}$  & 0.69$^{-0.60}_{+1.67}$  & 0.21$^{-0.11}_{+0.42}$  & 0.98$^{-0.77}_{+0.44}$  & 0.37$^{-0.28}_{+1.13}$     & 2.38$^{-2.31}_{+1.64}$  & 0.77$^{-0.56}_{+0.34}$ &		 \\
23-162  & 0.08$^{-0.08}_{+0.34}$  & 0.89$^{-0.33}_{+0.08}$  & 0.003$^{-0.003}_{+0.01}$  & 0.005$^{-0.004}_{+0.015}$  & 0.02$^{-0.01}_{+0.02}$  & 0.97$^{-0.02}_{+0.01}$  & 0.11$^{-0.09}_{+0.05}$ & 1010$^{-164}_{+405}$  & 190$^{-38}_{+131}$   \\
	& 1.03$^{-0.91}_{+3.94}$  & 1.17$^{-1.07}_{+3.83}$  & ---                       & 0.43$^{-0.33}_{+0.91}$     & 0.62$^{-0.46}_{+0.61}$  & 1.06$^{-0.96}_{+3.04}$  & 0.55$^{-0.39}_{+0.52}$ &		  \\
\\

01-580  & 0.54$^{-0.07}_{+0.08}$  & 0.42$^{-0.09}_{+0.06}$  & 0.02$^{-0.005}_{+0.01}$  & 0.02$^{-0.01}_{+0.01}$  & 0.002$^{-0.002}_{+0.01}$  & 0.96$^{-0.02}_{+0.01}$  & 0.005$^{-0.001}_{+0.001}$ &  1529$^{-57}_{+1}$  & 233$^{-8}_{+19}$   \\
	& 0.14$^{-0.04}_{+1.32}$  & 0.11$^{-0.01}_{+0.56}$  & 0.26$^{-0.12}_{+0.28}$   & 0.28$^{-0.18}_{+0.84}$  & 1.32$^{-1.49}_{+0.19}$  & 0.13$^{-0.03}_{+0.89}$  & 0.33$^{-0.18}_{+0.45}$    &		 \\
DG-481  & 0.36$^{-0.11}_{+0.18}$  & 0.58$^{-0.22}_{+0.13}$  & 0.03$^{-0.02}_{+0.02}$  & 0.01$^{-0.01}_{+0.03}$  & 0.01$^{-0.01}_{+0.02}$  & 0.94$^{-0.05}_{+0.03}$  & 0.03$^{-0.01}_{+0.01}$ &  1360$^{-178}_{+124}$  & 177$^{-9}_{+0}$   \\
	& 0.11$^{-0.01}_{+0.23}$  & 0.45$^{-0.34}_{+3.13}$  & 1.17$^{-0.39}_{+0.25}$  & 0.98$^{-0.84}_{+0.47}$  & 1.38$^{-0.94}_{+0.13}$  & 0.23$^{-0.13}_{+1.11}$  & 1.22$^{-0.37}_{+0.19}$ & 		\\
\enddata
\tablecomments{Parameters and properties of the silicate from the fitting of the 10 $\mu$m feature
(using the models described by Bouwman et al. 2006). Only the stars with good fits are included in the
list. The parameters are the fraction of mass in the different species (olivine, pyroxene, forsterite, enstatite, 
silica, total amorphous grains, in units of fraction of the total mass),
total mass in warm silicates (M$_t$, in M$_{Moon}$), continuum and grain temperature (T$_{cont}$,T$_{grain}$, in K),
and mass-weighted average size of the
grains of the different species (olivine, pyroxene, forsterite, enstatite, silica, total amorphous grains,
and total crystalline grains in $\mu$m). All the quantities include their 1-$\sigma$ errors. }
\end{deluxetable}

\begin{deluxetable}{lccccl}
\tabletypesize{\scriptsize}
\rotate
\tablenum{5}
\tablecolumns{6} 
\tablewidth{0pc} 
\tablecaption{Correlations between stellar, disk and silicate characteristics\label{correlation-tab}} 
\tablehead{
 \colhead{Parameters} & \colhead{Figure} & \colhead{N} &\colhead{r(low-mass)} & \colhead{r(all)}
&\colhead{Possible Implications/Comments} }
\startdata
F$_{8.6}$/F$_{9.8}$ vs. F$_{11.3}$/F$_{9.8}$ &  \ref{normfluxratio}  & 20 & 0.67 & 0.67 & Similar to Bouwman et al. (2006) \\
F$_{8.6}$/F$_{9.8}$ vs. F$_{8.6}$/F$_{11.3}$ &     & 20 & 0.37 & 0.37 & '' \\
F$_{8.6}$/F$_{9.8}$ vs. Spec. Type &     & 20 & 0.40 & 0.21 & Flux ratios correlation does not seem due to spectral type differences\\
F$_{8.6}$/F$_{11.3}$ vs. Spec. Type &     & 20 & 0.44 & 0.25 & '' \\
F$_{11.3}$/F$_{9.8}$ vs. Spec. Type &     & 20 & 0.04 & 0.02 & '' \\
F$_{8.6}$/F$_{9.8}$ vs. log(\.{M})   &     & 17 & 0.40  & 0.28  &  Flux ratios are not significantly affected by the accretion rates\\
F$_{8.6}$/F$_{11.3}$ vs. log(\.{M})   &    & 17 &  0.35 & 0.30  & '' \\
F$_{11.3}$/F$_{9.8}$ vs. log(\.{M})   &     & 17 & 0.25 &  0.13 & '' \\
F$_{8.6}$/F$_{9.8}$ vs. F$_{peak}$ &  \ref{normfluxratio}  & 20 & 0.77 & 0.52 &  Weak correlation between the peak flux and silicate feature shape \\
F$_{8.6}$/F$_{11.3}$ vs. F$_{peak}$ &    & 20 & 0.52 & 0.52 & '' \\
F$_{11.3}$/F$_{9.8}$ vs. F$_{peak}$ & \ref{normfluxratio}   & 20 & 0.36 & 0.36 & '' \\
F$_{peak}$ vs. Spec. Type   & \ref{normfluxratio}   & 20 & 0.61 & 0.22 & Higher peak in earlier stars (illumination? disk area?) but DG-481 is out of trend\\
F$_{peak}$ vs \.{M}  &    &   17 & 0.13  & 0.15  &  The accretion rate does not affect the silicate strength. \\
F$_{peak}$ vs log(\.{M})  &    & 17 & 0.07  & 0.17  & '' \\
F$_{peak}$ vs. H-K  &  \ref{peakflux}  & 20 & 0.04 & 0.15 & No correlation peak flux - inner disk brightness \\
F$_{peak}$ vs H-[8] &      & 19 & 0.22 & 0.26  & The dependence of the silicate peak flux with the SED shape is very weak or none\\
F$_{peak}$ vs. H-[24] & \ref{peakflux}   & 20 & 0.16 & 0.14  & '' \\
F$_{peak}$ vs H-[31]  &    & 20 & 0.35  & 0.32  & '' \\
F$_{peak}$ vs [8]-[24]  &     & 19 & 0.24  & 0.30  & '' \\
F$_{peak}$ vs [24]-[31] & \ref{peakflux}     & 19 & 0.12  & 0.15  & '' \\
F$_{peak}$ vs. Age   &  \ref{normfluxratio}  & 20 & 0.46 & 0.29 & Weak correlation peak flux - age for low-mass only \\
Mass (total) vs. Spect. Type      &  \ref{Mcorr}  & 13 & 0.05  &  0.13 & Total silicate mass does not depend on spectral type\\
Mass (total) vs. Age             &  \ref{Mcorr}    & 13 & 0.12  &  0.10 & Total silicate mass does not depend significantly on age \\
Log. Mass (total) vs. log( \.{M}) &  \ref{Mcorr}   & 12 & 0.78 &  0.45 &  More mass in silicates if \.{M} is larger (more turbulence?) \\
Mass (amorphous) vs. H-K &     &   13 & 0.62 &  0.60 & Brighter inner disk have more mass in silicates \\
Mass (total) vs. H-K     &  \ref{Mcorr}  &   13 & 0.62 &  0.57 & '' \\
Mass (amorphous) vs. Stellar Mass &    & 13 & 0.15 &  0.22  & No correlation between mass in silicates and stellar mass\\
Mass (total) vs. Stellar Mass     &    & 13 & 0.13 &  0.16  & '' \\
Mass (total) vs.  \.{M}  &    & 10 & 0.71 &  0.71  & More mass in silicates if \.{M} is larger (more turbulence?); (1) \\
Mass (total) vs. log( \.{M}) &    & 12 & 0.54 &  0.43 & '' \\
Mass (total) vs. Slope (H-[24])   &    & 12 & 0.13  &  0.18 & No correlation silicate/SED shape (slope)\\
Size vs. Spec. Type           &  \ref{Sizecorr}  & 13 & 0.39  & 0.41  & \\
Size (amorphous) vs. Age          &  \ref{Sizecorr}  & 13 & 0.48  &  0.52 & Large grains in atmosphere only at young ages (sedimentation?) \\
Size (crystalline) vs. Age        &  \ref{Sizecorr}  & 13 & 0.13  &  0.12 & No correlation between the size of crystalline grains and the age  \\
Size vs. log( \.{M})    &  \ref{Sizecorr}  & 12(11) & 0.39(15)  & 0.11(0.35)  & (2)\\
Size vs. \.{M}           &    & 12 & 0.22  & 0.15  & \\
Size vs. F$_{peak}$  &  \ref{Sizecorr}  & 13 & 0.50 & 0.38  &  Smaller grains for larger peak fluxes (larger illumination or more flared disks?)\\
Size (amorphous) vs. H-K  &  \ref{SizeIRcorr}   & 13 & 0.24 &  0.11 & No correlation inner disk brightness/silicate size \\
Size (amorphous) vs. H-[8]  &  \ref{SizeIRcorr}   & 13 &  0.47 & 0.48  & Smaller grains in more flared disks, sedimentation? \\
Size (amorphous) vs. H-[24]  &    & 13 & 0.48 &  0.49 & '' \\
Size (amorphous) vs. [8]-[24]  &  \ref{SizeIRcorr}   & 13 & 0.27  & 0.12  & \\
Size (amorphous) vs. [24]-[31]  & \ref{SizeIRcorr}    &  13 & 0.48  & 0.48  & \\
Crystal. Fraction vs. Spec. Type   & \ref{Cryscorr}   & 13 & 0.50  & 0.23  & No clear dependence of crystallization on spectral type \\
Crystal. Fraction vs. Age  & \ref{Cryscorr}   &  13 & 0.27 & 0.19  & Crystallization does not change with age. Early formation? \\
Crystal. Fraction vs. log( \.{M}) &  \ref{Cryscorr}  & 13 & 0.18  & 0.03  & Crystallinity fraction does not correlate with other properties \\
Crystal. Fraction vs. H-K  &  \ref{Cryscorr}  & 13 & 0.05  & 0.06  & ''\\
 \.{M} vs. Age            &    & 11(12) & 0.53(0.46)  & 0.41(0.19)  & Accretion rate tends to decrease with time but there is large scatter (3) \\
log( \.{M}) vs. Age      &     & 11(12) & 0.46(0.18)  & 0.19(0.20)  & '' \\
\enddata
\tablecomments{Possible correlations and lacks of correlation between stellar, disk, 
and silicate properties investigated. If the correlation is displayed in any of the figures,
the figure number is given.
The significance of each checked relation is given by the correlation coefficient
r, (including only low-mass stars, 'r(low-mass)', or low- and high-mass
stars, 'r(all)'; note that the only high-mass star with measurable silicate is DG-481).
Values of r = 1 would indicate a perfect correlation, and r = 0 characterizes completely
independent quantities.
Note that all these relations are valid only within the limit of our data (for Class II
objects with or without inner gaps, moderate to very low accretion rates, ages 1-12 Myr).
The number of data points (N) includes only low-mass stars, add 1 if DG-481 is included.
The accretion rates include upper limits (see Paper IV), and we assume
\.{M}=10$^-12$ M$_\odot$/yr for DG-481 (given the lack of H$\alpha$ emission (Paper II).
Note that 'Mass' correspond to the total mass in silicates in the warm disk atmosphere
(given that only the warm dust gas in the upper disk layers produces the silicate feature).
For all the stars, the mass fraction of crystalline silicates represents only a small
fraction of the total mass; therefore we consider mostly the total mass. Similarly, we
do not see any correlation between the sizes of crystalline grains and any other
stellar/disk properties, so the parameter 'Size' refers only to mass-weigted amorphous
grain size. (1)=Star 13-157 not included; (2)= Star 73-758 not included; (3)= Values in
parenthesis exclude G9 star 82-272, since G stars have larger accretion rates than K and M. }
\end{deluxetable}

\begin{deluxetable}{lccccccc}
\tabletypesize{\scriptsize}
\tabletypesize{\scriptsize}
\rotate
\tablenum{6}
\tablecolumns{8} 
\tablewidth{0pc} 
\tablecaption{Excesses over the Photosphere of High- and Intermediate-Mass Stars\label{debris-tab}} 
\tablehead{
 \colhead{ID}& \colhead{Spec. Type} &\colhead{Age (Myr)}  & \colhead{log(F$_{10}$/F$_{10p}$)} & \colhead{log(F$_{24}$/F$_{24p}$)} & \colhead{log(F$_{30}$/F$_{30p}$)}
&\colhead{log(F$_{70}$/F$_{70p}$)} &\colhead{Disk Type} }
\startdata
MVA-1312	& B4 & 4	& 0.0	& 1.3	& 2.1	& --- & D \\
KUN-314S	& A  & 4	& 1.0	& 1.4	& 1.4	& --- & P: \\
CCDM+5734A    &B3+B5 & 4	& 0.3	& 1.0	& 1.1 	& --- & P \\
MVA-468		& B7 & 4	& 0.0	& 0.7	& 1.0 	& --- & D \\
MVA-426		& B7 & 4	& 2.0	& 3.3	& 3.9	& 4.1 & P \\
MVA-447		& F0 & 4	& 0.0	& 1.0	& 1.1 	& --- & D \\
BD+572356	& A4 & 4	& 0.1	& 0.4	& 0.9	& --- & D \\
KUN-196		& B9 & 4	& 0.7	& 1.4	& 2.3	& --- & D \\
\\
DG-481		& A7 & 12	& 0.7	& 1.0	& 1.2	& --- & P \\
DG-39 		& A0 & 12	& 0.0	& 0.9	& 1.1	& --- & D \\
DG-682		& A2 & 12	& 0.0	& 1.4	& 1.4	& --- & D \\
DG-62		& F5 & 12	& --- 	& ---	& ---	& --- & --- \\
DG-912		& F5.5 & 12	& 0.0	& 0.6	& 1.3	& --- & D \\
\enddata
\tablecomments{Flux excess over the photosphere at different wavelengths (log(F$_\lambda$/F$_{\lambda p}$,
where F$_\lambda$ is the measured flux and F$_{\lambda p}$ is the photospheric
flux at the given wavelength) for
the high- and intermediate-mass with disks. Note that the ages given are the average ages in the
cluster. Errors are 10\%. No values are given for DG-62 due to the uncertainty of the measure and the
possibility that the excess is not real. The disk type denotes debris-like disks
(``D'') or protoplanetary-like disks (``P'') based on the presence of silicate
emission and in the magnitude of the excess. The colon (``:'') denotes uncertain
values. }
\end{deluxetable}


\begin{thebibliography}{}

\bibitem[Alexander et al.(2006a)]{alexander06a} Alexander, R., Clarke, C., Pringle, J, 2006a MNRAS, 369, 216

\bibitem[Alexander et al.(2006b)]{alexander06b}Alexander, R., Clarke, C., Pringle, J, 2006b MNRAS, 369, 229

\bibitem[Apai et al.(2004)]{apai04} Apai, D., Pascucci, I., Sterzik, M., van der Bliek, N., Bouwman, J., Dullemond, C., Henning, T., 2004, AA, 426, L53

\bibitem[Apai et al.(2005)]{apai05} Apai, D.; Pascucci, I.; Bouwman, J.; Natta, A.; Henning, Th.; Dullemond, C. P., 2005, Sci, 310, 834

\bibitem[Armitage et al.(1999)]{armitage99} Armitage, P., Clarke, C., Tout, C., 1999, MNRAS, 304, 425

\bibitem[Armitage et al.(2003)]{armitage03} Armitage, Ph., Clarke, C., Palla, F., 2003, MNRAS, 342, 1139

\bibitem[Bouwman et al.(2001)]{bouwman01} Bouwman, J., Meeus, G., de Koter, A., Hony, S., Dominik, C., Waters, L., 2001, AA, 375, 950

\bibitem[Bouwman et al.(2006)]{bouwman06} Bouwman, J. , Henning, Th., Hillenbrand, L., et al., 2006, submitted to ApJ 

\bibitem[Bowey \& Adamson(2001)]{bowey01} Bowey \& Adamson, 2001, MNRAS, 320, 131

\bibitem[Bryden et al.(2000)]{bryden00}Bryden, G, Rozyczka, M., Lin, D., Bodenheimer, P., 2000, 540, 1091

\bibitem[Burningham et al.(2005)]{burningham05} Burningham, B., Naylor, T., Littlefair, S., Jeffries, R., 2005, MNRAS, 363, 1389

\bibitem[Calvet et al.(1992)]{calvetall92} Calvet, N., Magris, G., Pati\~{n}o, A., D'Alessio, P., 1992, RMxAA 24, 29

\bibitem[Calvet et al.(2002)]{calvet02} Calvet, N., D'Alessio, P., Hartmann, L., Wilner, D., Walsh, A. \& Sitko, M., 2002, \apj, 568, 1008

\bibitem[Carpenter(1990)]{carpenter90} Carpenter, J., 2000, AJ, 120, 3139

\bibitem[Chiang \& Goldreich(1997)]{chiang97} Chiang, E., Goldreich, P. 1997, ApJ, 490, 368

\bibitem[Choi and Herbst(1996)]{choi96} Choi, P.I., \& Herbst, W., 1996, \aj, 111, 283

\bibitem[Clarke et al.(2001)]{clarke01} Clarke, C., Gendrin, A., \& Sotomayor, M., 2001, MNRAS 328, 485

\bibitem[Cohen et al.(2003)]{cohen03}Cohen, M.; Megeath, S. T.; Hammersley, P. L.; Martin-Luis, F.; Stauffer, J., 2003, AJ, 126, 1090

\bibitem[Contreras et al.(2002)]{contreras02}  Contreras, M.E., Sicilia-Aguilar, A., Muzerolle, J., Calvet, N., Berlind, P., Hartmann, L. 2002, AJ, 124, 1585

\bibitem[D'Alessio et al. (2005)]{dalessio05b} D'Alessio, P., Hartmann, L., Calvet, N., Franco-Hern\'{an}dez, R., and 10 more coauthors, 2005, \apj, 621, 461

\bibitem[Dorschner et al.(1995)]{dorschner95} Dorschner, J., Begemann, B., Henning, Th., J\"{a}ger, C., \& Mutschke, H., 1995, AA, 300, 503

\bibitem[Dullemond et al.(2001)]{dullemond01} Dullemond, C., Dominik, C., Natta, A., 2001, ApJ, 560, 957

\bibitem[Dullemond \& Dominik(2005)]{dullemond05} Dullemond, C., \& Dominik, C., 2005, AA, 434, 975 

\bibitem[Dullemond et al.(2006)]{dullemond06} Dullemond, C., Apai, D., Walch, S., 2006, ApJ in press 

\bibitem[Edwards et al.(1993)]{edwards93} Edwards, S., Strom, S., Hartigan, P., Strom, K., Hillenbrand, L., Herbst, W., Attridge, J., Merrill, K., Probst, R., Gatley, I., 1993, AJ, 106, 372

\bibitem[Forrest et al.(2004)]{forrest04} Forrest, W.J., Sargent, B., Furlan, E., \& 18 more coauthors, 2004, \apjs, 154, 443

\bibitem[Gullbring et al.(1998)]{gullbring98} Gullbring, E., Hartmann, L., Brice\~{n}o, C., Calvet, N., 1998, ApJ 492, 323

\bibitem[Haisch et al.(2001)]{haisch01} Haisch, K., Lada, E., \& Lada, C., 2001, \apj, 553,153

\bibitem[Hartmann et al.(1998)]{hartal98} Hartmann, L., Calvet, N., Gullbring, E. \& D'Alessio, P, 1998, \apj, 495, 385

\bibitem[Hartmann(2003)]{hartmann03} Hartmann, L., 2003, \apj, 585, 398

\bibitem[Hayashi et al.(1985)]{hay85} Hayashi, C., Nakazawa, K., \& Nakagawa, Y.\ 1985, in Protostars and Planets II, eds. D. Black \& M. Matthews(Tucson: University of Arizona Press), 1100

\bibitem[Henning et al.(1995)]{henning95} Henning, Th., Begemann, B., Mutschke, H., Dorschner, J., 1995, AA, 112, 143

\bibitem[Henning et al.(2005)]{henning05} Henning, T., Mutschke, H., J\"{a}ger, C., 2005, IAU Symposium, 231 

\bibitem[Higdon et al.(2004)]{higdon04}Higdon, S. J. U.; Devost, D.; Higdon, J. L.; Brandl, B. R.; Houck, J. R.; Hall, P.; Barry, D.; Charmandaris, V.; Smith, J. D. T.; Sloan, G. C.; Green, J., 2004, PASP, 116,975

\bibitem[Honda et al.(2003)]{honda03} Honda, M., Kataza, H., Okamoto, Y., Miyata, T., Yamashita, T., Sako, S., Takubo, S., Onaka, T., 2003, ApJ, 585, 59

\bibitem[Honda et al.(2006)]{honda06} Honda, M,, Kataza, H, Okamoto, Y., Yamashita, T., Min, M., Miyata, T., Sako, S., Fujiyoshi, T., Sakon, I., Onaka, T., 2006, ApJ,. 646, 1024

\bibitem[J\"{a}ger et al.(1994)]{jaeger94} J\"{a}ger, C., Mutschke, H., Begemann, B., Dorschner, J., Henning, Th., 1994, AA, 292, 641 

\bibitem[J\"{a}ger et al.(1998)]{jaeger98} J\"{a}ger, C., Mutschke, H., Dorschner, J., 1998, AA, 311, 291

\bibitem[Kenyon \& Bromley(2005)]{ken05} Kenyon, S.J., \& Bromley, B.C., 2005, AJ, 130, 1

\bibitem[Kenyon \& Hartmann(1987)]{ken87}  Kenyon, S.J. \& Hartmann, L., 1987, ApJ, 323, 714

\bibitem[Kenyon \& Hartmann(1995)]{ken95}  Kenyon, S.J. \& Hartmann, L., 1995, ApJS , 101, 117

\bibitem[Kessler-Silacci et al.(2005)]{kessler05}Kessler-Silacci, J., Hillenbrand, L., Blake, G., Meyer, M., 2005, ApJ, 622, 404 

\bibitem[Lada et al.(2000)]{lada00} Lada, C.J., Muench, A.A., Haisch, K.E., Lada, E.A., Alves, J.F., Tollestrup, E.V. \& Willner, S.P., 2000, \aj, 120, 3162

\bibitem[Lin and Papaloizou(1986)]{lin86} Lin, D.N.C., \& Papaloizou, J., 1986, \apj, 309, 846 
 
\bibitem[McCabe et al.(2006)]{mccabe06}  McCabe, C., Ghez, A., Prato, L, Duchene, G., Fisher, R., Telesco, C., 2006, ApJ 636, 932

\bibitem[Meeus et al.(2003)]{meeus03} Meeus, G., Sterzik, M., Bouwman, J., Natta, A., 2003, AA, 409, L25

\bibitem[Megeath et al.(2004)]{megeath04} Megeath, S.T., Allen, L., et al., 2004, ApJ, 154, 367

\bibitem[Meyer et al.(2004)]{meyer04} Meyer, M., Hillenbrand, L., et al., 2004, ApJ, 154, 422 

\bibitem[Min et al.(2005)]{min05} Min, M., Hoveiner, J.W., de Koter, A., 2005, AA, 432, 909

\bibitem[Mohanty et al.(2004)]{mohanty04} Mohanty, S., Jayawardhana, R., et al., 2004, ApJ, 609, L33

\bibitem[Muzerolle et al.(2000)]{muzerolle00} Muzerolle, J., Calvet, N., Brice\~{n}o, C., Hartmann, L. \& Hillenbrand, L., 2000, \apj, 535, L47

\bibitem[Patel et al.(1995)]{patel95}  Patel, N.A. , Goldsmith, P.F., Snell, R.L., Hezel, T. \& Xie, T., 1995, ApJ , 447, 721

\bibitem[Patel et al.(1998)]{patel98} Patel, N.A.,  Goldsmith, P.F., Heyer, M.H. \& Snell, R.L., 1998, ApJ , 507, 241

\bibitem[Platais et al.(1998)]{platais98}Platais, I,  Kozhurina-Platais, V., van Leeuwen, F., 1998, AJ, 116, 2423

\bibitem[Porras et al.(2003)]{porras03} Porras, A., Christopher, M., Allen, L., DiFrancesco, J., Megeath, S.T., Myers, P., 2003, AJ, 126, 1916 

\bibitem[Porter \& Rivinius(2003)]{porter03} Porter, J., \& Rivinius, T., 2003, PASP, 115, 1153

\bibitem[Przygodda et al.(2003)]{przygodda03} Przygodda, F., van Boekel, R., Abraham, P., Melnikov, S., Waters, L., Leinert, C., 2003, AA, 412, L43 

\bibitem[Quillen et al.(2004)]{quillen04} Quillen, A., Blackman, E., Frank, A., Varniere, P., 2004, \apj, 612, L137

\bibitem[Rieke et al.(2005)]{rieke05} Rieke, G.H., Su, K.Y.L., Stansberry, J.A., \& 9 more coauthors, 2005, ApJ, 620, 1010

\bibitem[Sargent et al.(2006)]{sargent06} Sargent, B., Forrest, W., D'Alessio, P., and 12 more coauthors, 2006, ApJ, 645, 395

\bibitem[Schr\"{a}pler \& Henning(2004)]{schraepler04} Schr\"{a}pler, R. \& Henning, Th., 2004, ApJ, 614, 960

\bibitem[Servoin \& Piriou(1973)]{servoin73} Servoin, J., Piriou, B., 1973, phys. stat.sol. 55, 677
 
\bibitem[Sicilia-Aguilar et al.(2004)]{sic04} Sicilia-Aguilar, A., Hartmann, L., Brice\~{n}o, C., Muzerolle, J., \& Calvet, N., 2004, \aj 128, 805, Paper I

\bibitem[Sicilia-Aguilar et al. (2005)]{sic05onc} Sicilia-Aguilar, A., Hartmann, L., Szentgyorgyi, A., Roll, J., 
Conroy, M., Calvet, N., Fabricant, D., \& Hern\'{a}ndez, J.,  Paper A, 2005, AJ, 129, 363 

\bibitem[Sicilia-Aguilar et al.(2005)]{sic05opt} Sicilia-Aguilar, A., Hartmann, L., Hern\'{a}ndez, J., Brice\~{n}o, C., Calvet, N., 2005, \aj 130, 188, Paper II

\bibitem[Sicilia-Aguilar et al.(2006)]{sic05ir} Sicilia-Aguilar, A., Hartmann, L., Calvet, N., Megeath, S.T., Muzerolle, J., Allen, L., D'Alessio, P., Mer\'{\i}n, B., Stauffer, J., Young, E., Lada, C., 2006, ApJ 638, 897

\bibitem[Skrutskie et al.(1990)]{skrutskie90} Skrutskie, M., Dutkevitch, D., Strom, S., Edwards, S., \& Strom, K., 1990, \aj, 99, 1187

\bibitem[Spitzer \& Kleinman(1960)]{spitzer60} Spitzer, W., \& Kleinman, D., 1960, Physical Review, 121, 1324

\bibitem[Strom et al.(1989)]{strom89} Strom, K., Wilkin, F., Strom, S., \& Seaman, R., 1989, \aj, 98, 1444

\bibitem[Sylvester \& Mannings(2000)]{sylvester00} Sylvester, R. \& Mannings, V., 2000, MNRAS, 313, 73

\bibitem[Uchida et al.(2004)]{uch04} Uchida, K.~I., et al. 2004, \apjs, 154, 439

\bibitem[Van Boekel et al.(2003)]{boekel03} Van Boekel, R., Waters, L., Dominik, C., Bouwman, J., de Koter, A., Dullemond, C., Paresce, F., 2003, AA, 400, L21 

\bibitem[Van Boekel et al.(2005)]{boekel05} Van Boekel, R., Min, M, Waters, L., de Koter, A., Dominik, C., van den Ancker, M., Bouwman, J., 2005, AA 437, 189

\bibitem[White \& Basri(2003)]{white03} White, R., \& Basri, G., 2003, \apj, 582, 1109


\end{thebibliography}
\end{document}